# [Giorgi Japaridze](#)

# A Survey of

# [Computability Logic](#)

გამოთვლადობის ლოგიკა

Логика вычислимости

可计算性逻辑

# (CoL)

Computability is one of the most interesting and fundamental concepts in mathematics and computer science, and it is natural to ask what logic it induces. This is where **Computability Logic** (**CoL**) comes in. It is a formal theory of computability in the same sense as classical logic is a formal theory of truth. In a broader and more proper sense, CoL is not just a particular theory but an ambitious and challenging program for redeveloping logic following the scheme "from truth to computability".

Under the approach of CoL, logical operators stand for operations on computational problems, formulas represent such problems, and their "truth" is seen as algorithmic solvability. In turn, computational problems --- understood in their most general, *interactive* sense --- are defined as games played by a machine against its environment, with "algorithmic solvability" meaning existence of a machine which wins the game against any possible behavior of the environment. With this semantics, CoL provides a systematic answer to the question "what can be computed?", just like classical logic is a systematic tool for telling what is true. Furthermore, as it happens, in positive cases "*what* can be computed" always allows itself to be replaced by "*how* can be computed", which makes CoL of potential interest in not only theoretical computer science, but many applied areas as well, including constructive applied theories, interactive knowledge base systems, resource oriented systems for planning and action, or declarative programming languages.

Currently CoL is still at an early stage of development, with open problems prevailing over answered questions. For this reason it offers plenty of research opportunities, with good chances of interesting findings, for those with interests in logic and its applications in computer science.

This article presents a survey of the subject: its philosophy and motivations, main concepts and most significant results obtained so far. No proofs of those result are included.

## Contents

# 1 The philosophy of CoL

## 1.1 Syntax vs. semantics

A starting philosophical point of CoL is that *syntax* --- the study of axiomatizations or other string-manipulation systems --- exclusively owes its right on existence to *semantics*, and is thus secondary to it. Logic is meant to be the most basic, general-purpose formal tool potentially usable by intelligent agents in successfully navigating the real life. And it is semantic that establishes that ultimate real-life meaning of logic. Syntax is important, yet it is so not in its own right but only as much as it serves a meaningful semantics, allowing us to realize the potential of that semantics in some systematic and perhaps convenient or efficient way. Not passing the test for soundness with respect to the underlying semantics would fully disqualify any syntax, no matter how otherwise appealing it is. Note --- disqualify the syntax and not the semantics. Why this is so hardly requires any explanation. Relying on an unsound syntax might result in wrong beliefs, misdiagnosed patients or crashed spaceships. An incomplete syntax, on the other hand, potentially means missing benefits that should not have been missed.

A separate question, of course, is what counts as a semantics. The model example of a semantics with a capital 'S' is that of classical logic. But in the logical literature this term often has a more generous meaning than what CoL is ready to settle for. As pointed out, CoL views logic as a universal-utility tool. So, a capital-'S'-semantics should be non-specific enough, and applicable to the world in general rather than some very special and artificially selected (worse yet, artificially created) fragment of it. Often what is called a semantics is just a special-purpose apparatus designed to help analyze a given syntactic construction rather than understand and navigate the outside world. The usage of Kripke models as a derivability test for intuitionistic formulas, or as a validity criterion in various systems of modal logic is an example. An attempt to see more than a technical, syntax-serving instrument in this type of lowercase-'s'-semantics might create a vicious circle: a deductive system $L$ under question is "right" because it derives exactly the formulas that are valid in a such and such Kripke semantics; and then it turns out that the reason why we are considering the such and such Kripke semantics is that ... it validates exactly what $L$ derives.

## 1.2 Why game semantics?

For CoL, a game are not just a game. It is a foundational mathematical concept on which a powerful enough logic (=semantics) should be based. This is so because, as noted, CoL sees logic as a "real-life navigational tool", and it is games that appear to offer the most comprehensive, coherent, natural, adequate and convenient mathematical models for the very essence of all "navigational" activities of agents: their interactions with the surrounding world. An *agent* and its *environment* translate into game-theoretic terms as two *players*; their *actions* as *moves*; *situation*s arising in the course of interaction as *positions*; and *success* or *failure* as *wins* or *losses*.

It is natural to require that the interaction strategies of the party that we have referred to as an "agent" be limited to *algorithmic* ones, allowing us to henceforth call that player a *machine*. This is a minimum condition that any non-esoteric game semantics would have to satisfy. On the other hand, no restrictions can or should be imposed on the environment, who represents a capricious user, the blind forces of nature, or the devil himself. Algorithmic activities being synonymous to *computations*, games thus represent *computational problems* --- interactive tasks performed by computing agents, with *computability* meaning *winnability*, i.e. existence of a machine that wins the game against any possible (behavior of the) environment.

In the 1930s, in the form of the famous Church-Turing thesis, mankind came up with what has been perceived as an ultimate mathematical definition of the precise meaning of algorithmic solvability. Curiously or not, such a definition was set forth and embraced before really having attempted to answer the seemingly more basic question about what *computational problems* are --- the very entities that may or may not have algorithmic solutions in the first place. The tradition established since then in theoretical computer science by computability simply means Church-Turing computability of *functions*, as the task performed by every Turing machine is nothing but receiving an input $x$ and generating the output $f(x)$ for some function $f$.

Yet most tasks that computers and computer networks perform are interactive, with strategies not always allowing themselves to be adequately understood as functions. To understand this, it would be sufficient to just reflect on the behavior of one's personal computer. The job of your computer is to play one long game against you. Now, have you noticed your faithful servant getting slower every time you use it? Probably not. That is because the computer is smart enough to follow a non-functional strategy in this game. If its strategy was a function from positions (interaction histories) to moves, the response time would inevitably keep worsening due to the need to read the entire, continuously lengthening interaction history every time before responding. Defining strategies as functions of only the latest moves is also not a way out. The actions of your computer certainly depend on more than your last keystroke.

Two main concepts on which the semantics of CoL is based are those of *static games* (defined later) and their *winnability*. Correspondingly, the philosophy of CoL relies on two beliefs that, together, present what can be considered an interactive version of the Church-Turing thesis:

**Thesis 1.2.1**
    (1) The concept of static games is an adequate formal counterpart of our intuition of ("pure", speed-independent) interactive computational problems.
    (2) The concept of winnability is an adequate formal counterpart of our intuition of algorithmic solvability of such problems.

So far games in logic have been mostly used to find models and semantical justifications for syntactically introduced popular systems such as intuitionistic logic or linear logic. For instance: Lorenzen's [Lor61, Fel85] game semantics was created for the purpose of justifying intuitionistic logic; Hintikka's [Hin73] game-theoretic semantics was originally meant to provide an alternative semantics for classical logic; Blass' [Bla92] game semantics was mainly motivated by the needs of linear logic, and so were the game-semantical approaches elaborated by Abramsky, Jagadeessan [Abr94] and others. In this respect, CoL turns the tables around and views games as foundational entities in their own right. It starts with identifying the most basic and meaningful operations on games. Understanding those operations as logical operators, it then looks at the logics induced by the corresponding concept of validity, regardless of how unusual such logics may turn out. There is no target syntactic construction to serve.

# 1.3 CoL vs. classical logic

Computability in the traditional Church-Turing sense is a special case of winnability --- winnability restricted to two-step (input/output, question/answer) interactive problems. So is the classical concept of truth, which is nothing but winnability restricted to propositions, viewed by CoL as zero-step problems, i.e., games with no moves that are automatically won by the machine when true and lost when false. This way, the game semantics of CoL is a generalization, refinement and conservative extension of that of classical logic.

Thinking of a human user in the role of the environment, computational problems are synonymous to computational tasks --- tasks performed by a machine for the user/environment. What is a task for a machine is then a resource for the environment, and vice versa. So the CoL games, at the same time, formalize our intuition of *computational resources*. Logical operators are understood as operations on such tasks/ resources/games, atoms as variables ranging over tasks/resources/games, and validity of a logical formula as being "always winnable", i.e. as existence --- under every particular interpretation of atoms --- of a machine that successfully accomplishes/provides/wins the corresponding task/resource/game no matter how the environment behaves. It is this approach that makes CoL "[a formal theory of computability in the same sense as classical logic is a formal theory of truth]". Furthermore, as mentioned, the classical concept of truth is a special case of winnability, which eventually translates into classical logic's being nothing but a special fragment of computability logic.

## 1.4 CoL vs. linear logic

CoL is a semantically conceived logic, and so far its syntax has only been partially developed. In a sense, this situation is opposite to the case with linear logic [Gir87], where a "full" syntactic construction existed right at the beginning, but where a really good formal semantics convincingly justifying the resource intuitions traditionally associated with that construction is still being looked for. In fact, the semantics of CoL can be seen to be providing such a justification, although only in a limited sense explained below.

The set of valid formulas in a certain fragment of the otherwise more expressive language of CoL forms a logic which is similar to but by no means the same as linear logic. When no exponentials are involved, the two logics typically agree on short and simple formulas. For instance, both logics reject $P \rightarrow P \wedge P$ and accept $P \rightarrow P \sqcap P$, with classical-shape propositional connectives here and later understood as the corresponding multiplicative operators of linear logic, and square-shape operators as additives ($\sqcap$ = "with", $\sqcup$ = "plus"). Similarly, both logics reject $P \sqcup \neg P$ and accept $P \vee \neg P$. On the other hand, CoL disagrees with linear logic on many more evolved formulas. E.g., CoL validates the following two principles rejected even by *affine logic*, i.e., linear logic with the weakening rule:

$$(P \wedge Q) \vee (R \wedge S) \rightarrow (P \vee R) \wedge (Q \vee S) \quad \text{(Blass's [Bla92] principle)};$$

$$(P \wedge (R \sqcap S)) \sqcap (Q \wedge (R \sqcap S)) \sqcap ((P \sqcap Q) \wedge R) \sqcap ((P \sqcap Q) \wedge S) \rightarrow (P \sqcap Q) \wedge (R \sqcap S).$$

With ⇓ and ⇑, which are CoL's counterparts of the exponentials !,? of linear logic, disagreements can be observed already at the level of very short formulas, such as ⇑⇓$P \rightarrow$ ⇓⇑$P$, accepted by CoL but rejected by affine logic. Generally, every formula provable in affine logic is valid in CoL but not vice versa.

Neither the similarities nor the discrepancies are a surprise. The philosophies of CoL and linear logic overlap in their pursuit to develop a logic of resources. But the ways this philosophy is materialized are rather different. CoL starts with a mathematically strict and intuitively convincing semantics, and only after that, as a natural second step, asks what the corresponding logic and its possible axiomatizations are. On the other hand, it would be accurate to say that linear logic started directly from the second step. Even though certain companion semantics were provided for it from the very beginning, those are not quite what we earlier agreed to call capital-'S'. As a formal theory of resources (rather than that of phases or coherence spaces), linear logic has been introduced syntactically rather than semantically, essentially by taking classical sequent calculus and deleting the rules that seemed unacceptable from a certain intuitive, naive and never formalized "resource-semantical" point of view. In the absence of a clear formal concept of resource-semantical truth or validity, the question about whether the resulting system was complete could not even be meaningfully asked. In this process of syntactically rewriting classical logic some innocent, deeply hidden principles could have easily gotten victimized. CoL believes that this is exactly what happened, with the above-mentioned formulas separating it from linear logic, along with many other principles, viewed as babies thrown out with the bath water. Of course, many retroactive attempts have been made to find semantical (often game-semantical) justifications for linear logic. Technically it is always possible to come up with some sort of a formal semantics that matches a given target syntactic construction, but the whole question is how natural and meaningful such a semantics is in its own rights, and how adequately it corresponds to the logic's underlying philosophy and ambitions. Unless, by good luck, the target system really *is* "the right logic", the chances of a decisive success when following the odd scheme *from syntax to semantics* could be rather slim. The natural scheme is *from semantics to syntax*. It matches the way classical logic evolved and climaxed in Gödel's completeness theorem. And, as we now know, this is exactly the scheme that computability logic, too, follows.

With the two logics in a sense competing for the same market, the main --- or perhaps only --- advantage of linear logic over CoL is its having a nice and simple syntax. In a sense, linear logic *is* (rather than *has*) a beautiful syntax; and computability logic *is* (rather than *has*) a meaningful semantics. An axiomatization of CoL in the full generality of its language has not been found yet. Only certain fragments of it have been axiomatized, including the one corresponding to the multiplicative-additive fragment of linear logic. Such axiomatizations tend to be more involved than that of linear logic, so the syntactic simplicity advantage of linear logic will always remain. Well, CoL has one thing to say: simplicity is good, yet, if it is most important, then nothing can ever beat ... the empty logic. The latter, just like linear logic, is sound with respect to resource semantics, whatever such a semantics means; it is sound with respect to *any* semantics for that matter.

From CoL's perspective, classical logic and (loosely understood) linear logic can be seen as two extremes within its all-unifying resource- (game-) semantical vision. Specifically, the main difference between linear logic and classical logic is that the former sees all occurrences of the same atom in a formula as different copies of the same resource, while the latter sees such occurrences as the same single copy, simply written at different places for technical reasons. So, linear logic rejects $\$1\rightarrow \$1\wedge \$1$ because in the antecedent we have one dollar while in the consequent two, with the possibility to buy an apple with one and an orange with the other. On the other hand, classical logic accepts thus principle because it sees a single dollar in the consequent, written twice for some strange reason; if the first conjunct of the consequent is spent buying an apple, the second conjunct is also automatically spent on the same apple, with no money remaining for oranges. As for CoL, it allows us to write expressions where all occurrences of \$1 stand for the same one-dollar bill, or all stand for separate bills, or we have a mixture of these two, where some occurrences stand for the same bill while some other occurrences in the same expression stand for different bills.

Blass [Bla92] was the first to attempt a game-semantical justification for linear logic. This goal was only partially achieved, as the resulting logic, just like the above-discussed fragment of CoL, turned out to validate more principles than linear logic does. It should be pointed out that the multiplicative-additive fragment of the logic induced by Blass' semantics coincides with the corresponding fragment of CoL. This does not extend to the full language of linear logic though. For instance, Blass' semantics validates the following principle which is invalid in CoL: [Jap12a]

$$P \wedge \wr((P \rightarrow P \wedge P) \wedge (P \vee P \rightarrow P)) \rightarrow \wr P.$$

In full generality, the "linear-logic fragment" of CoL is strictly between affine linear logic and the logic induced by Blass' semantics.[Jap09a]

# 1.5 CoL vs. intuitionistic logic

From CoL's perspective, the situation with intuitionistic logic [Hey71] is rather similar to what we had with linear logic. Intuitionistic logic is another example of a syntactically conceived logic. Despite decades of efforts, no fully convincing semantics has been found for it. Lorenzen's [Lor61] game semantics, which has a concept of validity without having a concept of truth, has been perceived as a technical supplement to the existing syntax rather than as having independent importance. Some other semantics, such as Kleene's realizability [Kle52] or Gödel's Dialectica interpretation [Göd58], are closer to what we might qualify as capital-'S'. But, unfortunately, they validate certain principles unnegotiably rejected by intuitionistic logic.

Just like this was the case with linear logic, the set of valid formulas in a certain fragment of the language of CoL forms a logic which is properly stronger[Mez10, Jap07b] than Heyting's intuitionistic logic. However, the two come "much closer" to each other than CoL and linear logic do. The shortest known formula separating intuitionistic logic from the corresponding fragment of CoL is

$$(\circ\!-\!P \circ\!-\! Q \sqcup R) \sqcap (\circ\!-\! \circ\!-\! P \circ\!-\! S \sqcup T) \circ\!-\! (\circ\!-\! P \circ\!-\! Q) \sqcup (\circ\!-\! P \circ\!-\! R) \sqcup (\circ\!-\! \circ\!-\! P \circ\!-\! S) \sqcup (\circ\!-\! \circ\!-\! P \circ\!-\! T),$$

where $\circ\!-\!, \circ\!\neg, \sqcup, \sqcap$ are CoL's counterparts of the intuitionistic implication, negation, disjunction and conjunction, respectively.

Just like the resource philosophy of CoL overlaps with that of linear logic, so does its algorithmic philosophy with the constructivistic philosophy of intuitionism. The difference, again, is in the ways this philosophy is materialized. Intuitionistic logic has come up with a "constructive syntax" without having an adequate underlying formal semantics, such as a clear concept of truth in some constructive sense. This sort of a syntax was essentially obtained from the classical one by banning the offending law of the excluded middle. But, as in the case of linear logic, the critical question immediately springs out: where is a guarantee that together with excluded middle some innocent principles would not be expelled just as well? The constructivistic claims of CoL, on the other hand, are based on the fact that it defines truth as algorithmic solvability. The philosophy of CoL does not find the term *constructive syntax* meaningful unless it is understood as soundness with respect to some *constructive semantics*, for only a semantics may or may not be constructive in a reasonable sense. The reason for the failure of $P \sqcup \neg P$ in CoL is not that this principle ... is not included in its axioms. Rather, the failure of this principle is exactly the reason why this principle, or anything else entailing it, would not be among the axioms of a sound system for CoL. Here "failure" has a precise semantical meaning. It is non-validity, i.e. existence of a problem $A$ for which $A \sqcup \neg A$ is not algorithmically solvable.

It is also worth noting that, while intuitionistic logic irreconcilably defies classical logic, computability logic comes up with a peaceful solution acceptable for everyone. The key is the expressiveness of its language, that has (at least) two versions for each traditionally controversial logical operator, and particularly the two versions $\vee$ and $\sqcup$ of disjunction. The semantical meaning of $\vee$ conservatively extends --- from moveless games to all games --- its classical meaning, and the principle $P \vee \neg P$ survives as it represents an always-algorithmically-solvable combination of problems, even if solvable in a sense that some constructivistically-minded might pretend to fail to understand. And the semantics of $\sqcup$, on the other hand, formalizes and conservatively extends a different, stronger meaning which apparently every constructivist associates with disjunction. As expected, then $P \sqcup \neg P$ turns out to be semantically invalid. CoL's proposal for settlement between classical and constructivistic logics then reads: 'If you are open (=classically) minded, take advantage of the full expressive power of CoL; and if you are constructivistically minded, just identify a collection of the operators whose meanings seem constructive enough to you, mechanically disregard everything containing the other operators, and put an end to those fruitless fights about what deductive methods or principles should be considered right and what should be deemed wrong'.

## 1.6 CoL vs. independence-friendly logic

Relatively late developments [Jap06c, Jap11b, Xu14] in CoL made it possible to switch from formulas to the more flexible and general means of expression termed *cirquents*. The main distinguishing feature of the latter is that they allow to account for various sorts of sharing between subexpressions. After such a generalization, independence-friendly (IF) logic [Hin73] became a yet another natural fragment of CoL.[Jap11b, Xu14, Xu16] As such, it is a conservative fragment, just like classical logic and unlike linear or intuitionistic logics. This is no surprise because, just like CoL and unlike linear or intuitionistic logics, IF logic is a semantically conceived logic.

In fact, for a long time, IF logic remained a pure semantics without a syntax. In its full first-order language, IF logic was simply known to be unaxiomatizable. As for the propositional fragment, there was none because the traditional approaches to IF logic had failed to offer any non-classical semantics for propositional connectives. Under CoL's cirquent-based approach to IF logic this is no longer so, and "independence-friendly" propositional connectives are just as meaningful as their quantifier counterparts. Based on this new treatment, a sound and complete axiomatization for propositional IF logic has been later found.[Xu14, Xu16]

## 1.7 The 'from semantics to syntax' paradigm

CoL's favorite 'from semantics to syntax' paradigm for developing a logic can be characterized as consisting of three main stages. The first one can be called the Informal semantics stage, at which the main intuitions are elaborated along with the underlying motivations, and the formal language of the logic is agreed upon. The second one is the Formal semantics stage, at which those intuitions are formalized as a mathematically strict semantics for the adopted language, and definitions of truth and validity are given. Finally, the third one is the Syntax stage, at which one constructs axiomatizations for the corresponding set of valid principles, whether in the full language of the logic or some natural fragments of it, and proves the adequacy (soundness and completeness) of such constructions.

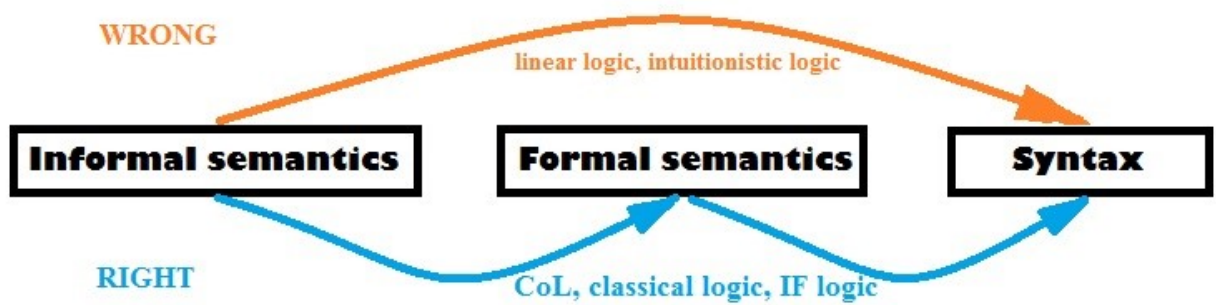

**Figure 1.7.1:** Three stages of developing a logic

CoL and classical logic went through all three stages in sequence. So did IF logic, even though for a long time it was stuck at the second stage. As for linear and intuitionistic logics, they essentially jumped from the first stage directly to the third stage, skipping the inermediary stage. It is the absence of formal rather than informal semantics that we meant when saying that the two logics were conceived syntactically rather than semantically. Why is such a jump wrong? It is impossible to directly "prove" that the proposed syntax adequately corresponds to the informal-semantical intuitions underlying the logic. After all, Syntax is mathematically well defined while Informal semantics is from the realm of philosophy or intuitions, so an adequacy claim lacks any mathematical meaning. Of course, the same is the case with Formal semantics vs. Informal semantics. But, at least, both are "semantics", which makes it qualitatively easier to convincingly argue (albeit not prove) that one adequately corresponds to the other. Once such a correspondence claim is accepted, one can prove the adequacy of the syntax by showing that it is sound and complete with respect to the formal semantics.

The intermediary role of Formal semantics can be compared with that of Turing machines. Since the intuitive concept of a machine (algorithm) and the mathematical concept of a Turing machine are both about "machines", it is relatively easy to argue in favor of the (mathematically unprovable) Church-Turing thesis, which claims an adequate correspondence between the two. Once this thesis is accepted, one can relatively easily show --- this time mathematically --- that recursive functions or lambda calculus, too, adequately correspond to our intuitions of machine-computability. Arguing in favor of such a claim directly, without having the intermediary Turing machine model, would not be easy, as recursive definitions or lambda terms do not at all resemble what our intuition perceives as machines.

Based directly on the resource intuitions associated with linear logic, can anyone tell whether, for instance, the principle $P \wedge \flat(P \rightarrow P \wedge P) \rightarrow \flat P$ should be accepted? An orthodox linear logician might say 'No, because it is not provable in Girard's canonical system'. But the whole point is that we are just trying to understand what should be provable and what should not. From similar circularity suffer the popular attempts to "semantically" justify intuitionistic provability in terms of … intuitionistic provability.

# 2 Games

## 2.1 The two players

The CoL games are between two players, called Machine and Environment (not always capitalized, and may take articles). On the picture below we see a few other names for these players.

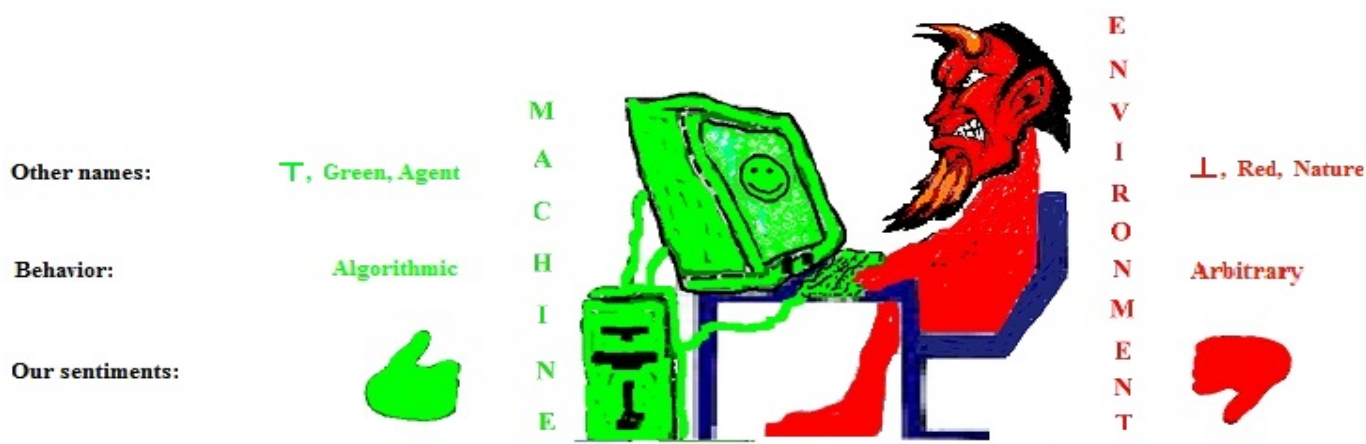

We will be using ⊤ and ⊥ as symbolic names for Machine and Environment, respectively. ⊤ is a deterministic mechanical device only capable of following algorithmic strategies. ⊥'s strategies, on the other hand, are arbitrary. Throughout this article, we shall consistently use the green color to represent Machine, and red for Environment. As seen from the picture, our sympathies are with the machine. Why should we be fans of the machine even when it is us who act in the role of its "evil" environment? Because the machine is a tool, and what makes it valuable as such is exactly its being a good player. In fact, losing a game by the machine means that it is malfunctioning. Imagine Bob using a computer for computing the "28% of $x$" function in the process of preparing his US federal tax return. This is a game where the first move is by Bob, consisting in inputting a number *m* and meaning asking his machine the question "what is 28% of *m*?". The machine wins iff it answers by the move/output *n* such that $n=0.28m$. Of course, Bob does not want the machine to tell him that 27,000 is 28% of 100,000. In other words, he does not want to win against the machine. For then he could lose the more important game against Uncle Sam.

## 2.2 Moves, runs and positions

Machine and Environment interact with each other through mutually observable *actions*. We will be using the term "move" instead of "action". Looking back at the ordinary Turing machine model, it is about games where only two moves are made: the first move, called "input", is by the environment, and the second move, called "output", is by the machine.

We agree on the following:

- A *move* is any finite string over the keyboard alphabet. We will be using $\alpha, \beta, \gamma$ as metavariables for moves.
- A *colored move* is a move $\alpha$ together with one of the two colors *green* or *red*, with the color indicating which player has made the move. We shall write such a move as $\alpha$ or $\alpha$, depending on its color (in black-and-white presentations, $\top\alpha$ and $\bot\alpha$ will be written instead). Often we omit the word "colored" and simply say "move". The letter $\lambda$ will be used as a metavariable for colored moves.
- A *run* is a (finite or infinite) sequence of colored moves. We will be using $\Gamma, \Delta$ as metavariables for runs.
- A *position* is a finite run. We will be using $\Phi, \Psi, \Xi, \Omega$ as metavariables for positions.
- We will be writing runs and positions as $\langle \alpha, \beta, \gamma \rangle$, $\langle \Phi \rangle$, $\langle \Phi, \Psi, \beta, \Gamma \rangle$, etc. The meanings of such expressions should be clear. For instance, $\langle \Phi, \Psi, \beta, \Gamma \rangle$ is the run consisting of the (colored) moves of the position $\Phi$, followed by the moves of the position $\Psi$, followed by the move $\beta$, and then by the moves of the (possibly infinite) run $\Gamma$.
- $\langle \rangle$ thus stands for the *empty position*.

## 2.3 Constant games

A *gamestructure* is a nonempty set **Lr** of positions, called *legal positions*, such that, whenever a position is in **Lr**, so are all initial segments of it. The empty position $\langle \rangle$ is thus always legal. We extend gamestructures to include infinite runs as well, by stipulating that an infinite run $\Gamma$ is in **Lr** iff so is every finite initial segment of $\Gamma$. Intuitively, **Lr** is the set of *legal runs*. The runs that are not in **Lr** are *illegal*. An *illegal move* in a given position $\langle \Phi \rangle$ is a move $\lambda$ such that $\langle \Phi, \lambda \rangle$ is illegal. The player who made the first illegal move in a given run is said to be the *offender*. Intuitively, illegal moves can be thought of as moves that cannot or should not be made. Alternatively, they can be seen as actions that cause the system crash (e.g., requesting to divide a number by **0**).

Gamestructures can be visualized as upside-down trees where the nodes represent positions. Each edge is labeled with a colored move. Such an edge stands for a legal move in the position represented by the upper node of it. Here is an example:

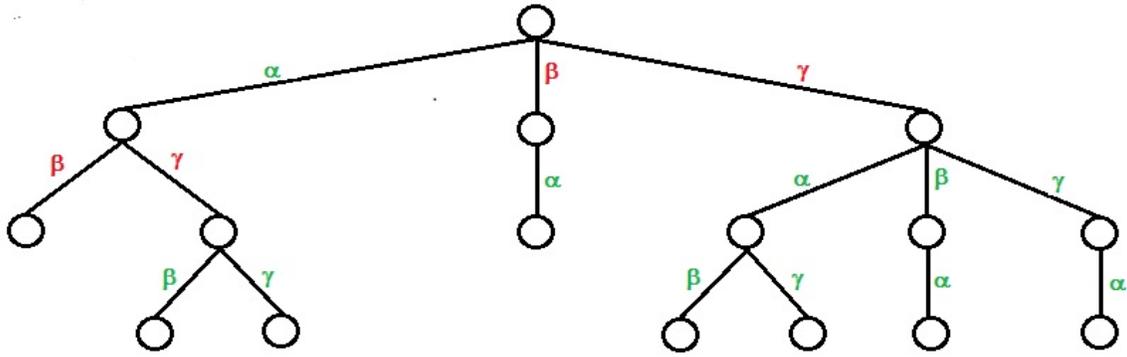

**Figure 2.3.1:** A gamrestructure

This gamestructure consists of the following 16 runs: ⟨ ⟩, ⟨α⟩, ⟨β⟩, ⟨γ⟩, ⟨α, β⟩, ⟨α, γ⟩, ⟨β, α⟩, ⟨γ, α⟩, ⟨γ,β⟩, ⟨γ, γ⟩, ⟨α, γ,β⟩, ⟨α, γ,γ⟩, ⟨γ, α, β⟩, ⟨γ, α, γ⟩, ⟨γ, β, α⟩, ⟨γ, γ, α⟩. All of the infinitely (in fact, uncountably) many other runs are illegal. An example of an illegal run is ⟨γ, γ, β, β, α⟩. The offender in it is Environment, because the offending third move is red.

Let **Lr** be a gamestructure. A *content on* **Lr** is a function **Wn**: **Lr** → {⊤,⊥}. When **Wn**⟨Γ⟩ = ⊤, we say that Γ is *won* by the machine (and *lost* by the environment); and when **Wn**⟨Γ⟩ =⊥, we say that Γ is won by the environment (and lost by the machine). We extend the domain of **Wn** to all runs by stipulating that an illegal run is always lost by the offender. Since we are fans of Machine, the default meaning of just "won" (or "lost") is "won (or lost) by the machine".

**Definition 2.3.2** A *constant game* $G$ is a pair (**Lr**$^G$,**Wn**$^G$), where **Lr**$^G$ is a gamestructure, and **Wn**$^G$ is a content on **Lr**$^G$.

Figure 2.3.3 below shows a constant game of the structure of Figure 2.3.1:

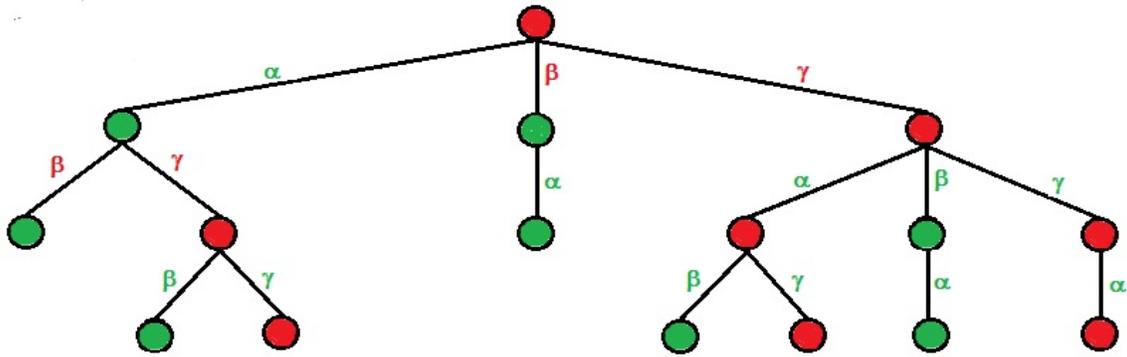

**Figure 2.3.3:** A constant game

Here the winner in each position is indicated by the color of the corresponding node. So, for instance, the run ⟨γ, α, β⟩ is won by the machine; but if this run had stopped after its second move, then the environment would be the winner. Of course, such a way of indicating winners is not sufficient if there are infinitely long branches (legal runs), for such branches do not have "the corresponding nodes".

We say that a constant game is *strict* iff, in every legal position, at most one of the two players has legal moves. Our games generally are not strict. For instance, in the start position of the game of Figure 2.3.3, both players have legal moves. We call such (not-necessarily-strict) games *free*. Free games model real-life situations more directly and naturally than strict games do. Hardly many tasks that humans, computers or robots perform are strict. Imagine you are playing chess over the Internet on two boards against two independent adversaries that, together, form the (one) environment for you. Let us say you play white on both boards. Certainly in the initial position of this game only you have legal moves. However, once you make your first move --- say, on board #1 --- the picture changes. Now both you and the environment have legal moves, and who will be the next to move depends on who can or wants to act sooner. Namely, you are free to make another opening move on board #2, while the environment --- adversary #1 --- can make a reply move on board #1. A strict-game approach would have to impose some not-very-adequate supplemental conditions uniquely determining the next player to move, such as not allowing you to move again until receiving a response to your previous move. Let alone that this is not how the real two-board game would proceed, such regulations defeat the very purpose of the idea of parallel/distributed computations with all the known benefits it offers.

Because our games are free, strategies for them cannot be defined as functions from positions to moves, because, in some positions (such as the root position in the game of Figure 2.3.3) both players may have legal moves and, if both are willing to move, which of them acts sooner will determine what will be the next move.

The exact meaning of "strategy" will be defined later, but whatever it means, we can see that the machine has a winning strategy in the game of Figure 2.3.3, which can be put this way:  *Regardless of what the adversary is doing or has done, go ahead and make move α; make β as your second move if and when you see that the adversary has made move γ, no matter whether this happened before or after your first move*".  This strategy obviously guarantees that the machine will not offend. There are 5 possible legal runs consistent with it, i.e., legal runs that could be generated when it is followed by the machine: ⟨α⟩, ⟨α, β⟩, ⟨β, α⟩, ⟨α, γ, β⟩ and ⟨γ, α, β⟩. All are won by the machine.

Let $G$ be a constant game. $G$ is said to be ***finite-depth*** iff there is a (smallest) integer $d$, called the ***depth*** of $G$, such that the length of every legal run of $G$ is $\leq d$. And $G$ is ***perifinite-depth*** iff every legal run of it is finite, even if there are arbitrarily long legal runs. Let us call a legal run $\Gamma$ of $G$ ***maximal*** iff $\Gamma$ is not a proper initial segment of any other legal run of $G$. Then we say that $G$ is ***finite-breadth*** iff the total number of maximal legal runs of $G$, called the ***breadth*** of $G$, is finite. Note that, in a general case, the breadth of a game may be not only infinite, but even uncountable. $G$ is said to be (simply) ***finite*** iff it only has a finite number of legal runs. Of course, $G$ is finite only if it is finite-breadth, and when $G$ is finite-breadth, it is finite iff it is finite-depth iff it is perifinite-depth.

As noted in Section 2.2, computational problems in the traditional sense are nothing but functions (to be computed). Such problems can be seen as the following types of games of depth 2:

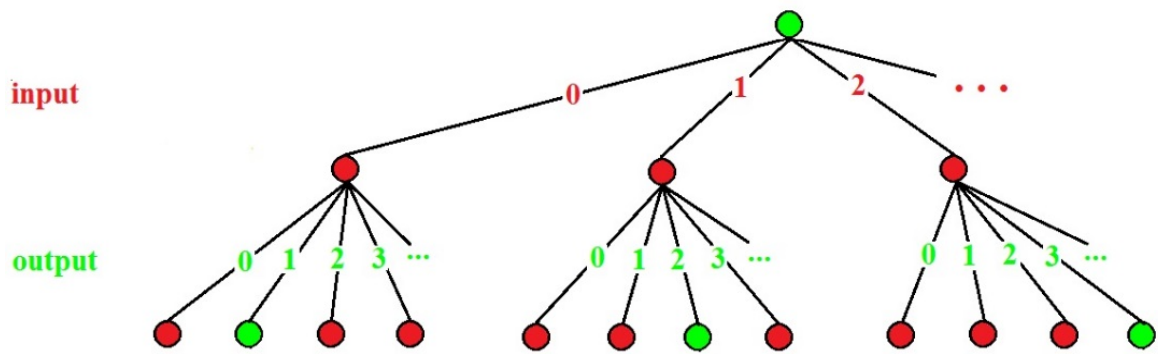

**Figure 2.3.4:** The "successor" game

- Why is the root green here?  Because it corresponds to the situation where there was no input. The machine has nothing to answer for, so it wins.
- Why are the 2[nd] level nodes red? Because they correspond to situations where there was an input but no output was generated by the machine. So the machine loses.
- Why does each group of 3[rd] level nodes has exactly one green node? Because a function has exactly one ("correct") value for each argument.

- What particular function is this game about? The successor function: $f(n)=n+1$.

Once we agree that computational problems are nothing but games, the difference in the degrees of generality and flexibility between the traditional approach to computational problems and our approach becomes apparent and appreciable. What we see in Figure 2.3.4 is indeed a very special sort of games, and there is no good call for confining ourselves to its limits. In fact, staying within those limits would seriously retard any more or less advanced and systematic study of computability.

First of all, one would want to get rid of the "one green node per sibling group" restriction for the third-level nodes. Many natural problems, such as the problem of finding a prime integer between $n$ and $2n$, or finding an integral root of $x^2-2n=0$, may have more than one as well as less than one solution. That is, there can be more than one as well as less than one "right" output on a given input $n$.

And why not further get rid of any remaining restrictions on the colors of whatever-level nodes and whatever-level arcs. One can easily think of natural situations where, say, some inputs do not obligate the machine to generate an output and thus the corresponding $2^{nd}$ level nodes should be green. An example would be the case where the machine is computing a partially-defined function $f$ and receives an input $n$ on which $f$ is undefined.

So far we have been talking about generalizations within the depth-2 restriction, corresponding to viewing computational problems as very short dialogues between the machine and its environment. Permitting longer-than-2 or even infinitely long runs would allow us to capture problems with arbitrarily high degrees of interactivity and arbitrarily complex interaction protocols. The task performed by a network server is an example of an infinite dialogue between the server and its environment --- the collection of clients, or let us just say the rest of the network.

It also makes sense to consider "dialogues" of lengths less than 2. "Dialogues" of length 0, i.e. games of depth 0 are said to be ***elementary***. There are exactly two elementary constant games, denoted by ⊤ and ⊥:

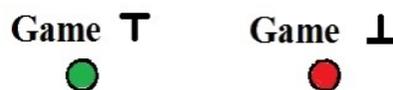

Note that the symbols ⊤ and ⊥ thus have dual meanings: they may stand for the two elementary games as above, or for the corresponding two players. It will usually be clear from the context which of these two meanings we have in mind.

Just like classical logic, CoL sees no extensional distinction between "snow is white" and ⊤, or between "snow is black" and ⊥: All true propositions, as games, are ⊤, and all false propositions are ⊥. In other words, a proposition is a game automatically won by the machine when true and lost when false. This way, CoL's concept of a constant game is a generalization of the classical concept of a proposition.

## 2.4 Not-necessarily-constant games

We fix an infinite set *Variables* = {$var_0$, $var_1$, $var_2$,... } of ***variables***. As usual, lowercase letters near the end of the Latin alphabet will be used to stand (as metavariables) for variables. We further fix the set *Constants* = {0,1,2,3,...} of decimal numerals, and call its elements ***constants***. With some abuse of concepts, we shall often identify constants with the numbers they represent.

A *universe* (of discourse) is a pair $U = (\mathbf{Dm}, \mathbf{Dn})$, where **Dm**, called the *domain* of $U$, is a nonempty set, and **Dn**, called the *denotation function* of $U$, is a (total) function of type *Constants* → **Dm**. The elements of **Dm** will be referred to as the *individuals* of $U$. The intuitive meaning of $d=\mathbf{Dn}(c)$ is that the individual $d$ is the *denotat* of the constant $c$ and thus $c$ is a *name* (or *code*) of $d$. So, the function **Nm** from **Dm** to the powerset of *Constants* satisfying the condition $c \in \mathbf{Nm}(d) \Leftrightarrow d=\mathbf{Dn}(c)$ can be called the *naming function* of $U$. Of course, whenever convenient, a universe can be characterized in terms of its naming function rather than denotation function.

A universe $(\mathbf{Dm}, \mathbf{Dn})$ is said to be *ideal* iff **Dn** is a bijection. Generally, in a not-necessarily-ideal universe, some individuals may have unique names, some have many names, and some have no names at all. Real-world universes are seldom ideal: not all people living or staying in the United States have social security numbers; most stars and planets of the Galaxy have no names at all, while some have several names (Morning Star = Evening Star = Venus); etc. A natural example of an inherently non-ideal universe from the world of mathematics would be the one whose domain is the set of real numbers, only some of whose elements have names, such as 5, 1/3, $\sqrt{2}$ or $\pi$. Generally, even if the set of constants was not fixed, no universe with an uncountable domain would be "ideal" for the simple reason that there can only be countably many names. This is so because names, by their very nature and purpose, have to be finite objects. Observe also that many properties of common interest, such as computability or decidability, are usually sensitive with respect to how objects (individuals) are named, as they deal with the names of those objects rather than the objects themselves. For instance, strictly speaking, computing a function $f(x)$ means the ability to tell, after seeing a (the) name of an arbitrary object $a$, a (the) name of the object $b$ with $b=f(a)$. Similarly, an algorithm that decides a predicate $p(x)$ on a set $S$, strictly speaking, takes not elements of $S$ --- which may be abstract objects such as numbers or graphs --- but rather names of those elements, such as decimal numerals or codes. It is not hard to come up with a nonstandard naming of the natural numbers through decimal numerals where the predicate "$x$ is even" is undecidable. On the other hand, for any undecidable arithmetical predicate $p(x)$, one can come up with a naming such that $p(x)$ becomes decidable --- for instance, one that assigns even-length names to all $a$ with $p(a)$ and assigns odd-length names to all $a$ with $\neg p(a)$. Classical logic exclusively deals with individuals of a universe without a need to also consider names for them, as it is not concerned with decidability or computability. CoL, on the other hand, with its computational semantics, inherently calls for being more careful about differentiating between individuals and their names, and hence for explicitly considering universes in the form $(\mathbf{Dm}, \mathbf{Dn})$ rather than just **Dm** as classical logic does.

Where $Vr$ is a set of variables and $Dm$ is the domain of some universe, by a $Vr \to Dm$ *valuation* we mean a (total) function $e$ of type $Vr \to Dm$. When $Vr$ and $Dm$ are clear from the context, we may omit an explicit reference to them and simply say "valuation". References to a universe $U$ or its components can be similarly omitted when talking about individuals, denotats, names or some later-defined concepts such as those of a game or a function.

**Definition 2.4.1** Let $n$ be a natural number. An *n-ary game* is a tuple $G=(\mathbf{Dm}^G, \mathbf{Dn}^G, \mathbf{Vr}^G, \mathbf{Mp}^G)$, where $(\mathbf{Dm}^G, \mathbf{Dn}^G)$ is a universe, $\mathbf{Vr}^G$ is a set of $n$ distinct variables, and $\mathbf{Mp}^G$ is a mapping of $\mathbf{Vr}^G \to \mathbf{Dm}^G$ valuations to constant games.

We refer to the elements of $\mathbf{Vr}^G$ as the variables on which the game $G$ *depends*. We further refer to the pair $(\mathbf{Dm}^G, \mathbf{Dn}^G)$ as the *universe of* $G$, and denote it by $\mathbf{Un}^G$. Correspondingly, a game sometimes can be written as the triple $(\mathbf{Un}^G, \mathbf{Vr}^G, \mathbf{Mp}^G)$ rather than quadruple $(\mathbf{Dm}^G, \mathbf{Dn}^G, \mathbf{Vr}^G, \mathbf{Mp}^G)$. We further refer to $\mathbf{Dm}^G$ as the *domain of* $G$, refer to $\mathbf{Dn}^G$ as the *denotation function of* $G$, and refer to $\mathbf{Vr}^G \to \mathbf{Dm}^G$ valuations as *G-valuations*.

In classical logic, under an intensional (variable-sensitive) understanding, the definition of the concept of an $n$-ary predicate would look exactly like our definition of an $n$-ary game after omitting the (now redundant) **Dn** component, with the only difference that there the **Mp** function returns propositions rather than constant games. And, just like propositions are nothing but 0-ary predicates, constant games are nothing but 0-ary games. Thus, games generalize constant games in the same way as predicates generalize propositions. And, as constant games are generalized propositions, games are generalized predicates.

In formal contexts, we choose a similar intensional approach to functions. The definition of a function $f$ below is literally the same as our definition of a game $G$, with the only difference that $\mathbf{Mp}^f$ is now a mapping of $\mathbf{Vr}^f \to \mathbf{Dm}^f$ valuations to $\mathbf{Dm}^f$ (rather than to constant games).

**Definition 2.4.2** Let $n$ be a natural number. An $n$-ary ***function*** is a tuple $f=(\mathbf{Dm}^f, \mathbf{Dn}^f \mathbf{Vr}^f, \mathbf{Mp}^f)$, where $(\mathbf{Dm}^G, \mathbf{Dn}^G)$ is a universe, $\mathbf{Vr}^f$ is a set of $n$ distinct variables, and $\mathbf{Mp}^f$ is a mapping of $\mathbf{Vr}^f \to \mathbf{Dm}^f$ valuations (called $f$-**valuations**) to $\mathbf{Dm}^f$.

Just like in the case of games, we refer to the elements of $\mathbf{Vr}^f$ as the variables on which the function $f$ ***depends***, refer to $\mathbf{Dm}^f$ as the ***domain of $f$***, etc.

Given a game $G$ and an $X \to \mathbf{Dm}^G$ valuation $e$ with $\mathbf{Vr}^G \subseteq X$, we write $e[G]$ to denote the constant game $\mathbf{Mp}^G(e')$ to which $\mathbf{Mp}^G$ maps $e'$, where $e'$ is the restriction of $e$ to $\mathbf{Vr}^G$ (i.e., the $G$-valuation that agrees with $e$ on all variables from $\mathbf{Vr}^G$). Such a constant game $e[G]$ is said to be an ***instance*** of $G$. Also, for readability, we usually write $\mathbf{Lr}_e^G$ and $\mathbf{Wn}_e^G$ instead of $\mathbf{Lr}^{e[G]}$ and $\mathbf{Wn}^{e[G]}$. Similarly, given a function $f$ and an $X \to \mathbf{Dm}^f$ valuation $e$ with $\mathbf{Vr}^f \subseteq X$, we write $e[f]$ to denote the individual $\mathbf{Mp}^f(e')$ to which $\mathbf{Mp}^f$ maps $e'$, where $e'$ is the restriction of $e$ to $\mathbf{Vr}^f$.

We say that a game is ***elementary*** iff so are all of its instances. Thus, games generalize elementary games in the same sense as constant games generalize ⊤ and ⊥. Further, since the "legal run" component of all instances of elementary games is trivial (the empty run $\langle\rangle$ is the only legal run), and since depending on runs is the only thing that differentiates constant games from propositions, we can and will use "predicate" and "elementary game" as synonyms. Specifically, we understand a predicate $p$ as the elementary game $G$ which depends on the same variables as $p$ such that, for any valuation $e$, $\mathbf{Wn}_e^G\langle\rangle=⊤$ iff $p$ is true at $e$. And vice versa: An elementary game $G$ will be understood as the predicate $p$ which depends on the same variables as $G$ and which is true at a given valuation $e$ iff $\mathbf{Wn}_e^G\langle\rangle=⊤$.

**Convention 2.4.3** Let $U$ be a universe, $c$ a constant and $x$ a variable. We shall write $c^U$ to mean the function $f$ with $\mathbf{Un}^f=U$, $\mathbf{Vr}^f=\emptyset$ and with $\mathbf{Mp}^f$ such that, for (the only) $f$-valuation $e$, $\mathbf{Mp}^f(e)=\mathbf{Dn}^U(c)$. And we shall write $x^U$ to mean the function $f$ with $\mathbf{Un}^f=U$, $\mathbf{Vr}^f=\{x\}$ and with $\mathbf{Mp}^f$ such that, for any $f$-valuation $e$, $\mathbf{Mp}^f(e)=e[x]$.

**Convention 2.4.4** Assume $F$ is a function (resp. game). Following the standard readability-improving practice established in the literature for functions and predicates, we may fix a tuple $(x_1,\ldots,x_n)$ of pairwise distinct variables for $F$ when first mentioning it, and write $F$ as $F(x_1,\ldots,x_n)$. When doing so, we do not necessarily mean that $\{x_1,\ldots,x_n\}=\mathbf{Vr}^F$. Representing $F$ as $F(x_1,\ldots,x_n)$ sets a context in which, where $g_1,\ldots,g_n$ are functions with the same universe as $F$, we can write $F(g_1,\ldots,g_n)$ to mean the function (resp. game) $H$ with $\mathbf{Un}^H=\mathbf{Un}^F$, $\mathbf{Vr}^H=(\mathbf{Vr}^H-\{x_1,\ldots,x_n\})\cup\mathbf{Vr}^{g_1}\cup\ldots\cup\mathbf{Vr}^{g_n}$ and with $\mathbf{Mp}^H$ such that the following condition is satisfied:

- For any $H$-valuation $e$, $e[H]=e'[F]$, where $e'$ is the $F$-valuation with $e'[x_1]=e[g_1]$, ..., $e'[x_n]=e[g_n]$.

Further, we allow for any of the above "functions" $g_i$ to be just a constant $c$ or just a variable $x$. In such a case, $g_i$ should be correspondingly understood as the 0-ary function $c^U$ or the unary function $x^U$, where $U=\mathbf{Un}^F$. So, for instance, $F(0,x)$ is our lazy way to write $F(0^F,x^U)$.

The entities that in common language we call games are at least as often non-constant as constant. Board games such as chess and checkers are examples of constant games. On the other hand, almost all card games are more naturally represented as non-constant games: each session/instance of such a game is set by a particular permutation of the card deck, and thus the game can be understood as a game that depends on a variable $x$ ranging over the possible settings of the deck. Even the game of checkers has a natural non-constant generalization *Checkers*($x$) with $x$ ranging over positive even integers, meaning a play on the board of size $x \times x$ where, in the initial position, the first $1.5x$ black cells are filled with white pieces and the last $1.5x$ black cells with black pieces. Then the ordinary checkers can be written as *Checkers*(8). Furthermore, the numbers of pieces of either color also can be made variable, getting an even more general game *Checkers*($x,y,z$), with the ordinary checkers being the instance *Checkers*(8,12,12) of it. By allowing rectangular-shape boards, we would get a game that depends on four variables, etc. Computability theory also often appeals to non-constant games to illustrate certain complexity-theory concepts such as alternating computation or PSPACE-completeness. The so called Formula Game or Generalized Geography are typical examples. Both can be understood as games that depend on a variable $x$, with $x$ ranging over quantified Boolean formulas in Formula Game and over directed graphs in Generalized Geography.

Consider a game $G$. What we call a **unilegal** run of $G$ is a run which is a legal run of all instances of $G$. And we say that $G$ is **unistructural** iff all legal runs of all of its instances are unilegal runs of $G$. The class of unistructural games is known to be closed under all of the game operations studies in CoL.[Jap03] While natural examples of non-unistructural games exist such as the games mentioned in the above paragraph, virtually all of the other examples of particular games discussed elsewhere in the present article are unistructural.

Non-constant games, as long as they are unistructural, can be visualized as trees in the earlier seen style, with the difference that the nodes of the tree can now be labeled with any predicates rather than only propositions (colors) as before. For any given valuation $e$, each such label $L$ is telling us the color of the node. Namely, the $L$-labeled node is green if $L$ is true at $e$, and red if $L$ is false.

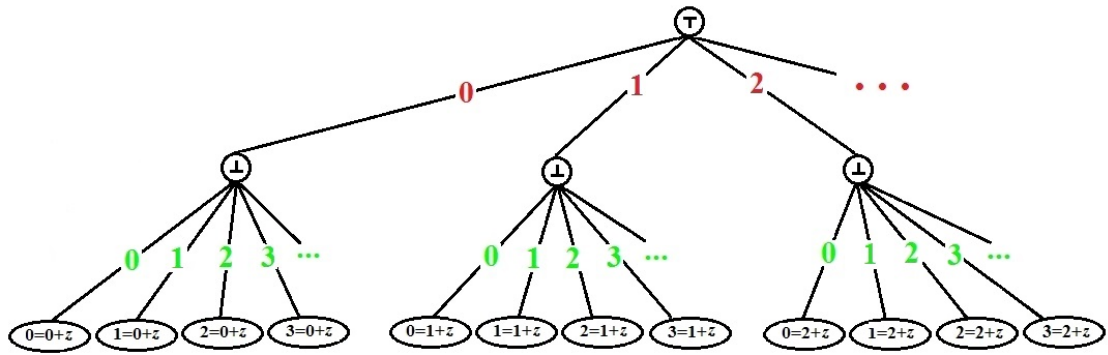

**Figure 2.3.5:** The "generalized successor" game

The above figure shows a game which depends on $x$. Specifically, for every valuation $e$, the game is about computing the function $f_e$, defined by $f_e(n) = n+e(z)$ ("the $z$th successor of $n$"). Note that we have different functions and thus different constant games for different valuations $e$ here.

Denoting the game of Figure 2.3.5 by $G(x)$, the game of Figure 2.3.4 can be seen to be the instance $G(1)$ of it. The latter results from replacing $z$ by 1 in Figure 2.3.5. This replacement turns every label $m=n+z$ into the proposition $m=n+1$, i.e., into ⊤ (green filling) or ⊥ (red filling).

## 2.5 Static games

In the particular games that we have seen so far or will see in the subsequent sections, when talking about the existence of a winning strategy or the absence of thereof, the question about the (relative) speeds of the players was never relevant. That is because all those games share one nice property, called the ***static*** property. Below are some intuitive characterizations of this important class of games.

- Static games are games where the speed of the adversary is not an issue: if a player has a winning strategy, the strategy will remain good no matter how fast its adversary is in making moves. And if a player does not have a winning strategy, it cannot succeed no matter how slow the adversary is.
- Static games are games where "it never hurts a player to postpone making moves" (Blass' words from his AMS review of [Jap03]).
- Static games are contests of intellect rather than speed. In such games, what matters is *what* to do (strategy) rather than *how fast* to do (speed).

The games that are not static we call ***dynamic***. The following games are dynamic:

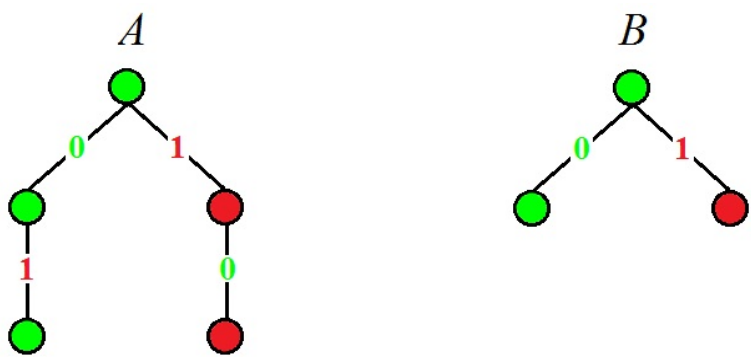

In either game, the player who is quick enough to make the first move becomes the winner. And asking whether Machine has a winning strategy is not really meaningful: whether Machine can win or not depends on the relative speeds of the two players. In a sense, *B* is even "worse" than *A*. Just as in *A*, it would not be a good idea for Machine to wait, because, unless Environment is stupid enough to also decide to wait, Machine will lose. Let us now say that Machine goes ahead and initiates the first move. What may happen in *B* is that Environment moves before Machine actually completes its move. This will make Machine not only the loser but also the offender. Machine cannot even be sure that its move will be legal! If communication happens by exchanging messages through a (typically) asynchronous network, that often has some unpredictable delays, this can also be a contest of luck: assuming that the arbitration is done by Environment or a third party who is recording the order of moves, it is possible that Machine makes a move earlier than Environment does, but the message carrying that move is delivered to the arbiter only after Environment's move arrives, and the arbiter will be unable to see that it was Machine who moved first. An engineering-style attempt to neutralize this problem could be to let all moves carry timestamps. But this would not save the case: even though timestamps would correctly show the real order of moves by each particular player, they could not be used to compare two moves by two different players, for the clocks of the two players would never be perfectly synchronized.

Another attempt to deal with problems in the above style could be to assign to each player, in a strictly alternating order, a constant-length time slot during which the player has exclusive access to the communication medium. Let alone that this could introduce some unfair asymmetry in favor of the player who gets the first slot, the essence of the problem would still not be taken care of: some games would still essentially depend on the relative speeds of the two players, even if arguably no longer on the speed of the network.

Formally, the concept of static games is defined in terms of "delays". We say that run Δ is a ***green*** (resp. ***red***) ***delay*** of run Γ iff the following two conditions are satisfied:

1. The moves of either color are arranged in Γ in the same order as in Δ.
2. For any $n,k \geq 1$, if the $k$th red (resp. green) move is made earlier than the $n$th green (resp. red) move in Γ, then so is it in Δ.

In other words, Δ is the result of shifting to the right ("delaying") some green (resp. red) moves in Γ without violating their order.

Example: ⟨0,2,1,3,4,5,7,8,6,10,9⟩ is a green delay of ⟨0,1,2,3,4,5,6,7,8,9,10⟩. The former is obtained from the latter by shifting to the right some green moves. When doing so, a green move can jump over a red move, but one green move cannot jump over another green move.

Now, we say that a constant game $G$ is **static** iff, for either player (color) $P$ and any runs Γ and Δ such that Δ is a $P$ delay of Γ, in the context of $G$, the following two conditions are satisfied:

1. If Γ is won by $P$, then so is Δ.
2. If $P$ does not offend in Γ, then neither does it in Δ.

This concept extends to all games by stipulating that a not-necessarily-constant game is static iff so are all of its instances.

All game operations studied in CoL have been shown to preserve the static property of games.[Jap03] So, as long as the atoms of an expression represent static games, so does the whole expression. One natural subset of all static games is the closure of elementary games under the operations of CoL.

As already noted, CoL restricts its attention to static games only (at least, at its present stage of development). To see why static games are really what CoL is willing to call "pure" computational problems, imagine a play over the delay-prone Internet. If there is a central arbiter (which can be located either in one of the players' computer or somewhere on a third, neutral territory) recording the order of moves, then the players have full access to information about the official version of the run that is being generated, even though they could suffer from their moves being delivered with delays. But let us make the situation even more dramatic: assume, as this is a typical case in distributed systems, that there is no central arbiter. Rather, each players' machine records moves in the order it receives them, so that we have what is called distributed arbitration. Every time a player makes a move, the move is appended to the player's internal record of the run and, at the same time, mailed to the adversary. And every time a message from the adversary arrives, the move contained in it is appended to the player's record of the run. The game starts. Seeing no messages from Environment, Machine decides to go ahead and make an opening move α. As it happens, Environment also decides to make an "opening" move β.

The messages carrying α and β cross. So, after they are both delivered, Machine's internal records show the position ⟨α,β⟩, while Environment thinks that the current position is ⟨β,α⟩. Both of the players decide to make two consecutive new moves: γ,δ and ε,ω, respectively, and the two pairs of messages, again, cross.

After making their second series of moves and receiving a second series of "replies" from their adversaries, both players decide to make no further moves. The game thus ends. According to Machine's records, the run was ⟨α,β,γ,δ,ε,ω⟩, while Environment thinks that the run was ⟨β,α,ε,ω,γ,δ⟩. As for the "real run", i.e. the real order in which these six moves were made (if this concept makes sense at all), it can be yet something different, such as, say, ⟨β,α,γ,ε,δ,ω⟩. A little thought can convince us that in any case the real run, as well as the version of the run seen by Environment, will be a green delay of the version of the run seen by Machine. Similarly, the real run, as well as the version of the run seen by Machine, will be a red delay of the version of the run seen by Environment. Hence, provided that the game is static, either player can fully trust its own version of the run and only care about making good moves for this version, because regardless of whether it shows the true or a distorted picture of the real run, the latter is guaranteed to be successful as long as the former is. Moreover: for similar reasons, the player will remain equally successful if, instead of immediately appending the adversary's moves to its own version of the run, it simply queues those moves in a buffer as if they had not arrived yet, and fetches them only later at a more convenient time, after perhaps making and appending to its records some of its own moves first. The effect will amount to having full control over the speed of the adversary, thus allowing the player to select its own pace for the play and worry only about what moves to make rather than how quickly to make them.

Thus, static games allow players to make a full abstraction from any specific assumptions regarding the type of arbitration (central or distributed), the speed of the communication network and the speed of the adversary: whatever strategy they follow, it is always safe to assume or pretend that the arbitration is fair and unique (central), the network is perfectly accurate (zero delays) and the adversary is "slow enough". On the other hand, with some additional thought, we can see that if a game is not static, there will always be situations when no particular one of the above three sorts of abstractions can be made. Specifically, such situations will emerge every time when a player $P$'s strategy generates a $P$-won run that has some $P$-lost $P$-delays.

# 3 The CoL zoo of game operations

## 3.1 Preview

As we already know, logical operators in CoL stand for operations on games. There is an open-ended pool of operations of potential interest, and which of those to study may depend on particular needs and taste. Below is an incomplete list of the operators that have been officially introduced so far.

***Negation:*** ¬

***Conjunctions:*** ∧ (*parallel*), ⊓ (*choice*), △ (*sequential*), ⩚ (*toggling*)

***Disjunctions:*** ∨ (*parallel*), ⊔ (*choice*), ▽ (*sequential*), ⩛ (*toggling*)

***Recurrences:*** ↓ (*branching*), ⋏ (*parallel*), ⩟ (*sequential*), ⩞ (*toggling*)

***Corecurrences:*** ↑ (*branching*), ⋎ (*parallel*), ⩝ (*sequential*), ⩜ (*toggling*)

***Universal quantifiers:*** ∀ (*blind*), ⋀ (*parallel*), ⊓ (*choice*), △ (*sequential*), ⩚ (*toggling*)

***Existential quantifiers:*** ∃ (*blind*), ⋁ (*parallel*), ⊔ (*choice*), ▽ (*sequential*), ⩛ (*toggling*)

***Implications:*** → (*parallel*), ⊐ (*choice*), ▷ (*sequential*), ⪢ (*toggling*)

***Rimplications:*** ∘− (*branching*), ≻− (*parallel*), ▷− (*sequential*), ≻− (*toggling*)

***Refutations:*** ∘⇁ (*branching*), ≻⇁ (*parallel*), ▷⇁ (*sequential*), ≻⇁ (*toggling*)

Among these we see all operators of classical logic, and our choice of the classical notation for them is no accident. It was pointed out <u>earlier</u> that classical logic is nothing but the elementary, zero-interactivity fragment of computability logic. Indeed, after analyzing the relevant definitions, each of the classically-shaped operations, when restricted to <u>elementary games</u>, can be easily seen to be virtually the same as the corresponding operator of classical logic. For instance, if $A$ and $B$ are elementary games, then so is $A \wedge B$, and the latter is exactly the classical conjunction of $A$ and $B$ understood as an (elementary) game. In a general --- not-necessarily-elementary --- case, however, ¬, ∧, ∨, → become more reminiscent of (yet not the same as) the corresponding multiplicative operators of linear logic. Of course, here we are comparing apples with oranges for, as noted earlier, linear logic is a syntax while computability logic is a semantics, and it may be not clear in what precise sense one can talk about similarities or differences.

In the same apples and oranges style, our operations ⊓, ⊔, ⊓, ⊔ can be perceived as relatives of the additive connectives and quantifiers of linear logic; ∧,∨ as "multiplicative quantifiers"; ↓,↑,⋏,⋎ as "exponentials", even though it is hard to guess which of the two groups --- ↓,↑ or ⋏,⋎ --- would be closer to an orthodox linear logician's heart. On the other hand, the blind, sequential and toggling groups of operators have no counterparts in linear logic.

In this section we are going to see intuitive explanations as well as formal definitions of all of the above-listed operators. We agree that throughout those definitions, Φ ranges over positions, and Γ over runs. Each such definition has two clauses, one telling us when a position is a legal position of the compound game, and the other telling us who wins any given legal run. The run Γ seen in the second clause of the definition is always implicitly assumed to be a legal legal run of the game that is being defined.

This section also provides many examples of particular games. Let us agree that, unless otherwise suggested by the context, in all those cases we have the ideal universe in mind. Often we let non-numerals such as people, Turing machines, etc. act in the roles of "constants". These should be understood as abbreviations of the corresponding decimal numerals that encode these objects in some fixed reasonable encoding. It should also be remembered that algorithmicity is a minimum requirement on ⊤'s strategies. Some of our examples implicitly assume stronger requirements, such as efficiency or ability to act with imperfect knowledge. For instance, the problem of telling whether there is or has been life on Mars is, of course, decidable, for this is a finite problem. Yet our knowledge does not allow us to actually solve the problem. Similarly, chess is a finite game and (after ruling out the possibility of draws) one of the players does have a winning strategy in it. Yet we do not know specifically what (and which player's) strategy is a winning one.

When omitting parentheses in compound expressions, we assume that all unary operators (negation, refutations, recurrences, corecurrences and quantifiers) take precedence over all binary operators (conjunctions, disjunctions, implications, rimplications), among which implications and rimplications have the lowest precedence. So, for instance, $A\to\neg B\vee C$ should be understood as $A\to((\neg B)\vee C)$ rather than, say, as $(A\to\neg B)\vee C$ or as $A\to(\neg(B\vee C))$.

**Theorem 3.1.1** (proven in [Jap03, Jap08b, Jap11a]) All operators listed in this subsection preserve the static property of games (i.e., when applied to static games, the resulting game is also static).

## 3.2 Prefixation

Unlike the operators listed in the preceding subsection, the operation of prefixation is not meant to act as a logical operator in the formal language of CoL. Yet, it is very useful in characterizing and analyzing games, and we want to start our tour of the zoo with it.

**Definition 3.2.1** Assume $A$ is a game and Ψ is a unilegal position of $A$ (otherwise the operation is undefined). The Ψ–*prefixation* of $A$, denoted $\langle\Psi\rangle A$, is defined as the game $G$ with $\mathbf{Un}^G=\mathbf{Un}^A$, $\mathbf{Vr}^G=\mathbf{Vr}^A$ and with $\mathbf{Mp}^G$ such that, for any $G$-valuation $e$, we have:

- $\mathbf{Lr}_e^G=\{\Phi|\ \langle\Psi,\Phi\rangle\in\mathbf{Lr}_e^A\}$;

- $\mathbf{Wn}_e^G \langle \Gamma \rangle = \mathbf{Wn}_e^A \langle \Psi, \Gamma \rangle$.

Intuitively, $\langle \Psi \rangle A$ is the game playing which means playing $A$ starting (continuing) from position $\Psi$. That is, $\langle \Psi \rangle A$ is the game to which $A$ **evolves** (is **brought down**) after the moves of $\Psi$ have been made. Visualized as a tree, $\langle \Psi \rangle A$ is nothing but the subtree of $A$ rooted at the node corresponding to position $\Psi$. Below is an illustration.

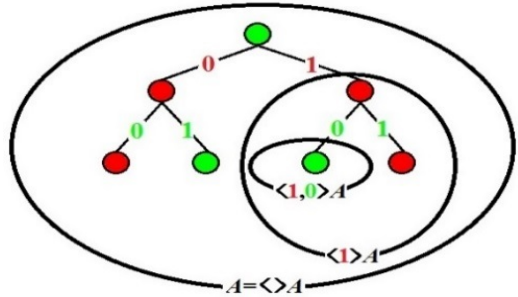

## 3.3 Negation

For a run $\Gamma$, by $\neg\Gamma$ we mean the "negative image" of $\Gamma$ (green and red interchanged). For instance, $\neg\langle\alpha,\beta,\gamma\rangle = \langle\alpha,\beta,\gamma\rangle$.

**Definition 3.3.1** Assume $A$ is a game. $\neg A$ is defined as the game $G$ with $\mathbf{Un}^G = \mathbf{Un}^A$, $\mathbf{Vr}^G = \mathbf{Vr}^A$ and with $\mathbf{Mp}^G$ such that, for any $G$-valuation $e$, we have:

- $\mathbf{Lr}_e^G = \{\neg\Phi : \Phi \in \mathbf{Lr}_e^A\}$;
- $\mathbf{Wn}_e^G \langle \Gamma \rangle = \top$ iff $\mathbf{Wn}_e^A \langle\neg\Gamma\rangle = \bot$.

In other words, $\neg A$ is $A$ with the roles of the two players interchanged: Machine's (legal) moves and wins become Environment's moves and wins, and vice versa. So, when visualized as a tree, $\neg A$ is the exact negative image of $A$, as illustrated below:

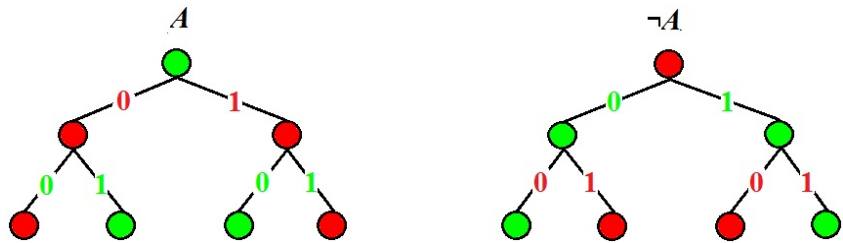

**Figure 3.3.2:** Negation

Let *Chess* be the game of chess (with draws ruled out) from the point of view of the white player. Then $\neg$*Chess* is *Chess* "upside down", i.e., the game of chess from the point of view of the black player:

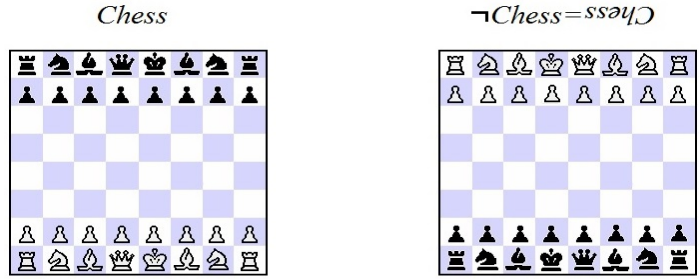

Observe that the classical double negation principle $\neg\neg A = A$ holds: interchanging the players' roles twice restores the original roles of the players. It is also easy to see that we always have $\neg\langle\Psi\rangle A = \langle\neg\Psi\rangle\neg A$. So, for instance, if α is Machine's legal move in the empty position of $A$ that brings $A$ down to $B$, then the same α is Environment's legal move in the empty position of $\neg A$, and it brings $\neg A$ down to $\neg B$. Test the game $A$ of Figure 3.3.2 to see that this is indeed so.

## 3.4 Choice operations

*Choice conjunction* ("*chand*") ⊓ and *choice disjunction* ("*chor*") ⊔ combine games in the way seen below:

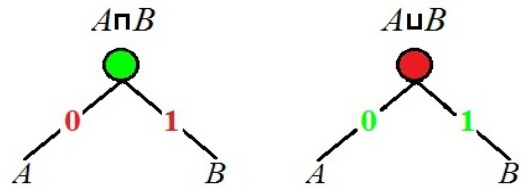

$A \sqcap B$ is a game where, in the initial (empty) position, only Environment has legal moves. Such a move should be either '0' or '1'. If Environment moves 0, the game continues as $A$, meaning that $\langle 0\rangle(A \sqcap B) = A$; if it moves 1, then the game continues as $B$, meaning that $\langle 1\rangle(A \sqcap B) = B$; and if it fails to make either move ("choice"), then it loses. $A \sqcup B$ is similar, with the difference that here it is Machine who has initial moves and who loses if no such move is made.

**Definition 3.4.1** Assume $A_0$ and $A_1$ are games with a common universe $U$.

a) $A_0 \sqcap A_1$ is defined as the game $G$ with $\mathbf{Un}^G = U$, $\mathbf{Vr}^G = \mathbf{Vr}^{A_0} \cup \mathbf{Vr}^{A_1}$ and with $\mathbf{Mp}^G$ such that, for any $G$-valuation $e$, we have:

- $\mathbf{Lr}_e^G = \{\langle\,\rangle\} \cup \{\langle i,\Phi\rangle: i \in \{0,1\}, \Phi \in \mathbf{Lr}_e^{A_i}\}$;
- $\mathbf{Wn}_e^G \langle\,\rangle = \top$; $\mathbf{Wn}_e^G \langle i,\Gamma\rangle = \mathbf{Wn}_e^{A_i} \langle\Gamma\rangle$.

b) $A_0 \sqcup A_1$ is defined as the game $G$ with $\mathbf{Un}^G = U$, $\mathbf{Vr}^G = \mathbf{Vr}^{A_0} \cup \mathbf{Vr}^{A_1}$ and with $\mathbf{Mp}^G$ such that, for any $G$-valuation $e$, we have:

- $\mathbf{Lr}^G = \{\langle\,\rangle\} \cup \{\langle i,\Phi\rangle: i \in \{0,1\}, \Phi \in \mathbf{Lr}^{A_i}\}$;
- $\mathbf{Wn}^G \langle\,\rangle = \bot$; $\mathbf{Wn}^G \langle i,\Gamma\rangle = \mathbf{Wn}^{A_i} \langle\Gamma\rangle$.

The game $A$ of Figure 3.3.2 can now be easily seen to be $(\bot \sqcup \top) \sqcap (\top \sqcup \bot)$, and its negation be $(\top \sqcap \bot) \sqcup (\bot \sqcap \top)$. The De Morgan laws familiar from classical logic persist: we always have $\neg (A \sqcap B) = \neg A \sqcup \neg B$ and $\neg (A \sqcup B) = \neg A \sqcap \neg B$. Together with the earlier observed double negation principle, this means that $A \sqcup B = \neg (\neg A \sqcap \neg B)$ and $A \sqcap B = \neg (\neg A \sqcup \neg B)$. Similarly for the quantifier counterparts $\sqcap$ and $\sqcup$ of $\sqcap$ and $\sqcup$.

Given a game $A(x)$, the *choice universal quantification* ("chall") $\sqcap x A(x)$ of it is nothing but the "infinite choice conjunction" $A(0) \sqcap A(1) \sqcap A(2) \sqcap \ldots$, and the *choice existential quantification* ("chexists") $\sqcup x A(x)$ of $A(x)$ is the "infinite choice disjunction" $A(0) \sqcup A(1) \sqcup A(2) \sqcup \ldots$:

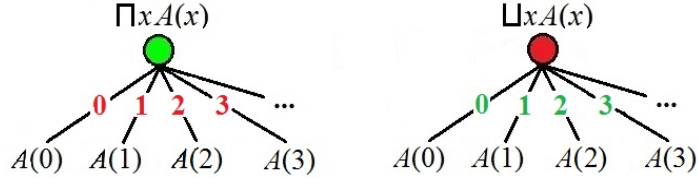

Specifically, $\sqcap x A(x)$ is a game where, in the initial position, only Environment has legal moves, and such a move should be one of the constants. If Environment moves $c$, then the game continues as $A(c)$, and if Environment fails to make an initial move/choice, then it loses. $\sqcup x A(x)$ is similar, with the difference that here it is Machine who has initial moves and who loses if no such move is made. So, we always have $\langle c \rangle \sqcap x A(x) = A(c)$ and $\langle c \rangle \sqcup x A(x) = A(c)$. Below is a formal definition of all choice operations.

**Definition 3.4.2** Assume $x$ is a variable and $A=A(x)$ is a game.

a) $\sqcap x A(x)$ is defined as the game $G$ with $\mathbf{Un}^G = \mathbf{Un}^A$, $\mathbf{Vr}^G = \mathbf{Vr}^A - \{x\}$ and with $\mathbf{Mp}^G$ such that, for any $G$-valuation $e$, we have:

- $\mathbf{Lr}_e^G = \{\langle \, \rangle\} \cup \{\langle c, \Phi \rangle : c \in Constants, \Phi \in \mathbf{Lr}_e^{A(c)}\}$;
- $\mathbf{Wn}_e^G \langle \, \rangle = \top$; $\mathbf{Wn}_e^G \langle c, \Gamma \rangle = \mathbf{Wn}_e^{A(c)} \langle \Gamma \rangle$.

b) $\sqcup x A(x)$ is defined as the game $G$ with $\mathbf{Un}^G = \mathbf{Un}^A$, $\mathbf{Vr}^G = \mathbf{Vr}^A - \{x\}$ and with $\mathbf{Mp}^G$ such that, for any $G$-valuation $e$, we have:

- $\mathbf{Lr}_e^G = \{\langle \, \rangle\} \cup \{\langle c, \Phi \rangle : c \in Constants, \Phi \in \mathbf{Lr}_e^{A(c)}\}$;
- $\mathbf{Wn}_e^G \langle \, \rangle = \bot$; $\mathbf{Wn}_e^G \langle c, \Gamma \rangle = \mathbf{Wn}_e^{A(c)} \langle \Gamma \rangle$.

Now we are already able to express traditional computational problems using formulas. Traditional problems come in two forms: the problem of computing a function $f(x)$, or the problem of deciding a predicate $p(x)$. The former can be written as $\sqcap x \sqcup y A(y=f(x))$, and the latter as $\sqcap x (\neg p(x) \sqcup p(x))$. So, for instance, the constant "successor" game of Figure 2.3.4 will be written as $\sqcap x \sqcup y A(y=x+1)$, and the unary "generalized successor" game of Figure 2.3.5 as $\sqcap x \sqcup y A(y=x+z)$. The following game, which is about deciding the "evenness" predicate, could be written as $\sqcap x (\neg \exists y(x=2y) \sqcup \exists y(x=2y))$ ($\exists$ will be officially defined later, but, as promised, its meaning is going to be exactly classical when applied to an elementary game like $x=2y$).

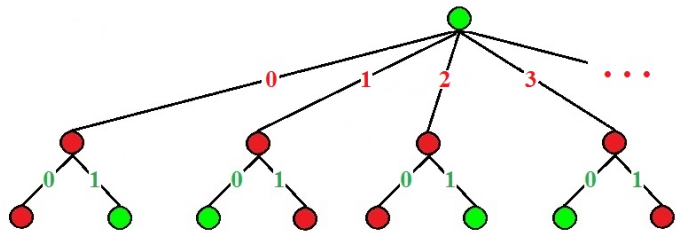

Classical logic has been repeatedly criticized for its operators not being constructive. Consider, for example, ∀x∃y(y=f(x)). It is true in the classical sense as long as f is a total function. Yet its truth has little (if any) practical import, as "∃y" merely signifies *existence* of y, without implying that such a y can actually be *found*. And, indeed, if f is an incomputable function, there is no method for finding y. On the other hand, the choice operations of CoL *are* constructive. Computability ("truth") of ⊓x⊔yA(y=f(x)) means more than just existence of y; it means the possibility to actually *find* (compute, construct) a corresponding y for every x.

Similarly, let *Halts*(x,y) be the predicate "Turing machine x halts on input y". Consider ∀x∀y(¬Halts(x,y) ∨ Halts(x,y)). It is true in classical logic, yet not in a constructive sense. Its truth means that, for all x and y, either ¬Halts(x,y) or Halts(x,y) is true, but it does not imply existence of an actual way to tell which of these two is true after all. And such a way does not really exist, as the halting problem is undecidable. This means that ⊓x⊓y(¬Halts(x,y) ⊔ Halts(x,y)) is not computable. Generally, as pointed out [earlier](), the principle of the excluded middle "¬A OR A", validated by classical logic and causing the indignation of the constructivistically-minded, is not valid in computability logic with OR understood as choice disjunction. The following is an example of a constant game of the form ¬A ⊔ A with no algorithmic solution (why, by the way?):

$$\neg \sqcap x \sqcap y(\neg Halts(x,y) \sqcup Halts(x,y)) \sqcup \sqcap x \sqcap y(\neg Halts(x,y) \sqcup Halts(x,y)).$$

*Chess* ⊔ ¬ *Chess*, on the other hand, is an example of a computable-in-principle yet "practically incomputable" problem, with no real computer anywhere close to being able to handle it.

There is no need to give a direct definition for the remaining choice operation of **choice implication** ("**chimplication**"), for it can be defined in terms of ¬, ⊔ in the "standard" way:

**Definition 3.4.3** A⊐B =_def ¬A ⊔ B.

Each of the other sorts of disjunctions (parallel, sequential and toggling) generates the corresponding implication the same way.

## 3.5 Parallel operations

The **parallel conjunction** ("**pand**") A∧B and the **parallel disjunction** ("**por**") A∨B of A and B are games playing which means playing the two games simultaneously. In order to win in A∨B (resp. A∧B), ⊤ needs to win in both (resp. at least one) of the components A,B. For instance, ¬Chess∨Chess is a two-board game, where ⊤ plays black on the left board and white on the right board, and where it needs to win in at least one of the two parallel sessions of chess. A win can be easily achieved here by just mimicking in *Chess* the moves that the adversary makes in ¬*Chess*, and vice versa. This *copycat strategy* guarantees that the positions on the two boards always remain symmetric, as illustrated below, and thus ⊤ eventually loses on one board but wins on the other.

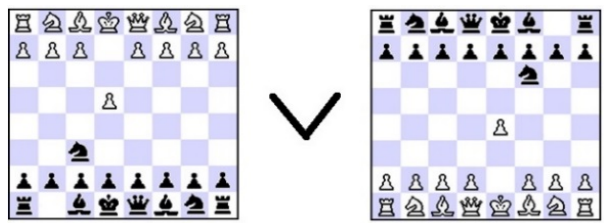

This is very different from ¬*Chess* ⊔ *Chess*. In the latter ⊤ needs to choose between the two components and then win the chosen one-board game, which makes ¬*Chess* ⊔ *Chess* essentially as hard to win as either ¬*Chess* or *Chess*. A game of the form $A \lor B$ is generally easier (at least, not harder) to win than $A \sqcup B$, the latter is easier to win than $A \sqcap B$, and the latter in turn is easier to win than $A \land B$.

Technically, a move α in the left (resp. right) ∧-conjunct or ∨-disjunct is made by prefixing α with '0.'. For instance, in (the initial position of) $(A \sqcup B) \lor (C \sqcap D)$, the move '1.0' is legal for ⊥, meaning choosing the left ⊓-conjunct in the second ∨-disjunct of the game. If such a move is made, the game continues as $(A \sqcup B) \lor C$. The player ⊤, too, has initial legal moves in $(A \sqcup B) \lor (C \sqcap D)$, which are '0.0' and '0.1'.

In the formal definitions of this section and throughout the rest of this webpage, we use the important notational convention according to which:

**Notation 3.5.1** For a run Γ and string α, $\Gamma^\alpha$ means the result of removing from Γ all moves except those of the form αβ, and then deleting the prefix 'α' in the remaining moves.

So, for instance, $\langle 1.2, 1.0, 0.33 \rangle^{1.} = \langle 2, 0 \rangle$ and $\langle 1.2, 1.0, 0.33 \rangle^{0.} = \langle 33 \rangle$. Another example: where Γ is the leftmost branch of the tree for $(\top \sqcap \bot) \lor (\bot \sqcap \top)$ shown in Figure 3.5.3, we have $\Gamma^{0.} = \langle 1 \rangle$ and $\Gamma^{1.} = \langle 1 \rangle$. Intuitively, we see this Γ as consisting of two subruns, one ($\Gamma^{0.}$) being a run in the first ∨-disjunct of $(\top \sqcap \bot) \lor (\bot \sqcap \top)$, and the other ($\Gamma^{1.}$) being a run in the second disjunct.

**Definition 3.5.2** Assume $A_0$ and $A_1$ are games with a common universe U.

a) $A_0 \land A_1$ is defined as the game G with $\mathbf{Un}^G = U$, $\mathbf{Vr}^G = \mathbf{Vr}^{A_0} \cup \mathbf{Vr}^{A_1}$ and with $\mathbf{Mp}^G$ such that, for any G-valuation e, we have:

- $\Phi \in \mathbf{Lr}_e^G$ iff every move of Φ has the prefix '0.' or '1.' and, for both $i \in \{0,1\}$, $\Phi^{i.} \in \mathbf{Lr}_e^{A_i}$;
- $\mathbf{Wn}_e^G \langle \Gamma \rangle = \top$ iff, for both $i \in \{0,1\}$, $\mathbf{Wn}_e^{A_i} \langle \Gamma^{i.} \rangle = \top$.

b) $A_0 \lor A_1$ is defined as the game G with $\mathbf{Un}^G = U$, $\mathbf{Vr}^G = \mathbf{Vr}^{A_0} \cup \mathbf{Vr}^{A_1}$ and with $\mathbf{Mp}^G$ such that, for any G-valuation e, we have:

- $\Phi \in \mathbf{Lr}_e^G$ iff every move of Φ has the prefix '0.' or '1.' and, for both $i \in \{0,1\}$, $\Phi^{i.} \in \mathbf{Lr}_e^{A_i}$;
- $\mathbf{Wn}_e^G \langle \Gamma \rangle = \bot$ iff, for both $i \in \{0,1\}$, $\mathbf{Wn}_e^{A_i} \langle \Gamma^{i.} \rangle = \bot$.

When A and B are (constant) finite games, the depth of $A \land B$ or $A \lor B$ is the sum of the depths of A and B, as seen below.

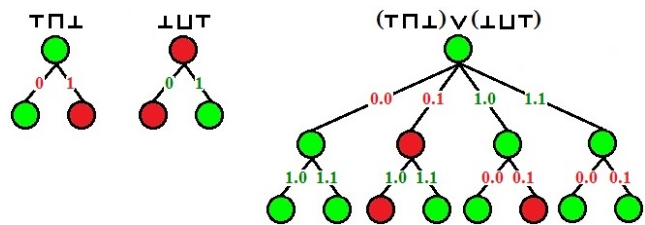

This signifies an exponential growth of the breadth, meaning that, once we have reached the level of parallel operations, continuing drawing trees in the earlier style becomes no fun. Not to be disappointed though: making it possible to express large- or infinite-size game trees in a compact way is what our game operators are all about after all.

Whether trees are or are not helpful in visualizing parallel combinations of unistructural games, prefixation is still very much so if we think of each unilegal position $\Phi$ of $A$ as the game $\langle\Phi\rangle A$. This way, every unilegal run $\Gamma$ of $A$ becomes a sequence of games as illustrated in the following example.

**Example 3.5.4**   To the (uni)legal run $\Gamma = \langle 1.7, 0.7, 0.49, 1.49 \rangle$ of game $A = \sqcup x \sqcap y(y \neq x^2) \vee \sqcap x \sqcup y(y=x^2)$ induces the following sequence, showing how things evolve as $\Gamma$ runs, i.e., how the moves of $\Gamma$ affect/modify the game that is being played:

- $\sqcup x \sqcap y(y \neq x^2) \vee \sqcap x \sqcup y(y=x^2)$     i.e. $A$
- $\sqcup x \sqcap y(y \neq x^2) \vee \sqcup y(y=7^2)$     i.e. $\langle 1.7 \rangle A$
- $\sqcap y(y \neq 7^2) \vee \sqcup y(y=7^2)$     i.e. $\langle 1.7, 0.7 \rangle A$
- $49 \neq 7^2 \vee \sqcup y(y=7^2)$     i.e. $\langle 1.7, 0.7, 0.49 \rangle A$
- $49 \neq 7^2 \vee 49 = 7^2$     i.e. $\langle 1.7, 0.7, 0.49, 1.49 \rangle A$

The run hits the true proposition $49 \neq 7^2 \vee 49 = 7^2$, and hence is won by the machine.

As we may guess, the **parallel universal quantification** ("*pall*") $\wedge x A(x)$ of $A(x)$ is nothing but $A(0) \wedge A(1) \wedge A(2) \wedge \ldots$ and the **parallel existential quantification** ("*pexists*") $\vee x A(x)$ of $A(x)$ is nothing but $A(0) \vee A(1) \vee A(2) \vee \ldots$

**Definition 3.5.5** Assume $x$ is a variable and $A=A(x)$ is a game.

a) $\wedge x A(x)$ is defined as the game $G$ with $\mathbf{Un}^G = \mathbf{Un}^A$, $\mathbf{Vr}^G = \mathbf{Vr}^A - \{x\}$ and with $\mathbf{Mp}^G$ such that, for any $G$-valuation $e$, we have:

- $\Phi \in \mathbf{Lr}_e^G$ iff every move of $\Phi$ has the prefix '$c.$' for some $c \in Constants$ and, for all such $c$, $\Phi^{c.} \in \mathbf{Lr}_e^{A(c)}$;
- $\mathbf{Wn}_e^G \langle \Gamma \rangle = \top$ iff, for all $c \in Constants$, $\mathbf{Wn}_e^A \langle \Gamma^{c.} \rangle = \top$.

b) $\vee x A(x)$ is defined as the game $G$ with $\mathbf{Un}^G = \mathbf{Un}^A$, $\mathbf{Vr}^G = \mathbf{Vr}^A - \{x\}$ and with $\mathbf{Mp}^G$ such that, for any $G$-valuation $e$, we have:

- $\Phi \in \mathbf{Lr}_e^G$ iff every move of $\Phi$ has the prefix '$c.$' for some $c \in Constants$ and, for all such $c$, $\Phi^{c.} \in \mathbf{Lr}_e^{A(c)}$;
- $\mathbf{Wn}_e^G \langle \Gamma \rangle = \bot$ iff, for all $c \in Constants$, $\mathbf{Wn}_e^A \langle \Gamma^{c.} \rangle = \bot$.

The next group of parallel operators are **parallel recurrence** ("*precurrence*") $\lambda$ and **parallel corecurrence** ("*coprecurrence*") $\curlyvee$. $\lambda A$ is nothing but the infinite parallel conjunction $A \wedge A \wedge A \wedge \ldots$, and $\curlyvee A$ is the infinite parallel disjunction $A \vee A \vee A \vee \ldots$. Equivalently, $\lambda A$ and $\curlyvee A$ can be respectively understood as $\wedge x A$ and $\vee x A$, where $x$ is a dummy variable on which $A$ does not depend. Intuitively, playing $\lambda A$ means simultaneously playing in infinitely many "copies" of $A$, and $\top$ is the winner iff it wins $A$ in all copies. $\curlyvee A$ is similar, with the only difference that here winning in just one copy is sufficient.

**Definition 3.5.6** Assume $A$ is a game.

a) $\lambda A$ is defined as the game $G$ with $\mathbf{Un}^G = \mathbf{Un}^A$, $\mathbf{Vr}^G = \mathbf{Vr}^A$ and with $\mathbf{Mp}^G$ such that, for any $G$-valuation $e$, we have:

- $\Phi \in \mathbf{Lr}_e^G$ iff every move of $\Phi$ has the prefix '$c.$' for some $c \in Constants$ and, for all such $c$, $\Phi^{c.} \in \mathbf{Lr}_e^A$;
- $\mathbf{Wn}_e^G \langle \Gamma \rangle = \top$ iff, for all $c \in Constants$, $\mathbf{Wn}_e^A \langle \Gamma^{c.} \rangle = \top$.

b) $\curlyvee A$ is defined as the game $G$ with $\mathbf{Un}^G = \mathbf{Un}^A$, $\mathbf{Vr}^G = \mathbf{Vr}^A$ and with $\mathbf{Mp}^G$ such that, for any $G$-valuation $e$, we have:

- $\Phi \in \mathbf{Lr}_e^G$ iff every move of $\Phi$ has the prefix '$c.$' for some $c \in Constants$ and, for all such $c$, $\Phi^{c.} \in \mathbf{Lr}_e^A$;
- $\mathbf{Wn}_e^G \langle \Gamma \rangle = \bot$ iff, for all $c \in Constants$, $\mathbf{Wn}_e^A \langle \Gamma^{c.} \rangle = \bot$.

As was the case with choice operations, we can see that the definition of each parallel operations seen so far in this section can be obtained from the definition of its dual by just interchanging ⊤ with ⊥. Hence it is easy to verify that we always have ¬($A \wedge B$) = ¬$A \vee$¬$B$, ¬($A \vee B$) = ¬$A \wedge$¬$B$, ¬$\wedge x A(x) = \vee x$¬$A(x)$, ¬$\vee x A(x) = \wedge x$¬$A(x)$, ¬$\curlywedge A = \curlyvee$¬$A$, ¬$\curlyvee A = \curlywedge$¬$A$, This, in turn, means that each parallel operation is definable in the standard way in terms of its dual operation and negation. For instance, $A \wedge B$ can be defined as ¬(¬$A \vee$¬$B$), and ¬$\curlywedge A$ as ¬$\curlyvee$¬$A$. Three more parallel operations defined in terms of negation and other parallel operations are *parallel implication* ("*pimplication*") →, *parallel rimplication* ("*primplication*") ⤳ and *parallel refutation* ("*prefutation*") ⤳. Here the prefix "p", as before, stands for "parallel", and the prefix "r" in "rimplication" stands for "recurrence".

**Definition 3.5.7** a) $A \rightarrow B =_{\text{def}}$ ¬$A \vee B$
b) $A \succ\!\!- B =_{\text{def}} \curlywedge A \rightarrow B$
c) $\succ\!\!- A =_{\text{def}} A \succ\!\!- \bot$

Note that, just like negation and unlike choice operations, parallel operations preserve the elementary property of games. When restricted to elementary games, the meanings of $\wedge$, $\vee$ and $\rightarrow$ coincide with those of classical conjunction, disjunction and implication. Further, as long as all individuals of the universe have naming constants, the meanings of $\wedge$ and $\vee$ coincide with those of classical universal quantifier and existential quantifier. The same conservation of classical meaning (but without any conditions on the universe) is going to be the case with the blind quantifiers ∀,∃ defined later; so, at the elementary level, when all individuals of the universe have naming constants, $\wedge$ and $\vee$ are indistinguishable from ∀ and ∃, respectively. As for the parallel recurrence and corecurrence, for an elementary $A$ we simply have $\curlywedge A = \curlyvee A = A$.

While all classical tautologies automatically remain valid when parallel operators are applied to elementary games, in the general case the class of valid ("always computable") principles shrinks. For example, $P \rightarrow P \wedge P$, i.e. ¬$P \vee (P \wedge P)$, is not valid. Back to our chess example, one can see that the earlier copycat strategy successful for ¬$Chess \vee Chess$ would be inapplicable to ¬$Chess \vee (Chess \wedge Chess)$. The best that ⊤ can do in this three-board game is to synchronize ¬$Chess$ with one of the two conjuncts of $Chess \wedge Chess$. It is possible that then ¬$Chess$ and the unmatched $Chess$ are both lost, in which case the whole game will be lost.

The principle $P \rightarrow P \wedge P$ is valid in classical logic because the latter sees no difference between $P$ and $P \wedge P$. On the other hand, in virtue of its semantics, CoL is *resource-conscious*, and in it $P$ is by no means the same as $P \wedge P$ or $P \vee P$. Unlike $P \rightarrow P \wedge P$, $P \succ\!\!- P \wedge P$ is a valid principle. Here, in the antecedent, we have infinitely many "copies" of $P$. Pick any two copies and, via copycat, synchronize them with the two conjuncts of the consequent. A win is guaranteed.

# 3.6 Reduction

Intuitively, $A \rightarrow B$ is the problem of *reducing* $B$ to $A$: solving $A \rightarrow B$ means solving $B$ while having $A$ as a computational resource. Specifically, ⊤ may observe how $A$ is being solved (by the environment), and utilize this information in its own solving $B$. As already pointed out, resources are symmetric to problems: what is a problem to solve for one player is a resource that the other player can use, and vice versa. Since $A$ is negated in ¬ $A \lor B$ and negation means switching the roles, $A$ appears as a resource rather than problem for ⊤ in $A \to B$. Our copycat strategy for $\neg Chess \lor Chess$ was an example of reducing *Chess* to *Chess*. The same strategy was underlying Example 3.5.4, where $\sqcap x \sqcup y(y=x^2)$ was reduced to itself.

Let us look at a more meaningful example: reducing the acceptance problem to the halting problem. The former, as a decision problem, will be written as $\sqcap x \sqcup y\, (\neg Accepts(x,y) \sqcup Accepts(x,y))$, where $Accepts(x,y)$ is the predicate "Turing machine $x$ accepts input $y$". Similarly, the halting problem is written as $\sqcap x \sqcup y\, (\neg Halts(x,y) \sqcup Halts(x,y))$. Neither problem has an algorithmic solution, yet the following pimplication does:

$$\sqcap x \sqcup y\, (\neg Halts(x,y) \sqcup Halts(x,y)) \to \sqcap x \sqcup y\, (\neg Accepts(x,y) \sqcup Accepts(x,y))$$

Here is Machine's winning strategy for the above game. Wait till Environment makes moves $1.m$ and $1.n$ for some $m$ and $n$. Making these moves essentially means asking the question "*Does machine m accept input n?*". If such moves are never made, you win. Otherwise, the moves bring the game down to

$$\sqcap x \sqcup y\, (\neg Halts(x,y) \sqcup Halts(x,y)) \to \neg Accepts(m,n) \sqcup Accepts(m,n)$$

Make the moves $0.m$ and $0.n$, thus asking the counterquestion "*Does machine m halt on input n?*". Your moves bring the game down to

$$\neg Halts(m,n) \sqcup Halts(m,n) \to \neg Accepts(m,n) \sqcup Accepts(m,n)$$

Environment will have to answer this question, or else it loses (why?). If it answers by move $0.0$ ("*No, m does not halt on n*"), you make the move $1.0$ (say "*m does not accept n*"). The game will be brought down to $\neg Halts(m,n) \to \neg Accepts(m,n)$. You win, because if $m$ does not halt on $n$, then it does not accept $n$, either. Otherwise, if Environment answers by move $0.1$ ("*Yes, m halts on n*"), start simulating $m$ on $n$ until $m$ halts. If you see that $m$ accepted $n$, make the move $1.1$ (say "*m accepts n*"); otherwise make the move $1.0$ (say "*m does not accept n*"). Of course, it is a possibility that the simulation goes on forever. But then Environment has lied when saying "*m halts on n*"; in other words, the antecedent is false, and you still win.

Note that what Machine did when following the above strategy was indeed reducing the acceptance problem to the halting problem: it solved the former using an external (Environment-provided) solution of the latter.

There are many natural concepts of reduction, and a strong case can be made in favor of the thesis that the sort of reduction captured by $\to$ is most basic among them. For this reason, we agree that, if we simply say "reduction", it always means the sort of reduction captured by $\to$. A great variety of other reasonable concepts of reduction is expressible in terms of $\to$. Among those is Turing reduction. Remember that a predicate $q(x)$ is said to be *Turing reducible* to a predicate $p(x)$ if $q(x)$ can be decided by a Turing machine equipped with an oracle for $p(x)$. For a positive integer $n$, *n-bounded Turing reducibility* is defined in the same way, with the only difference that here the oracle is allowed to be used only $n$ times. It turns out that parallel rimplication $\succ\!\!-$ is a conservative generalization of Turing reduction. Namely, when $p(x)$ and $q(x)$ are elementary games (i.e. predicates), $q(x)$ is Turing reducible to $p(x)$ if and only if the problem $\sqcap x(\neg p(x) \sqcup p(x)) \succ\!\!- \sqcap x(\neg q(x) \sqcup q(x))$ has an algorithmic solution. If here we change $\succ\!\!-$ back to $\to$, we get the same result for 1-bounded Turing reducibility. More generally, as one might guess, $n$-bounded Turing reduction will be captured by

$$\sqcap x_1(\neg p(x_1) \sqcup p(x_1)) \land \ldots \land \sqcap x_n(\neg p(x_n) \sqcup p(x_n)) \to \sqcap x(\neg q(x) \sqcup q(x)).$$

If, instead, we write

$$\sqcap x_1 \ldots \sqcap x_n\bigl((\neg p(x_1) \sqcup p(x_1)) \land \ldots \land (\neg p(x_n) \sqcup p(x_n))\bigr) \to \sqcap x(\neg q(x) \sqcup q(x)),$$

then we get a conservative generalization of *n-bounded weak truth-table reduction*. The latter differs from *n*-bounded Turing reduction in that here all *n* oracle queries should be made at once, before seeing responses to any of those queries. What is called *mapping* (*many-one*) *reducibility* of $q(x)$ to $p(x)$ is nothing but computability of ⊓x⊔y(q(x)↔p(y)), where A↔B abbreviates (A→B)∧(B→A). We could go on and on with this list.

And yet many other natural concepts of reduction expressible in the language of CoL may have no established names in the literature. For example, from the previous discussion it can be seen that a certain reducibility-style relation holds between the predicates *Accepts*(*x,y*) and *Halts*(*x,y*) in an even stronger sense than the algorithmic winnability of

$$⊓x⊔y\,(¬Halts(x,y) ⊔ Halts(x,y)) → ⊓x⊔y\,(¬Accepts(x,y) ⊔ Accepts(x,y)).$$

In fact, not only the above problem has an algorithmic solution, but also the generally harder-to-solve problem

$$⊓x⊔y\,(¬Halts(x,y) ⊔ Halts(x,y) → ¬Accepts(x,y) ⊔ Accepts(x,y)).$$

Among the merits of CoL is that it offers a formalism and deductive machinery for systematically expressing and studying computation-theoretic relations such as reducibility, decidability, enumerability, etc., and all kinds of variations of such concepts.

Back to reducibility, while the standard approaches only allow us to talk about (a whatever sort of) reducibility as a *relation* between problems, in our approach reduction becomes an *operation* on problems, with reducibility as a relation simply meaning computability of the corresponding combination (such as $A→B$) of games. Similarly for other relations or properties such as the property of *decidability*. The latter becomes the operation of *deciding* if we define the problem of deciding a predicate (or any computational problem) $p(x)$ as the game ⊓x(¬p(x) ⊔ p(x)). So, now we can meaningfully ask questions such as "*Is the reduction of the problem of deciding $q(x)$ to the problem of deciding $p(x)$ always reducible to the [mapping reduction](#) of $q(x)$ to $p(x)$?*". This question would be equivalent to whether the following formula is valid in CoL:

$$⊓x⊔y(q(x) ↔ p(y)) → \bigl(⊓x(¬p(x) ⊔ p(x)) → ⊓x(¬q(x) ⊔ q(x))\bigr).$$

The answer turns out to be "Yes", meaning that mapping reduction is at least as strong as reduction. Here is a strategy which wins this game no matter what particular predicates $p(x)$ and $q(x)$ are:

1. Wait till, for some *m*, Environment brings the game down to

$$⊓x⊔y(q(x) ↔ p(y)) → \bigl(⊓x(¬p(x) ⊔ p(x)) → ¬q(m) ⊔ q(m)\bigr).$$

2. Bring the game down to

$$⊔y(q(m) ↔ p(y)) → \bigl(⊓x(¬p(x) ⊔ p(x)) → ¬q(x) ⊔ q(x)\bigr).$$

3. Wait till, for some *n*, Environment brings the game down to

$$(q(m) ↔ p(n)) → \bigl(⊓x(¬p(x) ⊔ p(x)) → ¬q(x) ⊔ q(x)\bigr).$$

4. Bring the game down to

$$(q(m) ↔ p(n)) → \bigl(¬p(n) ⊔ p(n) → ¬q(x) ⊔ q(x)\bigr).$$

5. Wait till Environment brings the game down to one of the following:

5a. $(q(m) \leftrightarrow p(n)) \rightarrow (\neg p(n) \rightarrow \neg q(x) \sqcup q(x))$. In this case, further bring it down to $(q(m) \leftrightarrow p(n)) \rightarrow (\neg p(n) \rightarrow \neg q(x))$, and you have won.

5b. $(q(m) \leftrightarrow p(n)) \rightarrow (p(n) \rightarrow \neg q(x) \sqcup q(x))$. In this case, further bring it down to $(q(m) \leftrightarrow p(n)) \rightarrow (p(n) \rightarrow q(x))$, and you have won, again.

We can also ask: "*Is the mapping reduction of q(x) to p(x) always reducible to the reduction of the problem of deciding q(x) to the problem of deciding p(x)?*". This question would be equivalent to whether the following formula is valid:

$$\bigl(\sqcap x(\neg p(x) \sqcup p(x)) \rightarrow \sqcap x(\neg q(x) \sqcup q(x))\bigr) \rightarrow \sqcap x \sqcup y(q(x) \leftrightarrow p(y)).$$

The answer here turns out to be "No", meaning that mapping reduction is properly stronger than reduction. This negative answer can be easily obtained by showing that the above formula is not provable in the deductive system **CL4** that we are going to see later. **CL4** is sound and complete with respect to validity. Its completeness implies that any formula which is not provable in it (such as the above formula) is not valid. And the soundness of **CL4** implies that every provable formula is valid. So, had our ad hoc attempt to come up with a strategy for $\sqcap x \sqcup y(q(x) \leftrightarrow p(y)) \rightarrow \bigl(\sqcap x(\neg p(x) \sqcup p(x)) \rightarrow \sqcap x(\neg q(x) \sqcup q(x))\bigr)$ failed, its validity could have been easily established by finding a **CL4**-proof of it.

To summarize, CoL offers not only a convenient language for specifying computational problems and relations or operations on them, but also a systematic tool for answering questions in the above style and beyond.

## 3.7 Blind operations

Our definition of the ***blind universal quantifier*** ("***blall***") $\forall x$ and ***blind existential quantifier*** ("***blexists***") $\exists x$ assumes that the game $A(x)$ to which they are applied satisfies the condition of unistructurality in $x$. This condition is weaker than the earlier seen unistructurality: every unistructural game is also unistructural in $x$, but not vice versa. Intuitively, unistructurality in $x$ means that the structure of the game does not depend on (how the valuation evaluates) the variable $x$. Formally, we say that $A(x)$ is ***unistructural in*** $x$ iff, for any valuation $e$ and any two constants $a$ and $b$, we have $\mathbf{Lr}_e^{A(a)} = \mathbf{Lr}_e^{A(b)}$. All constant or elementary games are unistructural in (whatever variable) $x$. And all operations of CoL are known to preserve unistructurality in $x$.

**Definition 3.7.1** Assume $x$ is a variable and $A=A(x)$ is a game unistructural in $x$.

a) $\forall x A(x)$ is defined as the game $G$ with $\mathbf{Un}^G = \mathbf{Un}^A$, $\mathbf{Vr}^G = \mathbf{Vr}^A - \{x\}$ and with $\mathbf{Mp}^G$ such that, for any $G$-valuation $e$, we have:

- $\mathbf{Lr}_e^G = \mathbf{Lr}_e^A$;
- $\mathbf{Wn}_e^G \langle \Gamma \rangle = \top$ iff, for all $a \in \mathbf{Dm}^G$, $\mathbf{Wn}_e^{A(a)} \langle \Gamma \rangle = \top$.

b) $\exists x A(x)$ is defined as the game $G$ with $\mathbf{Un}^G = \mathbf{Un}^A$, $\mathbf{Vr}^G = \mathbf{Vr}^A - \{x\}$ and with $\mathbf{Mp}^G$ such that, for any $G$-valuation $e$, we have:

- $\mathbf{Lr}_e^G = \mathbf{Lr}_e^A$;
- $\mathbf{Wn}_e^G \langle \Gamma \rangle = \bot$ iff, for all $a \in \mathbf{Dm}^G$, $\mathbf{Wn}_e^{A(a)} \langle \Gamma \rangle = \bot$.

Intuitively, playing $\forall xA(x)$ or $\exists xA(x)$ means just playing $A(x)$ "blindly", without knowing the value of $x$. In $\forall xA(x)$, Machine wins iff the play it generates is successful for every possible value of $x$ from the domain, while in $\exists xA(x)$ being successful for just one value is sufficient. When applied to elementary games, the blind quantifiers act exactly like the quantifiers of classical logic, even if not all individuals of the universe have naming constants.

From the definition one can see a perfect symmetry between $\forall$ and $\exists$. Therefore, as with the other quantifiers seen so far, the standard De Morgan laws and interdefinabilities hold.

Unlike $\wedge xA(x)$ which is a game on infinitely many boards, both $\sqcap xA(x)$ and $\forall xA(x)$ are one-board games. Yet, they are very different from each other. To see this difference, compare the problems $\sqcap x(Even(x) \sqcup Odd(x))$ and $\forall x(Even(x) \sqcup Odd(x))$. The former is an easily winnable game of depth 2: Environment selects a number, and Machine tells whether that number is even or odd. The latter, on the other hand, is a game which is impossible to win. This is a game of depth 1, where the value of $x$ is not specified by either player, and only Machine moves --- tells whether (the unknown) $x$ is even or odd. Whatever Machine says, it loses, because there is always a value for $x$ that makes the answer wrong.

This should not suggest that nontrivial $\forall$-games can never be won. For instance, the problem $\forall x\bigl(Even(x) \sqcup Odd(x) \to \sqcap y(Even(x+y) \sqcup Odd(x+y))\bigr)$ has and easy solution. The idea of a winning strategy is that, for any given $y$, in order to tell whether $x+y$ is even or odd, it is not really necessary to know the value of $x$. Rather, just knowing whether $x$ is even or odd is sufficient. And such knowledge can be obtained from the antecedent. In other words, for any known $y$ and unknown $x$, the problem of telling whether $x+y$ is even or odd is reducible to the problem of telling whether $x$ is even or odd. Specifically, if both $x$ and $y$ are even or both are odd, then $x+y$ is even; otherwise $x+y$ is odd. Below is the evolution sequence induced by the run $\langle 1.5, 0.0, 1.1 \rangle$ where Machine used such a strategy.

- $\forall x\bigl(Even(x) \sqcup Odd(x) \to \sqcap y(Even(x+y) \sqcup Odd(x+y))\bigr)$
- $\forall x\bigl(Even(x) \sqcup Odd(x) \to Even(x+5) \sqcup Odd(x+5)\bigr)$
- $\forall x\bigl(Even(x) \to Even(x+5) \sqcup Odd(x+5)\bigr)$
- $\forall x\bigl(Even(x) \to Odd(x+5)\bigr)$

Machine won because the play hit the true $\forall x(Even(x) \to Odd(x+5))$. Notice how $\forall x$ persisted throughout the sequence. Generally, the ($\forall,\exists$)-structure of a game will remain unchanged in such sequences. The same is the case with the parallel operations such as $\to$ in the present case.

**Exercise 3.7.2** Are the following problems always computable?

$\forall xA(x) \to \sqcap xA(x)$ answer
$\sqcap xA(x) \to \forall xA(x)$ answer
$\sqcap xA(x) \to \exists xA(x)$ answer
$\forall xA(x) \to \wedge xA(x)$ answer
$\wedge xA(x) \to \forall xA(x)$ answer
$\wedge xA(x) \to \exists xA(x)$ answer
$\wedge xA(x) \to \sqcap xA(x)$ answer
$\sqcap xA(x) \to \wedge xA(x)$ answer
$\sqcap xA(x) \to \vee xA(x)$ answer
$\forall x(A(x) \wedge B(x)) \to \forall xA(x) \wedge \forall xB(x)$ answer
$\sqcap x(A(x) \wedge B(x)) \to \sqcap xA(x) \wedge \sqcap xB(x)$ answer
$\sqcap xA(x) \wedge \sqcap xB(x) \to \sqcap x(A(x) \wedge B(x))$ answer

## 3.8 Branching operations

There are only four branching operations that we consider: recurrence ⩘, corecurrence ⩗, rimplication ⟜ and refutation ⟜. Let us talk about recurrence first.

What is common for the members of the family of game operations called **recurrence operations** is that, when applied to a game *A*, they turn it into a game playing which means *repeatedly playing A*. In terms of resources, recurrence operations generate multiple "copies" of *A*, thus making *A* a reusable/recyclable resource. In classical logic, recurrence-style operations would be meaningless, because classical logic, as we know, is resource-blind and thus sees no difference between one and multiple copies of *A*. In the resource-conscious CoL, however, recurrence operations are not only meaningful, but also necessary to achieve a satisfactory level of expressiveness and realize its potential and ambitions. Hardly any computer program is used only once; rather, it is run over and over again. Loops within such programs also assume multiple repetitions of the same subroutine. In general, the tasks performed in real life by computers, robots or humans are typically recurring ones or involve recurring subtasks.

There is more than one naturally emerging recurrence operation. The differences between various recurrence operations are related to how "repetition" or "reusage" is exactly understood. Imagine a computer that has a program successfully playing chess. The resource that such a computer provides is obviously something stronger than just *Chess*, for it permits to play *Chess* as many times as the user wishes, while *Chess*, as such, only assumes one play. The simplest operating system would allow to start a session of *Chess*, then --- after finishing or abandoning and destroying it --- start a new play again, and so on. The game that such a system plays --- i.e. the resource that it supports/provides --- is ⩘*Chess*, which assumes an unbounded number of plays of *Chess* in a sequential fashion. A formal definition of the operation ⩘, called *sequential recurrence*, will be given later is [Section 3.9](#).

A more advanced operating system, however, would not require to destroy the old sessions before starting new ones; rather, it would allow to run as many parallel sessions as the user needs. This is what is captured by ⋏ *Chess*, meaning nothing but the infinite parallel conjunction *Chess* ∧ *Chess* ∧ *Chess* ∧ ... As we remember from [Section 3.5](#), ⋏ is called *parallel recurrence*.

Yet a really good operating system would not only allow the user to start new sessions of *Chess* without destroying old ones; it would also make it possible to branch/replicate any particular stage of any particular session, i.e., create any number of "copies" of any already reached position of the multiple parallel plays of *Chess*, thus giving the user the possibility to try different continuations from the same position. What corresponds to this intuition is the **branching recurrence** ("**brecurrence**") ⩘*Chess* of *Chess*.

At the intuitive level, the difference between ⩘ and ⋏ is that in ⩘*A*, unlike ⋏*A*, Environment does not have to restart *A* from the very beginning every time it wants to reuse it (as a resource); rather, Environment is allowed to backtrack to any of the previous --- not necessarily starting --- positions and try a new continuation from there, thus depriving the adversary of the possibility to reconsider the moves it has already made in that position. This is in fact the type of reusage every purely software resource allows or would allow in the presence of an advanced operating system and unlimited memory: one can start running process *A*; then fork it at any stage thus creating two threads that have a common past but possibly diverging futures (with the possibility to treat one of the threads as a "backup copy" and preserve it for backtracking purposes); then further fork any of the branches at any time; and so on.

The less flexible type of reusage of *A* assumed by ⋏*A*, on the other hand, is closer to what infinitely many autonomous physical resources would naturally offer, such as an unlimited number of independently acting robots each performing task *A*, or an unlimited number of computers with limited memories, each one only capable of and responsible for running a single thread of process *A*. Here the effect of forking an advanced stage of $A$ cannot be achieved unless, by good luck, there are two identical copies of the stage, meaning that the corresponding two robots or computers have so far acted in precisely the same ways.

The formal definitions of $⩔A$ and its dual ***branching corecurrence*** ("*cobrecurrence*") $⩗A$ ($=¬⩔¬A$) in early papers on CoL [Jap03, Jap09a] were direct formalizations of the above intuitions, with an explicit presence of "replicative" moves used by players to fork a given thread of $A$ and create two threads out of one. Later, in [Jap12b], another definition was found which was proven to be equivalent to the old one in the sense of mutual reducibility of the old and the new versions of $⩔A$. The new definition less directly corresponds to the above intuitions, but is technically simpler, and we choose it as our "canonical" definition of branching operations. To be able to state it, we agree on the following:

**Notation 3.8.1** Where $⟨Γ⟩$ is a run and $w$ is a ***bitstring*** (finite or infinite sequence of 0s and 1s), $Γ^{≤w}$ means the result of deleting from $Γ$ all moves except those that look like $u.β$ for some initial segment $u$ of $w$, and then further deleting the prefix "$u.$" from such moves.

**Definition 3.8.2** Assume $A$ is a game.

a) $⩔A$ is defined as the game $G$ with $\mathbf{Un}^G=\mathbf{Un}^A$, $\mathbf{Vr}^G=\mathbf{Vr}^A$ and with $\mathbf{Mp}^G$ such that, for any $G$-valuation $e$, we have:

- $Φ∈\mathbf{Lr}_e^G$ iff every move of $Φ$ has the prefix '$u.$' for some finite bitstring $u$, and, for every infinite bitstring $w$, $Φ^{≤w} ∈ \mathbf{Lr}_e^A$;
- $\mathbf{Wn}_e^G⟨Γ⟩ = ⊤$ iff, for every infinite bitstring $w$, $\mathbf{Wn}_e^A⟨Γ^{≤w}⟩ = ⊤$.

b) $⩗A$ is defined as the game $G$ with $\mathbf{Un}^G=\mathbf{Un}^A$, $\mathbf{Vr}^G=\mathbf{Vr}^A$ and with $\mathbf{Mp}^G$ such that, for any $G$-valuation $e$, we have

- $Φ∈\mathbf{Lr}_e^G$ iff every move of $Φ$ has the prefix '$u.$' for some finite bitstring $u$, and, for every infinite bitstring $w$, $Φ^{≤w} ∈ \mathbf{Lr}_e^A$;
- $\mathbf{Wn}_e^G⟨Γ⟩ = ⊥$ iff, for every infinite bitstring $w$, $\mathbf{Wn}_e^A⟨Γ^{≤w}⟩ = ⊥$.

The direct intuitions underlying the above definition are as follows. To play $⩔A$ or $⩗A$ means to simultaneously play in multiple parallel copies/threads of $A$. Each such thread is denoted by an infinite bitstring $w$ (so, there are in fact uncountably many threads). Every legal move by either player looks like $u.β$ for some finite bitstring $u$, and the effect/meaning of such a move is simultaneously making the move $β$ in all threads $w$ such that $u$ is an initial segment of $w$. So, where $Γ$ is the overall run of $⩔A$ or $⩗A$, the run in a given thread $w$ of $A$ is $Γ^{≤w}$. In order to win $⩔A$, Machine needs to win $A$ in all threads, while for winning $⩗A$ it is sufficient to win in just one thread.

It is obvious that $¬⩔A=⩗¬A$ and $¬⩗A=⩔¬A$, hence $⩔A=¬⩗¬A$ and $⩗A=¬⩔¬A$.

Branching recurrence $⩔$ can be shown to be stronger than its parallel counterpart $⋏$, in the sense that the principle $⩔A→⋏A$ is valid while $⋏A→⩔A$ is not. The two groups of operators, in isolation from each other, also validate different principles. For instance, $A∧⋏(A→A∧A) → ⋏A$ is valid while $A∧⩔(A→A∧A) → ⩔A$ is not; $⩔(A⊔B) → ⩔A⊔⩔B$ is valid while $⋏(A⊔B) → ⋏A⊔⋏B$ is not; $⩗⊓x(¬A(x)⊔A(x))$ is valid while $⩗⊓x(¬A(x)⊔A(x))$ is not.

***Branching rimplication*** ("*brimplication*") ⚬− and ***branching refutation*** ("*brefutation*") ⚬⊣ are defined in terms of $⩔$, $→$ and $⊥$ the same way as parallel rimplication ⊃− and refutation ⊃⊣ are defined in terms of $⋏$, $→$ and $⊥$:

**Definition 3.8.3**

a) $A \circ\!\!-\! B =_{\text{def}} \wedge\!\!\!\downarrow A \to B$

b) $\circ\!\!-\! A =_{\text{def}} A \circ\!\!-\!\bot$

**Exercise 3.8.4** The *Kolmogorov complexity* $k(x)$ of a number $x$ is the size of the smallest Turing machine which outputs $x$ on input 0. The Kolmogorov complexity problem $\sqcap x \sqcup y(y=k(x))$ has no algorithmic solution. Nor is it reducible to the halting problem in the strong sense of $\to$, meaning that the problem $\sqcap x(\neg Halts(x) \sqcup Halts(x)) \to \sqcap x \sqcup y(y=k(x))$ has no algorithmic solution, either. The Kolmogorov complexity problem, however, is reducible to the halting problem in the weaker sense of $\circ\!\!-$, meaning that Machine has a winning strategy for $\sqcap x(\neg Halts(x) \sqcup Halts(x)) \circ\!\!-\! \sqcap x \sqcup y(y=k(x))$. Describe such a strategy, informally.

Just like $\succ\!\!-$, $\circ\!\!-$ is a conservative generalization of Turing reduction. Specifically, for any predicates $p(x)$ and $q(x)$, the problem $\sqcap x(\neg p(x) \sqcup p(x)) \circ\!\!-\! \sqcap x(\neg q(x) \sqcup q(x))$ is computable iff $q(x)$ is Turing reducible to $p(x)$ iff the problem problem $\sqcap x(\neg p(x) \sqcup p(x)) \succ\!\!-\! \sqcap x(\neg q(x) \sqcup q(x))$ is computable. This means that, when restricted to traditional sorts of problems such as decision problems, the behaviors of $\circ\!\!-$ and $\succ\!\!-$ are indistinguishable. This stops being the case when these operators are applied to problems with higher degrees of interactivity though. For instance, the following problem is computable, but becomes incomputable with $\succ\!\!-$ instead of $\circ\!\!-$:

$$\sqcup y \sqcap x(\neg Halts(x,y) \sqcup Halts(x,y)) \circ\!\!-\! \sqcup y \big(\sqcap x(\neg Halts(x,y) \sqcup Halts(x,y)) \wedge \sqcap x(\neg Halts(x,y) \sqcup Halts(x,y))\big).$$

Generally, $(A \succ\!\!-\! B) \to (A \circ\!\!-\! B)$ is valid but $(A \circ\!\!-\! B) \to (A \succ\!\!-\! B)$ is not.

While both $\succ\!\!-$ and $\circ\!\!-$ are weaker than $\to$ and hence more general, $\circ\!\!-$ is still a more interesting operation of reduction than $\succ\!\!-$. What makes it special is the following belief. The latter, in turn, is based on the belief that $\wedge\!\!\!\downarrow$ (and by no means $\wedge$) is the operation allowing to reuse its argument in the strongest algorithmic sense possible.

Let $A$ and $B$ be computational problems (games). We say that $B$ is **brimplicationally** (resp. primplicationally, pimplicationally, etc.) ***reducible*** to $A$ iff $A \circ\!\!-\! B$ (resp. $A \succ\!\!-\! B$, $A \to B$, etc.) has an algorithmic solution (winning strategy).

**Thesis 3.8.5** Brimplicational reducibility is an adequate mathematical counterpart of our intuition of reducibility in the weakest (and thus most general) algorithmic sense possible. Specifically:

   (a) Whenever a problem $B$ is brimplicationally reducible to a problem $A$, $B$ is also algorithmically reducible to $A$ according to anyone's reasonable intuition.

   (b) Whenever a problem $B$ is algorithmically reducible to a problem $A$ according to anyone's reasonable intuition, $B$ is also brimplicationally reducible to $A$.

This is pretty much in the same sense as (by the Church-Turing thesis), a function $f$ is computable by a Turing machine iff $f$ has an algorithmic solution according to everyone's reasonable intuition.

## 3.9 Sequential operations

The ***sequential conjunction*** ("***sand***") $A \triangle B$ of games $A$ and $B$ starts and proceeds as $A$. It will also end as $A$ unless, at some point, Environment decides to switch to the next component, in which case $A$ is abandoned, and the game restarts, continues and ends as $B$. The ***sequential disjunction*** ("***sor***") $A \triangledown B$ of $A$ and $B$ is similar, with the difference that it is Machine who decides whether and when to switch from $A$ to $B$.

The original formal definition of $A \triangle B$ and $A \triangledown B$ found in [Jap08b] was a direct formalization of the above description. Definition 3.9.1 given below, while less direct, still faithfully formalizes the above intuitions as long as only static games are considered, and we opt for it because it is technically simpler. Specifically, Definition 3.9.1 allows either player to continue making moves in $A$ even after a switch takes place; such moves are meaningless but harmless. Similarly, it allows either player to make moves in $B$ without waiting for a switch to take place, even though a smart player would only start making such moves if and when a switch happens.

**Definition 3.9.1** Assume $A_0$ and $A_1$ are games with a common universe $U$.

a) $A_0 \triangle A_1$ is defined as the game $G$ with $\mathbf{Un}^G = U$, $\mathbf{Vr}^G = \mathbf{Vr}^{A_0} \cup \mathbf{Vr}^{A_1}$ and with $\mathbf{Mp}^G$ such that, for any $G$-valuation $e$, we have:

- $\Phi \in \mathbf{Lr}^G$ iff $\Phi = \langle \Xi, \Omega \rangle$ or $\Phi = \langle \Xi, 1, \Omega \rangle$, where every move of $\langle \Xi, \Omega \rangle$ has the prefix '0.' or '1.' and, for both $i \in \{0,1\}$, $\langle \Xi, \Omega \rangle^{i.} \in \mathbf{Lr}^{A_i}$;
- If $\Gamma$ does not contain the ("*switch*") move 1, then $\mathbf{Wn}^G \langle \Gamma \rangle = \mathbf{Wn}^{A_0} \langle \Gamma^{0.} \rangle$; otherwise $\mathbf{Wn}^G \langle \Gamma \rangle = \mathbf{Wn}^{A_1} \langle \Gamma^{1.} \rangle$.

b) $A_0 \triangledown A_1$ is defined as the game $G$ with $\mathbf{Un}^G = U$, $\mathbf{Vr}^G = \mathbf{Vr}^{A_0} \cup \mathbf{Vr}^{A_1}$ and with $\mathbf{Mp}^G$ such that, for any $G$-valuation $e$, we have:

- $\Phi \in \mathbf{Lr}^G$ iff $\Phi = \langle \Xi, \Omega \rangle$ or $\Phi = \langle \Xi, 1, \Omega \rangle$, where every move of $\langle \Xi, \Omega \rangle$ has the prefix '0.' or '1.' and, for both $i \in \{0,1\}$, $\langle \Xi, \Omega \rangle^{i.} \in \mathbf{Lr}^{A_i}$;
- If $\Gamma$ does not contain the ("*switch*") move 1, then $\mathbf{Wn}^G \langle \Gamma \rangle = \mathbf{Wn}^{A_0} \langle \Gamma^{0.} \rangle$; otherwise $\mathbf{Wn}^G \langle \Gamma \rangle = \mathbf{Wn}^{A_1} \langle \Gamma^{1.} \rangle$.

Recall that, for a predicate $p(x)$, $\sqcap x (\neg p(x) \sqcup p(x))$ is the problem of deciding $p(x)$. Then what is the similar-looking $\sqcap x (\neg p(x) \triangledown p(x))$? If you've guessed that this is the problem of *semideciding* $p(x)$, you are right. Machine has a winning strategy in this game if and only if $p(x)$ is semidecidable, i.e., recursively enumerable. Namely, if $p(x)$ is recursively enumerable, a winning strategy by Machine is to wait until Environment brings the game down to $\neg p(n) \triangledown p(n)$ for some particular $n$. After that, Machine starts looking for a certificate of $p(n)$'s being true. If and when such a certificate is found (meaning that $p(n)$ is indeed true), Machine makes a switch move turning $\neg p(n) \triangledown p(n)$ into the true $p(n)$; and if no certificate exists (meaning that $p(n)$ is false), then Machine keeps looking for a non-existent certificate forever and thus never makes any moves, meaning that the game ends as $\neg p(n)$, which, again, is true. And vice versa: any effective winning strategy for $\sqcap x (\neg p(x) \triangledown p(x))$ can obviously be seen as a semidecision procedure for $p(x)$, which accepts an input $n$ iff the strategy ever makes a switch move in the scenario where Environment's initial choice of a value for $x$ is $n$.

Existence of effective winning strategies for games has been shown[Jap03] to be closed under the rules '*from $A \rightarrow B$ and $A$ conclude $B$*', '*from $A$ and $B$ conclude $A \wedge B$*', '*from $A$ conclude $\sqcap xA$*', '*from $A$ conclude $\natural A$*'. In view of these closures, the validity of the principles discussed below implies certain known facts from the theory of computation. Needless to say, those examples once again demonstrate how CoL can be used as a systematic tool for defining new interesting properties and relations between computational problems, and not only reproducing already known theorems but also discovering an infinite variety of new facts.

The valid formula $\sqcap x (\neg p(x) \triangledown p(x)) \wedge \sqcap x (p(x) \triangledown \neg p(x)) \rightarrow \sqcap x (\neg p(x) \sqcup p(x))$ "expresses" the well known fact that, if both a predicate $p(x)$ and its complement $\neg p(x)$ are recursively enumerable, then $p(x)$ is decidable. Actually, the validity of this formula means something more: it means that the problem of deciding $p(x)$ is reducible to the ($\wedge$-conjunction of) the problems of semideciding $p(x)$ and $\neg p(x)$. In fact, a reducibility in an even stronger sense (in a sense that has no name) holds, expressed by the formula $\sqcap x\bigl((\neg p(x) \triangledown p(x)) \wedge ((p(x) \triangledown \neg(x)) \rightarrow \neg p(x) \sqcup p(x)\bigr)$.

The formula $\sqcap x \sqcup y(q(x) \leftrightarrow p(y)) \wedge \sqcap x(\neg p(x) \triangledown p(x)) \rightarrow \sqcap x(\neg q(x) \triangledown q(x))$ is also valid, which implies the known fact that, if a predicate $q(x)$ is <u>mapping reducible</u> to a predicate $p(x)$ and $p(x)$ is recursively enumerable, then so is $q(x)$. Again, the validity of this formula, in fact, means something even more: it means that the problem of semideciding $q(x)$ is reducible to the problems of mapping reducing $q(x)$ to $p(x)$ and semideciding $p(x)$.

Certain other reducibilities hold only in the sense of rimplications rather than implications. Here is an example. Two Turing machines are said to be *equivalent* iff they accept exactly the same inputs. Let $Neq(x,y)$ be the predicate 'Turing machines $x$ and $y$ are not equivalent'. This predicate is neither semidecidable nor co-semidecidable. However, the problem of its semideciding <u>primplicationally</u> (and hence also <u>brimplicationally</u>) reduces to the halting problem. Specifically, Machine has an effective winning strategy for the following game:

$$\sqcap z \sqcap t(\neg Halts(z,t) \sqcup Halts(z,t)) \succ\!\!- \sqcap x \sqcap y(\neg Neq(x,y) \triangledown Neq(x,y)).$$

A strategy here is to wait till Environment specifies some values $m$ and $n$ for $x$ and $y$. Then, create a variable $i$, initialize it to 1 and do the following. Specify $z$ and $t$ as $m$ and $i$ in one yet-unused copy of the antecedent, and as $n$ and $i$ in another yet-unused copy. That is, ask Environment whether $m$ halts on input $i$ and whether $n$ halts on the same input. Environment will have to provide the correct pair of answers, or else it loses. (1) If the answers are "No,No", increment $i$ to $i+1$ and repeat the step. (2) If the answers are "Yes,Yes", simulate both $m$ and $n$ on input $i$ until they halt. If both machines accept or both reject, increment $i$ to $i+1$ and repeat the step. Otherwise, if one accepts and one rejects, make a switch move in the consequent and celebrate victory. (3) If the answers are "Yes,No", simulate $m$ on $i$ until it halts. If $m$ rejects $i$, increment $i$ to $i+1$ and repeat the step. Otherwise, if $m$ accepts $i$, make a switch move in the consequent and you win. (4) Finally, if the answers are "No,Yes", simulate $n$ on $i$ until it halts. If $n$ rejects $i$, increment $i$ to $i+1$ and repeat the step. Otherwise, if $n$ accepts $i$, make a switch move in the consequent and you win.

The *sequential universal quantification* ("*sall*") $\triangle x A(x)$ of $A(x)$ is essentially nothing but the infinite sequential conjunction $A(0) \triangle A(1) \triangle A(2) \triangle \ldots$; the *sequential existential quantification* ("*sexists*") $\triangledown x A(x)$ of $A(x)$ is $A(0) \triangledown A(1) \triangledown A(2) \triangledown \ldots$; the *sequential recurrence* ("*srecurrence*") $\lambda A$ of $A$ is $A \triangle A \triangle A \triangle \ldots$; and the *sequential corecurrence* ("*cosrecurrence*") $\curlyvee A$ of $A$ is $A \triangledown A \triangledown A \triangledown \ldots$ Formally, we have:

**Definition 3.9.2** Assume $x$ is a variable and $A=A(x)$ is a game.

a) $\triangle x A(x)$ is defined as the game $G$ with $\mathbf{Un}^G = \mathbf{Un}^A$, $\mathbf{Vr}^G = \mathbf{Vr}^A - \{x\}$ and with $\mathbf{Mp}^G$ such that, for any $G$-valuation $e$, we have:

- $\Phi \in \mathbf{Lr}^G$ iff $\Phi = \langle \Xi_0, 1, \Xi_1, 2, \Xi_2, \ldots, n, \Xi_n \rangle$ ($n \geq 0$), where every move of $\langle \Xi_0, \Xi_1, \Xi_2, \ldots, \Xi_n \rangle$ has the prefix '$c$.' for some constant $c$ and, for every constant $c$, $\langle \Xi_0, \Xi_1, \Xi_2, \ldots, \Xi_n \rangle^{c.} \in \mathbf{Lr}^{A(c)}$;
- Call the moves $1,2,\ldots$ *switch* moves. If $\Gamma$ does not contain any switch moves, then $\mathbf{Wn}^G \langle \Gamma \rangle = \mathbf{Wn}^{A(0)} \langle \Gamma^{0.} \rangle$; if $\Gamma$ contains infinitely many switch moves, then $\mathbf{Wn}^G \langle \Gamma \rangle = \top$; otherwise, where $n$ is the last switch move of $\Gamma$, $\mathbf{Wn}^G \langle \Gamma \rangle = \mathbf{Wn}^{A(n)} \langle \Gamma^{n.} \rangle$.

b) $\triangledown x A(x)$ is defined as the game $G$ with $\mathbf{Un}^G = U$, $\mathbf{Vr}^G = \mathbf{Vr}^A - \{x\}$ and with $\mathbf{Mp}^G$ such that, for any $G$-valuation $e$, we have:

- $\Phi \in \mathbf{Lr}^G$ iff $\Phi = \langle \Xi_0, 1, \Xi_1, 2, \Xi_2, \ldots, n, \Xi_n \rangle$ ($n \geq 0$), where every move of $\langle \Xi_0, \Xi_1, \Xi_2, \ldots, \Xi_n \rangle$ has the prefix '$c$.' for some constant $c$ and, for every constant $c$, $\langle \Xi_0, \Xi_1, \Xi_2, \ldots, \Xi_n \rangle^{c.} \in \mathbf{Lr}^{A(c)}$;

- Call the moves $1,2,3,\ldots$ *switch* moves. If $\Gamma$ does not contain any switch moves, then $\mathbf{Wn}^G \langle \Gamma \rangle = \mathbf{Wn}^{A(0)} \langle \Gamma^{0\cdot} \rangle$; if $\Gamma$ contains infinitely many switch moves, then $\mathbf{Wn}^G \langle \Gamma \rangle = \bot$; otherwise, where $n$ is the last switch move of $\Gamma$, $\mathbf{Wn}^G \langle \Gamma \rangle = \mathbf{Wn}^{A(n)} \langle \Gamma^{n\cdot} \rangle$.

**Definition 3.9.3** Assume $A$ is a game.

a) $\measuredangle A$ is defined as the game $G$ with $\mathbf{Un}^G = \mathbf{Un}^A$, $\mathbf{Vr}^G = \mathbf{Vr}^A$ and with $\mathbf{Mp}^G$ such that, for any $G$-valuation $e$, we have:

- $\Phi \in \mathbf{Lr}^G$ iff $\Phi = \langle \Xi_0, 1, \Xi_1, 2, \Xi_2, \ldots, n, \Xi_n \rangle$ ($n \geq 0$), where every move of $\langle \Xi_0, \Xi_1, \Xi_{10}, \Xi_{11}, \ldots, \Xi_n \rangle$ has the prefix '$c$.' for some constant $c$ and, for every constant $c$, $\langle \Xi_0, \Xi_1, \Xi_2, \ldots, \Xi_n \rangle^{c\cdot} \in \mathbf{Lr}^A$;
- Call the moves $1,2,\ldots$ *switch* moves. If $\Gamma$ does not contain any switch moves, then $\mathbf{Wn}^G \langle \Gamma \rangle = \mathbf{Wn}^A \langle \Gamma^{0\cdot} \rangle$; if $\Gamma$ contains infinitely many switch moves, then $\mathbf{Wn}^G \langle \Gamma \rangle = \top$; otherwise, where $n$ is the last switch move of $\Gamma$, $\mathbf{Wn}^G \langle \Gamma \rangle = \mathbf{Wn}^A \langle \Gamma^{n\cdot} \rangle$.

b) $\mathbb{Y} A$ is defined as the game $G$ with $\mathbf{Un}^G = U$, $\mathbf{Vr}^G = \mathbf{Vr}^A$ and with $\mathbf{Mp}^G$ such that, for any $G$-valuation $e$, we have:

- $\Phi \in \mathbf{Lr}^G$ iff $\Phi = \langle \Xi_0, 1, \Xi_1, 2, \Xi_2, \ldots, n, \Xi_n \rangle$ ($n \geq 0$), where every move of $\langle \Xi_0, \Xi_1, \Xi_2, \ldots, \Xi_n \rangle$ has the prefix '$c$.' for some constant $c$ and, for every constant $c$, $\langle \Xi_0, \Xi_1, \Xi_2, \ldots, \Xi_n \rangle^{c\cdot} \in \mathbf{Lr}^A$;
- Call the moves $1,2,\ldots$ *switch* moves. If $\Gamma$ does not contain any switch moves, then $\mathbf{Wn}^G \langle \Gamma \rangle = \mathbf{Wn}^A \langle \Gamma^{0\cdot} \rangle$; if $\Gamma$ contains infinitely many switch moves, then $\mathbf{Wn}^G \langle \Gamma \rangle = \bot$; otherwise, where $n$ is the last switch move of $\Gamma$, $\mathbf{Wn}^G \langle \Gamma \rangle = \mathbf{Wn}^A \langle \Gamma^{n\cdot} \rangle$.

For an illustration, remember the [Kolmogorov complexity](#) function $k(x)$. The value of $k(x)$ is known to be bounded, not exceeding $x$ or a certain constant *const*, whichever is greater. While $\sqcap x \sqcup y(y=k(x))$ is not computable, Machine does have an algorithmic winning strategy for the problem $\sqcap x \mathbb{Y} \sqcup y(y=k(x))$. It goes like this: Wait till Environment specifies a value $m$ for $x$, thus asking "what is the Kolmogorov complexity of $m$?" and bringing the game down to $\mathbb{Y} \sqcup y(y=k(m))$. Answer that it is $m$, i.e. specify $y$ as $m$, and after that start simulating, in parallel, all machines $n$ smaller than $m$ on input 0. Whenever you find a machine $n$ that returns $m$ on input 0 and is smaller than any of the previously found such machines, make a switch move and, in the new copy of $\mathbb{Y} \sqcup y(y=k(m))$, specify $y$ as the size $|n|$ of $n$ or as *const*, whichever is greater. This obviously guarantees success: sooner or later the real Kolmogorov complexity $c$ of $m$ will be reached and named; and, even though the strategy will never be sure that $k(m)$ is not something yet smaller than $c$, it will never really find a reason to further reconsider its latest claim that $c=k(m)$.

**Exercise 3.9.4** Describe a winning strategy for $\sqcap x \nabla y(k(x)=(x-y))$.

As expected, *sequential implication* ("*simplication*") $\triangleright$, *sequential rimplication* ("*srimplication*") $\vartriangleleft$ and *sequential refutation* ("*srefutation*") $\vartriangleleft\neg$ are defined as follows:

**Definition 3.9.5**    a) $A \triangleright B =_{def} \neg A \nabla B$
                            b) $A \vartriangleleft B =_{def} \measuredangle A \rightarrow B$
                            c) $\vartriangleleft\neg A =_{def} A \vartriangleleft \bot$

# 3.10 Toggling operations

As all other sorts of conjunctions and disjunctions, **toggling conjunction** ("*tand*") ⩕ and **toggling disjunction** ("*tor*") ⩖ are dual to each other, and the definition of one is obtained from the definition of the other by interchanging the roles of the two players. Let us just focus on disjunction. One of the ways to characterize $A⩖B$ is the following. This game starts and proceeds as a play of $A$. It will also end as an ordinary play of $A$ unless, at some point, Machine decides to switch to $B$, after which the game becomes and continues as $B$. It will also end as $B$ unless, at some point, Machine decides to switch back to $A$. In such a case the game again becomes $A$, where $A$ resumes from the position in which it was abandoned (rather than from its start position, as would be the case, say, in $A▽B▽A$). Later Machine may again switch to (the abandoned position of) $B$, and so on. Machine wins the overall play iff it switches from one component to another ("changes its mind", or "corrects its mistake") at most finitely many times and wins in its final choice, i.e., in the component which was chosen last to switch to.

An alternative way to characterize $A⩖B$ is to say that it is played exactly like $A∨B$, with the only difference that Machine is allowed to make a 'choose $A$' or 'choose $B$' move some finite number of times. If infinitely many choices are made, Machine loses. Otherwise, the winner in the play will be the player who wins in the component that was chosen last ("the eventual choice"). The case of Machine having made no choices at all is treated as if it had chosen $A$. Thus, as in sequential disjunction, the leftmost component is the "default", or "automatically made", initial choice.

It is important to note that the adversary (or perhaps even the machine itself) never knows whether a given choice of a component of $A⩖B$ is the last choice or not. Otherwise, if Machine was required to indicate that it has made its final choice, then the resulting operation, for static games, would essentially be the same as $A⊔B$. Indeed, as we remember, in static games it never hurts a player to postpone making moves, so Environment could just inactively wait till the last choice is declared, and start playing the chosen component only after that, as in the case of $A⊔B$; under these circumstances, making some temporary choices before making the final choice would not make any sense for Machine, either.

What would happen if we did not require that Machine can change its mind only finitely meany times? There would be no "final choice" in this case. So, the only natural winning condition in the case of infinitely many choices would be to say that Machine wins iff it simply wins in one of the components. But then the resulting operation would be essentially the same as ∨, as a smart machine would always opt for keeping switching between components forever. That is, allowing infinitely many choices would amount to not requiring any choices at all, as is the case with $A∨B$.

One may also ask what would happen if we allowed Machine to make an arbitrary initial choice between $A$ and $B$ and then reconsider its choice only (at most) once? Such an operation on games, albeit meaningful, would not be basic. That is because it can be expressed through our primitives as $(A▽B) ⊔ (B▽A)$.

The very weak sort of choice captured by ⩖ is the kind of choice that, in real life, one would ordinarily call *choice after trial and error*. Indeed, a problem is generally considered to be solved after trial and error (a correct choice/solution/answer found) if, after perhaps coming up with several wrong solutions, a true solution is eventually found. That is, mistakes are tolerated and forgotten as long as they are eventually corrected. It is however necessary that new solutions stop coming at some point, so that there is a last solution whose correctness determines the success of the effort. Otherwise, if answers have kept changing all the time, no answer has really been given after all. Or, imagine Bob has been married and divorced several times. Every time he said "I do", he probably honestly believed that this time, at last, his bride was "the one", with whom he would live happily ever after. Bob will be considered to have found his Ms. Right after all if and only if one of his marriages indeed turns out to be happy and final.

As we remember, for a predicate $p(x)$, $⊓x(¬p(x) ⊔ p(x))$ is the problem of deciding $p(x)$, and $⊓x(¬p(x) ▽ p(x))$ is the weaker (easier to solve) problem of semideciding $p(x)$. Not surprisingly, $⊓x(¬p(x) ⩔ p(x))$ is also a decision-style problem, but still weaker than the problem of semideciding $p(x)$. This problem has been studied in the literature under several names, the most common of which is ***recursively approximating*** $p(x)$. It means telling whether $p(x)$ is true or not, but doing so in the same style as semideciding does in negative cases: by correctly saying "Yes" or "No" at some point (after perhaps taking back previous answers several times) and never reconsidering this answer afterwards. Observe that semideciding $p(x)$ can be seen as always saying "No" at the beginning and then, if this answer is incorrect, changing it to "Yes" at some later time; so, when the answer is negative, this will be expressed by saying "No" and never taking back this answer, yet without ever indicating that the answer is final and will not change. Thus, the difference between semideciding and recursively approximating is that, unlike a semidecision procedure, a recursive approximation procedure can reconsider *both* negative and positive answers, and do so several times rather than only once.

It is known that a predicate $p(x)$ is recursively approximable (i.e., the problem of its recursive approximation has an algorithmic solution) iff $p(x)$ has the arithmetical complexity $\Delta_2$, i.e., if both $p(x)$ and its complement $¬p(x)$ can be written in the form $\exists z \forall y\, s(z,y,x)$, where $s(z,y,x)$ is a decidable predicate. Let us see that, indeed, algorithmic solvability of $⊓x(¬p(x) ⩔ p(x))$ is equivalent to $p(x)$'s being of complexity $\Delta_2$.

First, assume $p(x)$ is of complexity $\Delta_2$, specifically, for some decidable predicates $q(z,y,x)$ and $r(z,y,x)$ we have $p(x) = \exists z \forall y\, q(z,y,x)$ and $¬p(x) = \exists z \forall y\, r(z,y,x)$. Then $⊓x(¬p(x) ⩔ p(x))$ is solved by the following strategy. Wait till Environment specifies a value $m$ for $x$, thus bringing the game down to $¬p(m) ⩔ p(m)$. Then create the variables $i$ and $j$, initialize both to 1, choose the $p(m)$ component, and do the following:

*Step 1:* Check whether $q(i,j,m)$ is true. If yes, increment $j$ to $j+1$ and repeat Step 1. If not, switch to the $¬p(m)$ component, reset $j$ to 1, and go to Step 2.

*Step 2:* Check whether $r(i,j,m)$ is true. If yes, increment $j$ to $j+1$ and repeat Step 2. If not, switch to the $p(m)$ component, reset $j$ to 1, increment $i$ to $i+1$, and go to Step 1.

With a little thought, one can see that the above algorithm indeed solves $⊓x(¬p(x) ⩔ p(x))$.

For the opposite direction, assume a given algorithm $Alg$ solves $⊓x(¬p(x) ⩔ p(x))$. Let $q(z,y,x)$ be the predicate such that $q(i,j,m)$ is true iff, in the scenario where the environment specified $x$ as $m$ at the beginning of the play, so that the game was brought down to $¬p(m) ⩔ p(m)$, we have: (1) at the $i$th computation step, $Alg$ chose the $p(m)$ component; (2) at the $j$th computation step, $Alg$ did not move. Quite similarly, let $r(z,y,x)$ be the predicate such that $r(i,j,m)$ is true iff, in the scenario where the environment specified $x$ as $m$ at the beginning of the play, so that the game was brought down to $¬p(m) ⩔ p(m)$, we have: (1) either $i=1$ or, at the $i$th computation step, $Alg$ chose the $¬p(m)$ component; (2) at the $j$th computation step, $Alg$ did not move. Of course, both $q(z,y,x)$ and $r(z,y,x)$ are decidable predicates, and hence so are $y>z \rightarrow q(z,y,x)$ and $y>z \rightarrow r(z,y,x)$. Now, it is not hard to see that $p(x) = \exists z \forall y (y>z \rightarrow q(z,y,x))$ and $¬p(x) = \exists z \forall y (y>z \rightarrow r(z,y,x))$. So, $p(x)$ is indeed of complexity $\Delta_2$.

As a real-life example of a predicate which is recursively approximable but neither semidecidable nor co-semidecidable, consider the predicate $k(x)<k(y)$, saying that number $x$ is simpler than number $y$ in the sense of Kolmogorov complexity. As noted earlier, $k(z)$ (the Kolmogorov complexity of $z$) is bounded, never exceeding $\max(z,const)$ for a certain constant *const*. Here is an algorithm which recursively approximates the predicate $k(x)<k(y)$, i.e., solves the problem $⊓x⊓y(¬k(x)<k(y) ⩔ k(x)<k(y))$. Wait till Environment brings the game down to $¬k(m)<k(n) ⩔ k(m)<k(n)$ for some $m$ and $n$. Then start simulating, in parallel, all Turing machines $t$ satisying $t \leq \max(m,n,const)$ on input 0. Whenever you see that a machine $t$ returns $m$ and the size of $t$ is smaller than that of any other previously found machines that return $m$ or $n$ on input 0, choose $k(m)<k(n)$. Quite similarly, whenever you see that a machine $t$ returns $n$ and the size of $t$ is smaller than that of any other previously found machine that returns $n$ on input 0, as well as smaller or equal to the size of any other previously found machines that return $m$ on input 0, choose $\neg k(m)<k(n)$. Obviously, the correct choice between $\neg k(m)<k(n)$ and $k(m)<k(n)$ will be made sooner or later and never reconsidered afterwards. This will happen when the procedure hits --- in the role of $t$ --- a smallest-size machine that returns either $m$ or $n$ on input 0.

Anyway, here is our formal definition of the toggling conjunction and disjunction:

**Definition 3.10.1** Assume $A_0$ and $A_1$ are games with a common universe $U$.

a) $A_0 ⩚ A_1$ is defined as the game $G$ with $\mathbf{Un}^G = U$, $\mathbf{Vr}^G = \mathbf{Vr}^{A_0} \cup \mathbf{Vr}^{A_1}$ and with $\mathbf{Mp}^G$ such that, for any $G$-valuation $e$, we have:

- $\Phi \in \mathbf{Lr}^G$ iff $\Phi = \langle \Xi_0, i_1, \Xi_1, i_2, \Xi_2, \ldots, i_n, \Xi_n \rangle$ ($n \geq 0$), where $i_1, i_2, \ldots, i_n \in \{0,1\}$, every move of $\langle \Xi_0, \Xi_1, \Xi_2, \ldots, \Xi_n \rangle$ has the prefix '0.' or '1.' and, for both $i \in \{0,1\}$, $\langle \Xi_0, \Xi_1, \Xi_2, \ldots, \Xi_n \rangle^{i.} \in \mathbf{Lr}^{A_i}$;
- Call $0$ and $1$ *switch moves*. If $\Gamma$ does not contain any switch moves, then $\mathbf{Wn}^G \langle \Gamma \rangle = \mathbf{Wn}^{A_0} \langle \Gamma^{0.} \rangle$; if $\Gamma$ has infinitely many occurrences of switch moves, then $\mathbf{Wn}^G \langle \Gamma \rangle = \top$; otherwise, where $i$ is the last switch move, $\mathbf{Wn}^G \langle \Gamma \rangle = \mathbf{Wn}^{A_i} \langle \Gamma^{i.} \rangle$.

b) $A_0 ⩛ A_1$ is defined as the game $G$ with $\mathbf{Un}^G = U$, $\mathbf{Vr}^G = \mathbf{Vr}^{A_0} \cup \mathbf{Vr}^{A_1}$ and with $\mathbf{Mp}^G$ such that, for any $G$-valuation $e$, we have:

- $\Phi \in \mathbf{Lr}^G$ iff $\Phi = \langle \Xi_0, i_1, \Xi_1, i_2, \Xi_2, \ldots, i_n, \Xi_n \rangle$ ($n \geq 0$), where $i_1, i_2, \ldots, i_n \in \{0,1\}$, every move of $\langle \Xi_0, \Xi_1, \Xi_2, \ldots, \Xi_n \rangle$ has the prefix '0.' or '1.' and, for both $i \in \{0,1\}$, $\langle \Xi_0, \Xi_1, \Xi_2, \ldots, \Xi_n \rangle^{i.} \in \mathbf{Lr}^{A_i}$;
- Call $0$ and $1$ *switch moves*. If $\Gamma$ does not contain any switch moves, then $\mathbf{Wn}^G \langle \Gamma \rangle = \mathbf{Wn}^{A_0} \langle \Gamma^{0.} \rangle$; if $\Gamma$ has infinitely many occurrences of switch moves, then $\mathbf{Wn}^G \langle \Gamma \rangle = \bot$; otherwise, where $i$ is the last switch move, $\mathbf{Wn}^G \langle \Gamma \rangle = \mathbf{Wn}^{A_i} \langle \Gamma^{i.} \rangle$.

As expected, the **toggling universal quantification** ("tall") $⩚ x A(x)$ of $A(x)$ is essentially nothing but $A(0) ⩚ A(1) ⩚ A(10) ⩚ A(11) ⩚ \ldots$, and the **toggling existential quantification** ("texists") $⩛ x A(x)$ of $A(x)$ is $A(0) ⩛ A(1) ⩛ A(10) ⩛ A(11) ⩛ \ldots$. Formally, we have:

**Definition 3.10.2** Assume $x$ is a variable and $A = A(x)$ is a game.

a) $⩚ x A(x)$ is defined as the game $G$ with $\mathbf{Un}^G = \mathbf{Un}^A$, $\mathbf{Vr}^G = \mathbf{Vr}^A - \{x\}$ and with $\mathbf{Mp}^G$ such that, for any $G$-valuation $e$, we have:

- $\Phi \in \mathbf{Lr}^G$ iff $\Phi = \langle \Xi_0, c_1, \Xi_1, c_2, \Xi_2, \ldots, n, \Xi_n \rangle$ ($n \geq 0$), where every move of $\langle \Xi_0, \Xi_1, \Xi_2, \ldots, \Xi_n \rangle$ has the prefix '$c$.' for some constant $c$ and, for every constant $c$, $\langle \Xi_0, \Xi_1, \Xi_2, \ldots, \Xi_n \rangle^{c.} \in \mathbf{Lr}^{A(c)}$;
- Call the moves $0, 1, 2, \ldots$ *switch* moves. If $\Gamma$ does not contain any switch moves, then $\mathbf{Wn}^G \langle \Gamma \rangle = \mathbf{Wn}^{A(0)} \langle \Gamma^{0.} \rangle$; if $\Gamma$ has infinitely many occurrences of switch moves, then $\mathbf{Wn}^G \langle \Gamma \rangle = \top$; otherwise, where $n$ is the last switch move of $\Gamma$, $\mathbf{Wn}^G \langle \Gamma \rangle = \mathbf{Wn}^{A(n)} \langle \Gamma^{n.} \rangle$.

b) $⩛ x A(x)$ is defined as the game $G$ with $\mathbf{Un}^G = U$, $\mathbf{Vr}^G = \mathbf{Vr}^A - \{x\}$ and with $\mathbf{Mp}^G$ such that, for any $G$-valuation $e$, we have:

- $\Phi \in \mathbf{Lr}^G$ iff $\Phi = \langle \Xi_0, c_1, \Xi_1, c_2, \Xi_2, \ldots, n, \Xi_n \rangle$ ($n \geq 0$), where every move of $\langle \Xi_0, \Xi_1, \Xi_2, \ldots, \Xi_n \rangle$ has the prefix '$c$.' for some constant $c$ and, for every constant $c$, $\langle \Xi_0, \Xi_1, \Xi_2, \ldots, \Xi_n \rangle^{c.} \in \mathbf{Lr}^{A(c)}$;

- Call the moves $0,1,2,\ldots$ *switch* moves. If $\Gamma$ does not contain any switch moves, then $\mathbf{Wn}^G \langle\Gamma\rangle = \mathbf{Wn}^{A(0)} \langle\Gamma^{0\cdot}\rangle$; if $\Gamma$ has infinitely many occurrences of switch moves, then $\mathbf{Wn}^G \langle\Gamma\rangle = \bot$; otherwise, where $n$ is the last switch move of $\Gamma$, $\mathbf{Wn}^G \langle\Gamma\rangle = \mathbf{Wn}^{A(n)}\langle\Gamma^{n\cdot}\rangle$.

For an example illustrating toggling quantifiers at work, remember that [Kolmogorov complexity]() $k(x)$ is not a computable function, i.e., the problem $\sqcap x \sqcup y(y=k(x))$ has no algorithmic solution. However, replacing $\sqcup y$ with $\curlyvee y$ in it yields an algorithmically solvable problem. A solution for $\sqcap x \curlyvee y(y=k(x))$ goes like this. Wait till the environment chooses a number $m$ for $x$, thus bringing the game down to $\curlyvee y(y=k(m))$, which is essentially nothing but $0=k(m) \curlyvee 1=k(m) \curlyvee 2=k(m) \curlyvee \ldots$ Initialize $i$ to a sufficiently large number, such as $i=\max(|m|+const)$ where *const* is the constant mentioned [earlier](), and then do the following routine: Switch to the disjunct $i=k(m)$ of $0=k(m) \curlyvee 1=k(m) \curlyvee 2=k(m) \curlyvee \ldots$, and then start simulating on input 0, in parallel, all Turing machines whose sizes are smaller than $i$; if and when you see that one of such machines returns $m$, update $i$ to the size of that machine, and repeat the present routine.

We close this sections with formal definitions of ***toggling recurrence*** ("***trecurrence***") ⩚, ***toggling corecurrence*** ("***cotrecurrence***") ⩛, ***toggling implication*** ("***timplication***") ⩾, ***toggling rimplication*** ("***trimplication***") ⪰ and ***toggling refutation*** ("***trefutation***") ⪰¬:

**Definition 3.10.3** Assume $A$ is a game.

a) ⩚$A$ is defined as the game $G$ with $\mathbf{Un}^G = \mathbf{Un}^A$, $\mathbf{Vr}^G = \mathbf{Vr}^A$ and with $\mathbf{Mp}^G$ such that, for any $G$-valuation $e$, we have:

- $\Phi \in \mathbf{Lr}^G$ iff $\Phi = \langle \Xi_0, c_1, \Xi_1, c_2, \Xi_2, \ldots, c_n, \Xi_n \rangle$ ($n \geq 0$), where every move of $\langle \Xi_0, \Xi_1, \Xi_2, \ldots, \Xi_n \rangle$ has the prefix '$c.$' for some constant $c$ and, for every constant $c$, $\langle \Xi_0, \Xi_1, \Xi_2, \ldots, \Xi_n \rangle^{c\cdot} \in \mathbf{Lr}^A$;
- Call the moves $0,1,2,\ldots$ *switch* moves. If $\Gamma$ does not contain any switch moves, then $\mathbf{Wn}^G \langle\Gamma\rangle = \mathbf{Wn}^A \langle\Gamma^{0\cdot}\rangle$; if $\Gamma$ has infinitely many occurrences of switch moves, then $\mathbf{Wn}^G \langle\Gamma\rangle = \square$; otherwise, where $n$ is the last switch move of $\Gamma$, $\mathbf{Wn}^G \langle\Gamma\rangle = \mathbf{Wn}^A\langle\Gamma^{n\cdot}\rangle$.

b) ⩛$A$ is defined as the game $G$ with $\mathbf{Un}^G = U$, $\mathbf{Vr}^G = \mathbf{Vr}^A$ and with $\mathbf{Mp}^G$ such that, for any $G$-valuation $e$, we have:

- $\Phi \in \mathbf{Lr}^G$ iff $\Phi = \langle \Xi_0, c_1, \Xi_1, c_2, \Xi_2, \ldots, n, \Xi_n \rangle$ ($n \geq 0$), where every move of $\langle \Xi_0, \Xi_1, \Xi_2, \ldots, \Xi_n \rangle$ has the prefix '$c.$' for some constant $c$ and, for every constant $c$, $\langle \Xi_0, \Xi_1, \Xi_2, \ldots, \Xi_n \rangle^{c\cdot} \in \mathbf{Lr}^A$;
- Call the moves $0,1,2,\ldots$ *switch* moves. If $\Gamma$ does not contain any switch moves, then $\mathbf{Wn}^G \langle\Gamma\rangle = \mathbf{Wn}^A\langle\Gamma^{0\cdot}\rangle$; if $\Gamma$ has infinitely many occurrences of switch moves, then $\mathbf{Wn}^G \langle\Gamma\rangle = \square$; otherwise, where $n$ is the last switch move of $\Gamma$, $\mathbf{Wn}^G \langle\Gamma\rangle = \mathbf{Wn}^A\langle\Gamma^{n\cdot}\rangle$.

**Definition 3.10.4**  a) $A \geqslant B =_{def} \neg A \curlyvee B$
  b) $A \succcurlyeq B =_{def}$ ⩚$A \rightarrow B$
  c) $\succcurlyeq\neg A =_{def} A \succcurlyeq \bot$

## 3.11 Cirquents

The syntactic constructs called ***cirquents***, briefly mentioned earlier, take the expressive power of CoL to a qualitatively higher level, allowing us to form, in a systematic way, an infinite variety of game operations. Each cirquent is --- or can be seen as --- an independent operation on games, generally not expressible via composing operations taken from some fixed finite pool of primitives, such as the operations seen in the preceding subsections of the present section.

Cirquents come in a variety of versions, and the main common characteristic feature of them is having mechanisms for explicitly accounting for possible *sharing* of subcomponents between different components. Sharing is the main distinguishing feature of cirquents from more traditional means of expression such as formulas, sequents, hypersequents [Avr87], or structures of the calculus of structures [Gug07]. While the latter can be drawn as (their parse) trees, cirquents more naturally call for circuit- or graph-style constructs. The earliest cirquents were intuitively conceived as collections of sequents (sequences of formulas) that could share some formulas and, as such, could be drawn like *cirquits* rather than linear expressions. This explains the etimology of the word: CIRcuit+seQUENT. All Boolean circuits are thus cirquents, but not all cirquents are Boolean circuits. Firstly, because cirquents may have various additional sorts of *gates* (⊓-gates, △-gates, ⋏-gates, etc.). Secondly, because cirquents may often have more evolved sharing mechanisms than just child-sharing between different gates. For instance, in addition to (or instead of) the possibility of sharing children, a "cluster"[Jap11b] of ⊔-gates may also share choices associated with ⊔ in game-playing: if the machine choses the left or the right child for one gate of the cluster, then the same left or right choice automatically extends to all gates of the cluster.

For simplicity considerations, we are not going to introduce cirquents and their semantics in full generality and formal detail here. This is done in [Jap06c, Jap08a, Jap11b, Jap13a]. Instead, to get some intuitive insights, let us only focus on cirquents that look like Boolean circuits with ⋏- and ⋎-gates. Every such cirquent $C$ can be seen as an $n$-ary operation on games, where $n$ is the number of inputs of $C$. For instance, the cirquent represents the 3-ary game operation ♥ informally defined as follows. Playing ♥$(A,B,C)$, as is the case with all parallel operations, means playing simultaneously in all components of it. In order to win, Machine needs to win in at least two out of the three components. Any attempt to express this operation in terms of ⋏, ⋎ or other already defined operations is going to fail. For instance, the natural candidate $(A⋏B)⋎(A⋏C)⋎(B⋏C)$ is dramatically inadequate. This is a game on six rather than three boards, with $A$ played on boards #1 and #3, $B$ on boards #2 and #5, and $C$ on boards #4 and #6. A special case of it is $(A⋏A)⋎(A⋏A)⋎(A⋏A)$, which fails to indicate for instance that the 1st and the 3rd occurrences of $A$ stand for the same copy of $A$ while the 2nd occurrence for a different copy where a different run can be generated.

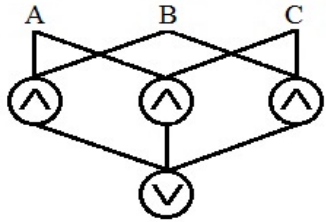

As we just had a chance to see, even at the most basic (⋏,⋎) level, cirquents are properly more expressive than formulas. It is this added expressiveness and flexibility that, for some fragments of CoL, makes a difference between axiomatizability and unaxiomatizability: even if one is only trying to set up a deductive system for proving valid formulas, intermediate steps in proofs of such formulas still inherently require using cirquents that cannot be written as formulas.[Jap06c, Jap13a, Jap13b] An example is the system **CL15** presented in Section 6.4.

In the present piece of writing we are exclusively focused on the formula-based version of CoL, seeing cirquents (in Section 6.4) only as technical servants to formulas. This explains why we do not attempt to define the semantics of cirquents formally. It should however be noted that cirquents are naturally called for not only within the specific formal framework of CoL, but also in the framework of all resource-sensitive approaches in logic. Such approaches intrinsically require the ability to account for the possibility of *resource sharing*, as sharing is a ubiquitous phenomenon in the world of resources. Yet formula- or sequent-based approaches do not and cannot possess such an ability. The following naive example may provide some insights. It is about the earlier discussed ♥(*A*,*A*,*A*) combination, seen as a combination of resources in the abstract/naive sense rather than the specific game-semantical sense of CoL.

Imagine a vending machine that has slots for 25-cent (25c) coins, with each slot taking a single coin. Coins can be true (authentic) or false (counterfeited). Inserting a false coin into a slot fills the slot up (so that no other coins can be inserted into it until the operation is complete), but otherwise does not fool the machine into thinking that it has received 25 cents. A candy costs 50 cents, and the machine will dispense a candy if at least two of its slots receive true coins. Pressing the "dispense" button while having inserted anything less than 50 cents, such as a single coin, or one true and two false coins, results in a non-recoverable loss.

Victor has three 25c-coins, and he knows that two of them are true while one is perhaps false (but he has no way to tell which one is false). Could he get a candy? Expected or not, the answer depends on the number of slots that the machine has. Consider two cases: machine *M2* with two slots, and machine *M3* with three slots. Victor would have no problem with *M3*: he can insert his three coins into the three slots, and the machine, having received ≥50c, will dispense a candy. With *M2*, however, Victor is powerless. He can try inserting arbitrary two of his three coins into the two slots of the machine, but there is no guarantee that one of those two coins is not false, in which case Victor will end up with no candy and only 25 cents remaining in his pocket.

Both *M2* and *M3* can be understood as resources --- resources turning coins into a candy. And note that these two resources are not the same: *M3* is obviously stronger ("better"), as it allows Victor to get a candy whereas *M2* does not, while, at the same time, anyone rich enough to be able to make *M2* dispense a candy would be able to do the same with *M3* as well. Yet, formulas fail to capture this important difference. With →, ∧, ∨ here understood as abstract multiplicative-style operations on resources, *M2* and *M3* can be written as *R2*→*Candy* and *R3*→*Candy*, respectively: they consume a certain resource *R2* or *R3* and produce *Candy*. What makes *M3* stronger than *M2* is that the subresource *R3* that it consumes is weaker (easier to supply) than the subresource *R2* consumed by *M2*. Specifically, with one false and two true coins, Victor is able to satisfy *R3* but not *R2*.

The resource *R2* can be represented as the cirquent

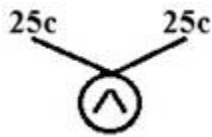

which, due to being tree-like, can also be adequately written as the formula 25c ∧ 25c. As for the resource *R3*, either one of the following two cirquents is an adequate representation of it, with one of them probably showing the relevant part of the actual physical circuitry used in *M3*:

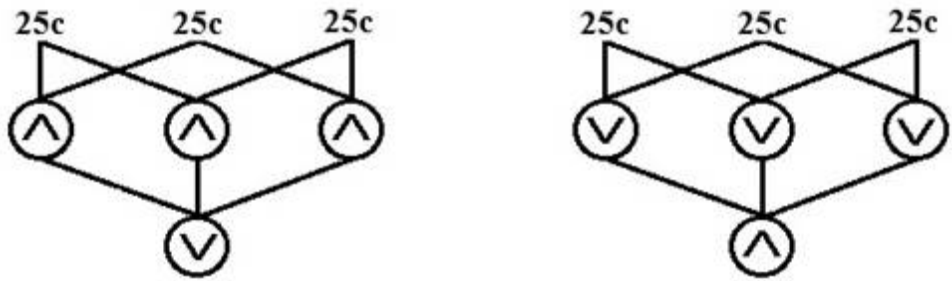

**Figure 3.11.1**: Cirquents for the "two out of three" combination of resources

Unlike *R2*, however, *R3* cannot be represented through a formula. 25c⋀25c does not fit the bill, for it represents *R2* which, as we already agreed, is not the same as *R3*. In another attempt to find a formula, we might try to rewrite one of the above two cirquents --- let it be the one on the right --- into an "equivalent" formula in the standard way, by duplicating and separating shared nodes, as we did earlier when tackling ♥(*A*,*A*,*A*). This results in (25c⋁25c)⋀(25c⋁25c)⋀(25c⋁25c), which, however, is not any more adequate than 25c⋀25c. It expresses not *R3* but the resource consumed by a machine with six coin slots grouped into three pairs, where (at least) one slot in each of the three pairs needs to receive a true coin. Such a machine thus dispenses a candy for ≥75 rather than ≥50 cents, which makes Victor's resources insufficient.

The trouble here is related to the inability of formulas to explicitly account for resource sharing or the absence thereof. The cirquent on the right of the above figure stands for a (multiplicative-style) conjunction of three resources, each conjunct, in turn, being a disjunction of two subresources of type 25c. However, altogether there are three rather than six 25c-type subresources, each one being shared between two different conjuncts of the main resource.

From the abstract resource-philosophical point of view of CoL, classical logic and linear logic are two imperfect extremes. In the former, all occurrences of a same subformula mean "the same" (represent the same resource), i.e., *everything is shared* that can be shared; and in the latter, each occurrence stands for a separate resource, i.e., *nothing is shared* at all. Neither approach does thus permit to account for mixed cases where certain occurrences are meant to represent the same resource while some other occurrences stand for different resources of the same type. It is a shame that linear logic fails to express simple, natural and unavoidable things such as the "two out of three" combination expressed by the cirquents of Figure 3.11.1.

# 4  Interactive machines

## 4.1 Interactive computability

In traditional game-semantical approaches, including those of Lorenzen [Lor61], Hintikka [Hin73], Blass [Bla92], Abramsky [Abr94] and others, players' strategies are understood as *functions* --- typically as functions from interaction histories (positions) to moves, or sometimes as functions that only look at the latest move of the history. This *strategies-as-functions* approach, however, is generally inapplicable in the context of CoL, whose relaxed semantics, in striving to get rid of "bureaucratic pollutants" and only deal with the remaining true essence of games, does not impose any regulations on which player can or should move in a given situation. Here, in many cases, either player may have (legal) moves, and then it is unclear whether the next move should be the one prescribed by ⊤'s strategy function or the one prescribed by the strategy function of ⊥. The advantages of CoL's approach become especially appreciable when one tries to bring complexity theory into interactive computation: hardly (m)any really meaningful and interesting complexity-theoretic concepts can be defined for games (particularly, games that may last long) with the strategies-as-functions approach.

In CoL, (⊤'s effective) strategies are defined in terms of interactive machines, where computation is one continuous process interspersed with --- and influenced by --- multiple "input" (Environment's moves) and "output" (Machine's moves) events. Of several, seemingly rather different yet equivalent, machine models of interactive computation studied in CoL, we will only discuss the most basic, **HPM** ("Hard-Play Machine") model.

An HPM is a Turing machine with the additional capability of making moves. The adversary can also move at any time, and such moves are the only nondeterministic events from the machine's perspective. Along with one or more ordinary read/write *work tapes*, the machine has an additional, read-only tape called the *run tape*. The latter, serving as a dynamic input, at any time spells the "current position" of the play. Its role is to make the evolving run fully visible to the machine. In these terms, an algorithmic solution (⊤'s winning strategy) for a given constant game $A$ is understood as an HPM $M$ such that, no matter how the environment acts during its interaction with $M$ (what moves it makes and when), the run incrementally spelled on the run tape is a ⊤-won run of $A$.

As for ⊥'s strategies, there is no need to define them: all possible behaviors by ⊥ are accounted for by the different possible nondeterministic updates of the run tape of an HPM.

In the above outline, we described HPMs in a relaxed fashion, without being specific about details such as, say, how, exactly, moves are made by the machine, how many moves either player can make at once, what happens if both players attempt to move "simultaneously", etc. As it happens, all reasonable design choices yield the same class of winnable games as long as we only consider static games, the only sort of games that we are willing to consider.

While design choices are largely unimportant and "negotiable", we still want to agree on some technical details for clarity. Just like an ordinary Turing machine, an HPM has a finite set of *states*, one of which has the special status of being the *start state*. There are no accept, reject, or halt states, but there are specially designated states called *move states*. Each tape of the machine has a beginning but no end, and is divided into infinitely many *cells*, arranged in the left-to-right order. At any time, each cell contains one symbol from a certain fixed finite set of *tape symbols*. The *blank* symbol is among the tape symbols. Each tape has its own *scanning head*, at any given time looking (located) at one of the cells of the tape.

For technical purposes, we additionally assume the (physical or imaginary) presence of a *move buffer*. The size of the latter is unlimited and, at any time, it contains some (possibly empty) finite string over the keyboard alphabet. The function of the buffer is to let the machine construct a move piece-by-piece before officially making such a move.

A transition from one *computation step* ("*clock cycle*", "*time*") to another happens according to the fixed *transition function* of the machine. The latter, depending on the current state and the symbols seen by the scanning heads on the corresponding tapes, deterministically prescribes: (1) the next state that the machine should assume; (2) the tape symbol by which the old symbol should be overwritten in the current cell (the cell currently scanned by the head), for each work tape individually; (3) the (finite, possibly empty) string over the keyboard alphabet that should be appended to the content of the move buffer; and (4) the (not necessarily the same) direction --- stay put, one cell to the left, or one cell to the right --- in which each scanning head should move. It is stipulated that when the head of a tape is looking at the first (leftmost) cell, an attempt to move to the left results in staying put. The same happens when the head tries to move to the right while looking at a blank cell (a cell containing the "blank" symbol).

When the machine starts working, it is in its start state, all scanning heads are looking at the leftmost cells of the corresponding tapes, all work tapes are blank, the run tape does not contain any green moves (but it may contain some red moves, signifying that Environment has made those moves "at the very beginning"), and the buffer is empty.

Whenever the machine enters a move state, the green-colored move $\alpha$ written in the buffer by that time is automatically appended to the contents of the run tape, and the buffer is simultaneously emptied. Also, on every transition, any finite sequence $\beta_1,\ldots,\beta_m$ of red moves may be nondeterministically appended to the contents of the run tape. If the above two events happen on the same clock cycle, then both $\alpha$ and $\beta_1,\ldots,\beta_m$ will be appended to the contents of the run tape, where α can (nondeterministically) go before, after or anywhere in between $β_1,…,β_m$.

When describing the work of a machine, we may use the jargon "***retire***". What will be meant by retiring is going into an infinite loop that makes no moves, puts nothing into the buffer, and does not reposition the scanning heads. Retiring thus achieves the same effect as halting would achieve if this was technically allowed.

A *configuration* is a full description of the situation in the machine at some given computation step. It consists of records of the ("current") contents of the work and run tapes, the content of the buffer, the location of each scanning head, and the state of the machine. A *computation branch* of the machine is an infinite sequence $C_0,C_1,C_2,…$ of configurations, where $C_0$ is an *initial configuration* (one described [earlier](#)), and every $C_{i+1}$ is a configuration that could have legally followed (again, in the sense explained earlier) $C_i$ according to the transition function of the machine. In less formal contexts, we may say "*play*" instead of "computation branch". For a computation branch $B$, *the run spelled by $B$* is the run Γ incrementally spelled on the run tape in the corresponding scenario of interaction. We say that such a Γ is a run *generated by* the machine.

**Definition 4.1.1** Let $A$ be a constant game. We say that an HPM $M$ ***computes*** (***solves***) $A$ --- and write $M ⊨ A$ --- iff every run generated by $M$ is a ⊤-won run of $A$. We say that $A$ is ***computable*** iff there is an HPM $M$ with $M ⊨ A$; such an HPM is said to be an (algorithmic) ***solution***, or ***winning strategy***, for $A$.

## 4.2 Interactive complexity

At present, the theory of interactive computation is far from being well developed, and even more so is the corresponding complexity theory. The studies of interactive computation in the context of complexity, while having been going on since long ago, have been relatively scattered and ad hoc: more often than not, interaction has been used for better understanding certain complexity issues for traditional, non-interactive problems rather than being treated as an object of systematic studies in its own rights (examples would be alternating computation [Cha81], or interactive proof systems and Arthur-Merlin games [Gol89]). As if complexity theory was not "complex" enough already, taking it to the interactive level would most certainly generate a by an order of magnitude greater diversity of species from the complexity zoo.

CoL has made its first modest steps[Jap15] towards elaborating some basic complexity-theoretic concepts for game-playing strategies, which we are going to present in this section. It should however be pointed out that our way of measuring the complexity of strategies presented in this section is merely one out of a huge and interesting potential variety of complexity measures meaningful and useful in the interactive context.

**Definition 4.2.1** In the context of a given computation branch (play) of a given HPM $M$:

     1. By the ***background*** of a clock cycle $c$ we mean the greatest of the sizes of Environment's moves made before time $c$, or is 0 if there are no such moves.

     2. By the ***timecost*** of a cycle $c$ we mean $c-d$, where $d$ is the greatest cycle with $d< c$ on which a move was made by Environment, or is 0 if there is no such cycle.

     3. By the ***spacecost*** of a cycle $c$ we mean the maximum number of cells ever visited by any (any *one*) work-tape scanning heads before time $c$.

**Definition 4.2.2** Let $M$ be an HPM, and $h$ an arithmetical function. Below, if $h$ is not unary, $h(b)$ should be understood as $h(b,…,b)$, with as many occurrences of $b$ in "$(b,…,b)$" as the arity of $h$. We say that:

    1. *M **runs in amplitude** h* iff, in every play by *M*, whenever *M* makes a move on a cycle *c*, the size of that move does not exceed *h*(*b*), where *b* is the background of *c*;

    2. *M **runs in space** h* iff, in every play by *M*, the spacecost of any given clock cycle *c* does not exceed *h*(*b*), where *b* is the background of *c*;

    3. *M **runs in time** h* iff, in every play by *M*, whenever *M* makes a move on a cycle *c*, the timecost of *c* does not exceed *h*(*b*), where *b* is the background of *c*.

Amplitude complexity is to keep track of (set bound on) the sizes of ⊤ 's moves relative to the sizes of ⊥'s moves. This nontraditional complexity measure is indispensable when it comes to interactive problems. Next, our time complexity concept can be seen to be in the spirit of what is usually called *response time*. The latter generally does not and should not depend on the length of the preceding interaction history. On the other hand, it is not and should not merely be a function of the adversary's last move, either. A similar characterization applies to our concept of space complexity. All three complexity measures are equally meaningful whether it be in the context of "short-lasting" games (such as the ones represented by the formulas of the later-defined logic **CL12**) or the context of games that may have "very long" and even infinitely long legal runs.

**Definition 4.2.3** Let *A* be a constant game, *h* an arithmetical function, and *M* an HPM. We say that *M **computes** (**solves**) A in time h*, or that *M* is an *h* time **solution** for *A*, iff *M* runs in time *h* and *M* computes *A*. We say that *A* is **computable** (**solvable**) in time *h* iff it has an *h* time solution. Similarly for space and amplitude.

When we say **polynomial time**, it is to be understood as "time *h* for some polynomial function *h*". Similarly for polynomial space, polynomial amplitude, logarithmic space, etc. More generally, using the asymptotic "Big-O" notation, where *g* is an arithmetical function, "time (space, amplitude) O(*g*)" should be understood as "time (space, amplitude) *h* for some function *h* with *h*∈O(*g*)".

We want to close this section with two terminological conventions.

**Convention 4.2.4** Assume *H* is a set of arithmetical functions, and *M* is an HPM. By saying that *M* runs (or plays) in time *H* we shall mean that *M* runs in time *h* for some *h*∈*H*. Similarly for "*M* is an *H* time solution of ...", "*M* is an *H* time machine", etc. Similarly for space and amplitude.

**Convention 4.2.5** Assume $h_1,h_2,h_3$ are arithmetical functions, and *M* is an HPM. By saying that *M* runs (or plays) in **tricomplexity** $(h_1,h_2,h_3)$ we shall mean that *M* runs in amplitude $h_1$, space $h_2$ and time $h_3$. Similarly for "*M* is a $(h_1,h_2,h_3)$ tricomplexity solution of ...", "*M* is a $(h_1,h_2,h_3)$ tricomplexity machine", etc. Similarly for $(H_1,H_2,H_3)$ instead of $(h_1, h_2,h_3)$, where $H_1,H_2,H_3$ are sets of arithmetical functions.

# 5 The language of CoL and its semantics

## 5.1 Formulas

It is not quite accurate to say "the language" of CoL because, as pointed out earlier, the latter has an open-ended formalism. Yet, in the present article, by "the language of CoL" we will mean the particular language defined below. It extends the language of classical logic by adding to it all operators listed in Section 3.1, and differentiating between two --- elementary and general --- sorts of nonlogical atoms.

The *variables* and *constants* of the language are those fixed earlier when discussing games. We also have infinitely many *nonlogical function letters*, *elementary game letters* and *general game letters*. With each of these is associated a fixed natural number called the *arity*. We usually use $f,g,h\ldots$ as metavariables for function letters, $p,q,r,\ldots$ for elementary game letters, and $P,Q,R,\ldots$ for general game letters. Other than these nonlogical letters, there are three *logical* letters, all elementary: ⊤ (nullary), ⊥ (nullary) and = (binary).

*Terms* are defined inductively as follows:

- All variables are terms.
- All constants are terms.
- If $t_1,\ldots,t_n$ are terms ($n \geq 0$) and $f$ is an $n$-ary function letter, then $f(t_1,\ldots,t_n)$ is a term.

*Atoms* are defined by:

- ⊤ and ⊥ are atoms.
- If $t_1$ and $t_2$ are terms, then $t_1 = t_2$ is an atom.
- If $t_1,\ldots,t_n$ are terms ($n \geq 0$) and $L$ is a nonlogical $n$-ary (elementary or general) game letter, then $L(t_1,\ldots,t_n)$ is an atom.

Finally, *formulas* are defined by:

- All atoms are formulas.
- If $E$ is a formula, then so are ¬(E), ⚬(E), ⊳(E), ⊲(E), ⊳(E), ⌓(E), ⋏(E), ⋌(E), ⋋(E), ⦵(E), ⵛ(E), ⵎ(E), ⵛ(E).
- If $E$ and $F$ are formulas, then so are $(E)\wedge(F)$, $(E)\sqcap(F)$, $(E)\wedge(F)$, $(E)\triangle(F)$, $(E)\wedge(F)$, $(E)\vee(F)$, $(E)\sqcup(F)$, $(E)\triangle(F)$, $(E)\vee(F)$, $(E)\rightarrow(F)$, $(E)\sqsupset(F)$, $(E)\triangleright(F)$, $(E)\triangleright(F)$, $(E)\multimap(F)$, $(E)\triangleright(F)$, $(E)\triangleright(F)$, $(E)\triangleright(F)$.
- If $E$ is a formula and $x$ is a variable, then $\forall x(E), \wedge x(E), \sqcap x(E), \triangle x(E), \wedge x(E), \exists x(E), \vee x(E), \sqcup x(E), \triangledown x(E), \vee x(E)$ are formulas.

Unnecessary parentheses will be usually omitted in formulas according to the standard conventions, with partial precedence order as agreed upon earlier for the corresponding game operations. The notions of free and bound occurrences of variables are also standard, with the only adjustment that now we have eight rather than two quantifiers.

A *sentence* (*closed formula*) is a formula with no free occurrences of variables.

## 5.2 Interpretations

For the following definition, recall Conventions 2.4.3 and 2.4.4. Also recall that $var_1,\ldots,var_n$ are the first $n$ variables from the lexicographic list of all variables.

**Definition 5.2.1** An *interpretation* is a mapping * which:

- Sends the word "universe" to a universe **Un***=(**Dm***, **Dn***). Such a **Un*** (resp. **Dm***, resp. **Dn***) is said to be the *universe* (resp. *domain*, resp. *denotation function*) of *.

- Sends every $n$-ary function letter $f$ to an $n$-ary function $f^*=f^*(var_1,\ldots,var_n)$ with $\mathbf{Dm}^{f^*}=\mathbf{Dm}^*$ and $\mathbf{Vr}^{f^*}=\{var_1,\ldots,var_n\}$.
- Sends every nonlogical $n$-ary game letter $L$ to an $n$-ary static game $L^*=L^*(var_1,\ldots,var_n)$ with $\mathbf{Dm}^{L^*}=\mathbf{Dm}^*$ and $\mathbf{Vr}^{L^*}=\{var_1,\ldots,var_n\}$, such that, if the letter $L$ is elementary, then so is the game $L^*$.

\* is said to be **admissible** for a formula $E$ (or $E$-**admissible**) iff, whenever $E$ has an occurrence of an atom $A(t_1,\ldots,t_n)$ in the scope of $\forall x$ or $\exists x$ and one of the terms $t_i$ ($1\leq i\leq n$) contains the variable $x$, $A^*$ is unistructural in $var_i$. We uniquely extend \* to a mapping which sends each term $t$ to a function $t^*$, and each formula $E$ for which it is admissible to a game $E^*$, by stipulating that:

- Where $c$ is a constant, $c^*$ is (the nullary function) $c^{\mathbf{Un}^*}$.
- Where $x$ is a variable, $x^*$ is (the unary function) $x^{\mathbf{Un}^*}$.
- Where $f$ is an $n$-ary function letter and $t_1,\ldots,t_n$ are terms, $(f(t_1,\ldots,t_n))^*$ is $f^*(t_1^*,\ldots,t_n^*)$.
- $\top^*$ is $\top$ and $\bot^*$ is $\bot$.
- Where $t_1$ and $t_2$ are terms, $(t_1=t_2)^*$ is $t_1^*=t_2^*$.
- Where $L$ is an $n$-ary game letter and $t_1,\ldots,t_n$ are terms, $L((t_1,\ldots,t_n))^*$ is $L^*(t_1^*,\ldots,t_n^*)$.
- \* commutes with all logical operators, seeing them as the corresponding game operations: $(\neg E)^*$ is $\neg(E^*)$, $(\wedge E)^*$ is $\wedge(E^*)$, $(E\sqcap F)^*$ is $(E^*)\sqcap(F^*)$, $(\forall xE)^*$ is $\forall x(E^*)$, etc.

When $O$ is a function letter, a game letter, a term or a formula and $O^*=W$, we say that \* **interprets** $O$ as $W$. We can also refer to such a $W$ as "$O$ under interpretation \*".

## 5.3 Validity

**Definition 5.3.1** Let $S$ be a sentence. We say that:

- $S$ is **logically** (or **uniformly**) **valid** iff there is an HPM $M$ such that, for every $S$-admissible interpretation \*, $M$ computes $S^*$. Such an $M$ is said to be a **logical** (or **uniform**) **solution** of $S$.
- $S$ is **nonlogically** (or **multiformly**) **valid** iff for every $S$-admissible interpretation \* there is an HPM $M$ such that $M$ computes $S^*$.

**Convention 5.3.2** When $S$ is a formula but not a sentence, its validity is understood as that of the $\sqcap$-closure of $S$, i.e., of the sentence $\sqcap x_1\ldots\sqcap x_n S$, where $x_1,\ldots,x_n$ are all free variables of $S$ listed according to their lexicographic order.

Every logically valid formula is, of course, also nonlogically valid. But some nonlogically valid formulas may fail to be logically valid. For instance, where $p$ is a 0-ary elementary game letter, the formula $p\sqcup\neg p$ is valid nonlogically but not logically. It is nonlogically valid for a trivial reason: given an interpretation \*, either $p$ or $\neg p$ is true under \*. If $p$ is true, then the strategy that chooses the left disjunct wins; and if $p$ is false, then the strategy that chooses the right disjunct wins. The trouble is that, even though we know that one of these two strategies succeeds, generally we have no way to tell which one does. Otherwise, we would have an answer to the question on whether there is life on Mars or not, or whether P=NP or not.

The nonlogical validity of $p\sqcup\neg p$ is not only non-constrive, but also a fragile sort of validity: this property, unlike logical validity, is not closed under substitution of atoms. For instance, where $q$ is a unary elementary game letter, the formula $q(x)\sqcup\neg q(x)$, while having the same form as $p\sqcup\neg p$, is no longer nonlogically valid. The papers on CoL written prior to 2016 had a more relaxed understanding of interpretations than our present understanding. Namely, there was no requirement that an interpretation should respect the arity of a game letter. In such a case, as it turns out, the extensional difference between logical and nonlogical validity disappears: while the class of logically valid principles remains the same, the class of nonlogically valid principles shrinks down to that of logically valid ones.

Intuitively, a logical solution $M$ for a sentence $S$ is an interpretation-independent winning strategy: since the "intended" or "actual" interpretation * is not visible to the machine, $M$ has to play in some standard, uniform way that would be successful for any possible interpretation of $F$. It is uniform rather than multiform validity that is of interest in applied areas of computer science such as substantial theories (Section 7) or knowledgebase systems (Section 8). In these sorts of applications we want a logic on which a universal problem-solving machine can be based. Such a machine would or should be able to solve problems represented by logical formulas without any specific knowledge of the meanings of their atoms, i.e. without knowledge of the actual interpretation. In other words, such a machine has to be a logical rather than nonlogical solution.

For the above reasons, in the subsequent sections we will only be focused on the logical sort of validity.

# 6 Axiomatizations

## 6.1 Outline

An axiomatization of the set of valid formulas of the full language of Section 5 has not been found so far. In fact, it is not known if such an axiomatization exists in principle. However, axiomatizations do exist for a number of fragments of the language, obtained by disallowing certain operators, disallowing one or another sort of game letters, or imposing some other restrictions on formulas.

In this section we explore six of such axiomatic systems: **CL7**, **Int**$^+$, **CL15**, **CL13**, **CL4** and **CL12**. The first four are *propositional* systems (only nullary game letters allowed), while the last two are *first-order* systems (no arity restrictions on game letters). **CL7**, **Int**$^+$ and **CL15** are *general-base*, meaning that only general game letters are allowed. **CL12** is *elementary-base*, meaning that only elementary game letters are allowed. The remaining systems **CL13** and **CL4** are *mixed-base*, allowing both sorts of game letters. The sets of operators allowed in formulas are: just $\{\rightarrow\}$ for **CL7**; $\{\sqcap,\sqcup,\circ\!\!-\}$ for **Int**$^+$; $\{\neg,\wedge,\vee,⦶,⦷\}$ for **CL15**; $\{\neg,\wedge,\vee,\sqcap,\sqcup,\triangle,\triangledown,⩓,⩔\}$ for **CL13**; $\{\neg,\wedge,\vee,\sqcap,\sqcup,\sqcap,\sqcup,\forall,\exists\}$ for **CL4**; and $\{\neg,\wedge,\vee,\sqcap,\sqcup,\sqcap,\sqcup,\forall,\exists,$ external $\circ\!\!-\}$ for **CL12**. We also divide these systems into the categories *Gentzen style* (**CL7**, **Int**$^+$), *cirquent calculus* (**CL15**) and *brute force* (**CL13**,**CL4**,**CL12**). Gentzen-style systems are sequent calculus systems in the traditional sense. It turns out that only very limited fragments of CoL can be axiomatized this or any other traditional (e.g. Hilbert-style) way, and some novel sorts of proof systems are necessary. Cirquent calculus and brute force systems are such novel sorts. Cirquent calculus operates with cirquents rather than formulas or sequents. Brute force systems operate with formulas or sequents, but in an unusual way, with their rules being relatively directly derived from the underlined game semantics and hence somewhat resembling games themselves.

All of the above systems have been proven to be adequate (sound and complete). Specifically, we have:

**Theorem 6.1.1** Let **S** be any one of the systems **CL7**, **Int**$^+$, **CL15**, **CL13**, **CL4** or **CL12** defined in the forthcoming Subsections 6.2-6.7. With "**S**-formula" meaning a formula of the language of **S** and "**S**-proof" meaning "proof in **S**", we have:

1. *Adequacy*: An **S**-formula $F$ is logically valid iff $F$ is provable in **S**.
2. *Uniform-constructive soundness*: There is an effective procedure that takes an arbitrary **S**-proof of an arbitrary **S**-formula $F$ and generates a logical solution for $F$.

In the sequel we shall refer to the property of satisfying the above two clauses together as ***uniform-constructive adequacy***.

## 6.2 The Gentzen-style system CL7

**CL7-*formulas*** are formulas of the language of CoL that do not contain any function letters, do not contain any game letters (including the logical letters ⊤ and ⊥) other than nonlogical 0-ary general game letters, and do not contain any operators other than →.

A **CL7**-*sequent* is a pair $\Gamma \Rightarrow F$, where $\Gamma$ is a (possibly empty) multiset of **CL7**-formulas, and $F$ is a **CL7**-formula.

As usual, when $\Gamma$ and $\Delta$ are multisets of formulas and $F$ is a formula, we shall write "$\Gamma,\Delta$" for $\Gamma \cup \Delta$ and "$\Gamma,F$" for $\Gamma \cup \{F\}$.

The ***axioms*** of **CL7** are all **CL7**-sequents of the form $\Gamma,F \Rightarrow F$. And the system only has the following two ***rules*** of inference:

$$\frac{\Gamma, F \Rightarrow G \quad \Delta \Rightarrow E}{\Gamma, \Delta, E \rightarrow F \Rightarrow G} \qquad \frac{\Gamma, E \Rightarrow F}{\Gamma \Rightarrow E \rightarrow F}$$

A **CL7-*proof*** of a sequent $S$ is a sequence $X_1,\ldots,X_n$ of sequents, where $S=X_n$, and where each $X_i$ is either an axiom or follows from some earlier sequents of the sequence by one of the rules of inference. A **CL7-*proof*** of a formula $F$ is a **CL7**-proof of the empty-antecedent sequent $\Rightarrow F$.

With → understood as linear implication, **CL7** proves nothing less and nothing more than the implicative fragment of affine logic does. Adding the (left) *contraction* rule

$$\frac{\Gamma, E, E \Rightarrow F}{\Gamma, E \Rightarrow F}$$

to **CL7** results in an unsound system. Let us call it **CL7**+Contraction. An example of a formula provable in the latter but not in **CL7** is $(E \rightarrow (F \rightarrow G)) \rightarrow ((E \rightarrow F) \rightarrow (E \rightarrow G))$. It can be shown however that **CL7**+Contraction is sound and complete if we rewrite → as either ∘– or ≻–. The logical behaviors of these two rimplications in isolation are thus undistinguishable. This stops being the case when the rimplications are taken in combination with other operators. For instance, the intuitionistically provable principle $(E \supset G)((F \supset G) \supset (E \sqcup F \supset G))$ is valid with ⊃ read as ∘– but not as ≻–. An example of a classical tautology invalid with all three operators ⊃∈{→,∘–A, ≻–} is Peirce's Law $((E \supset F) \supset E) \supset E$.

**CL7** was studied and proven uniform-constructively adequate in [Jap09b].

**Open problem 6.2.1** Does **CL7** remain complete with respect to multiform (rather than uniform) validity?

## 6.3 The Gentzen-style system $\mathbf{Int^+}$

**$\mathbf{Int^+}$-formulas** are formulas of the language of CoL that do not contain any function letters, do not contain any game letters (including the logical letters ⊤, ⊥) other than nonlogical 0-ary general game letters, and do not contain any operators other than ⊓, ⊔, ∘−. It turns out that logically valid **$\mathbf{Int^+}$**-formulas are exactly those provable in the positive (negation- and absurd-free) fragment of propositional intuitionistic logic $\mathbf{Int^+}$. As a deductive system, $\mathbf{Int^+}$ is the result of replacing → by ∘− in the earlier discussed **CL7**+Contraction, allowing ⊓ and ⊔ as additional operators in formulas, and adding the following rules to the system, with $i$ ranging over $\{1,2\}$:

$$\frac{\Gamma, E_1 \Rightarrow F \quad \Gamma, E_2 \Rightarrow F}{\Gamma, E_1 \sqcup E_2 \Rightarrow F} \qquad \frac{\Gamma \Rightarrow E_i}{\Gamma \Rightarrow E_1 \sqcup E_2}$$

$$\frac{\Gamma, E_i \Rightarrow F}{\Gamma, E_1 \sqcap E_2 \Rightarrow F} \qquad \frac{\Gamma \Rightarrow E_1 \quad \Gamma \Rightarrow E_2}{\Gamma \Rightarrow E_1 \sqcap E_2}$$

The full propositional intuitionistic logic **Int** is obtained from $\mathbf{Int^+}$ by allowing ⊥ in the language, and adding the axiom $\Gamma, \bot \Rightarrow F$. Such a system is sound but incomplete with respect to our semantics. Below is an example of a logically valid yet not **Int**-provable principle, with $\neg F$ (the intuitionistic negation of $F$) understood as an abbreviation of $F \circ\!\!- \bot$:

$$(\circ\!\!-\! E \circ\!\!-\! F \sqcup G) \sqcap (\circ\!\!-\! \circ\!\!-\! E \circ\!\!-\! H \sqcup J) \circ\!\!-\! (\circ\!\!-\! E \circ\!\!-\! F) \sqcup (\circ\!\!-\! E \circ\!\!-\! G) \sqcup (\circ\!\!-\! \circ\!\!-\! E \circ\!\!-\! H) \sqcup (\circ\!\!-\! \circ\!\!-\! E \circ\!\!-\! J).$$

The uniform-constructive adequacy of $\mathbf{Int^+}$ with respect to our semantics follows from slightly stronger results proven in [Jap07d, Mez10].

**Open problems 6.3.1**

1. Does $\mathbf{Int^+}$ remain complete with respect to multiform (rather than uniform) validity?
2. Add ⊥ to the language of $\mathbf{Int^+}$, and adequately axiomatize (if possible) the corresponding superintuitionistic logic. What kinds of (if any) Kripke models would such a logic have?
3. Is the first-order (with quantifiers ⊓,⊔) positive (negation- and absurd-free) intuitionistic logic also adequate with respect to the CoL semantics? If not, could it be axiomatized and how?
4. Add ⊥ to the language of $\mathbf{Int^+}$, but replace the general game letters of the latter with nonlogical elementary game letters. Adequately axiomatize (if possible) the corresponding logic.
5. If the preceding task is achieved, then try the first-order version of it as well.
6. In the language of $\mathbf{Int^+}$, replace ∘− with ⤳. What logic are we getting? How about the variations of this logic in the style of the above items 1-5?

## 6.4 The cirquent calculus system **CL15**

**CL15-formulas** are formulas of the language of CoL that do not contain any function letters, do not contain any game letters (including the logical letters ⊤,⊥) other than nonlogical 0-ary general game letters, and do not contain any operators other than ¬, ∧, ∨, ⊓, ⊔. Besides, it is required that ¬ is only applied to atoms. This condition can be violated if we understand ¬(E∧F), ¬(E∨F), ¬⊓E and ¬⊔E as abbreviations of ¬E∨¬F, ¬E∧¬F, ⊔¬E and ⊓¬E, respectively. We may as well write E→F or write E∘–F, understanding these expressions as abbreviations of ¬E∨F and ⊓E→F (i.e. ⊔¬E∨F), respectively.

**Definition 6.4.1** A **CL15-cirquent** (hencefore simply "cirquent") is a triple $C=(F,U,O)$ where:

- **F** is a nonempty finite sequence of **CL15**-formulas, whose elements are said to be the **oformulas** of C. Here the prefix "o" is for "occurrence", and is used to mean a formula together with a particular occurrence of it in **F**. So, for instance, if $F=\langle E,G,E\rangle$, then the cirquent has three oformulas even if only two formulas.
- Both **U** and **O** are nonempty finite sequences of nonempty sets of oformulas of C. The elements of **U** are said to be the **undergroups** of C, and the elements of **O** are said to be the **overgroups** of C. As in the case of oformulas, it is possible that two undergroups or two overgroups are identical as sets (have identical **contents**), yet they count as different undergroups or overgroups because they occur at different places in the sequence **U** or **O**. Simply "**group**" will be used as a common name for undergroups and overgroups.
- Additionally, every oformula is required to be in (the content of) at least one undergroup and at least one overgroup.

While oformulas are not the same as formulas, we may often identify an oformula with the corresponding formula and, for instance, say "the oformula E" if it is clear from the context which of the possibly many occurrences of E is meant. Similarly, we may not always be very careful about differentiating between groups and their contents.

We represent cirquents using three-level diagrams such as the one shown below:

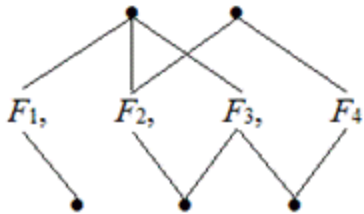

This diagram represents the cirquent with four oformulas $F_1$, $F_2$, $F_3$, $F_4$, three undergroups $\{F_1\}$, $\{F_2,F_3\}$, $\{F_3,F_4\}$ and two overgroups $\{F_1,F_2,F_3\}$, $\{F_2,F_4\}$. We typically do not terminologically differentiate between cirquents and diagrams: for us, a diagram *is* (rather than *represents*) a cirquent, and a cirquent *is* a diagram. Each group is represented by (and identified with) a •, where the **arcs** (lines connecting the • with oformulas) are pointing to the oformulas that the given group contains.

The **axioms** of **CL15** are all cirquents of the form

$$(\langle \neg F_1, F_1, \ldots, \neg F_n, F_n \rangle, \langle \{\neg F_1, F_1\}, \ldots, \{\neg F_n, F_n\} \rangle, \langle \{\neg F_1, F_1\}, \ldots, \{\neg F_n, F_n\} \rangle),$$

where $n$ is any positive integer, and $F_1,\ldots,F_n$ are any formulas. The diagram of such a cirquent looks like an array of $n$ "diamonds", as shown below for the case of $n=3$:

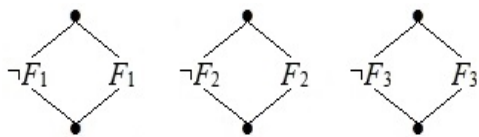

**CL15** has nine rules of inference. Below we explain them in a relaxed fashion, in terms of deleting arcs, swapping oformulas, etc. Such explanations are rather clear, and translating them into rigorous formulations in the style and terms of Definition 6.4.1, while possible, is hardly necessary. All rules take a single premise.

**1.** *Exchange* This rule comes in three flavors: ***Undergroup Exchange***, ***Oformula Exchange*** and ***Overgroup Exchange***. Each one allows us to swap any two adjacent objects (undergroups, oformulas or overgroups) of a cirquent, otherwise preserving all oformulas, groups and arcs.

Below we see three examples. In each case, the upper cirquent is the premise and the lower cirquent is the conclusion of an application of the rule. Between the two cirquents --- here and later --- is placed the name of the rule by which the conclusion follows from the premise.

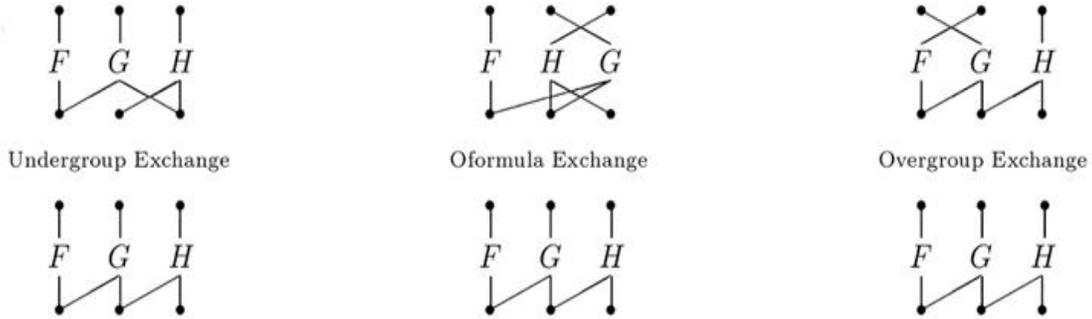

The presence of Exchange essentially allows us to treat all three components (***F***,***U***,***O***) of a cirquent as multisets rather than sequences.

**2.** *Weakening* The premise of this rule is obtained from the conclusion by deleting an arc between some undergroup *U* with ≥2 elements and some oformula *F*; if *U* was the only undergroup containing *F*, then *F* should also be deleted (to satisfy Condition 3 of Definition 6.4.1), together with all arcs between *F* and overgroups; if such a deletion makes some overgroups empty, then they should also be deleted (to satisfy Condition 2 of Definition 6.4.1).   Below are three examples:

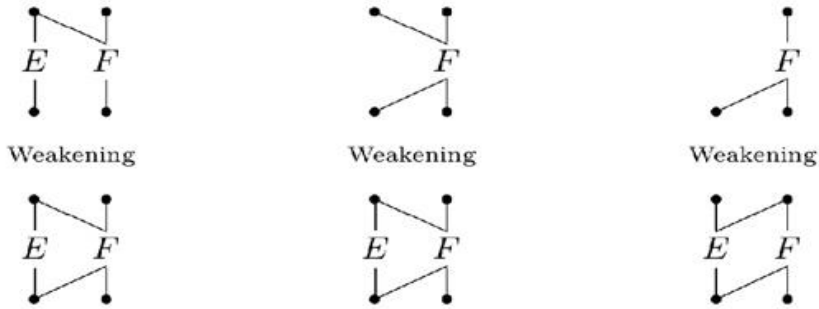

**3.** *Contraction* The premise of this rule is obtained from the conclusion through replacing an oformula ⚷*F* by two adjacent oformulas ⚷*F*,⚷*F*, and including them in exactly the same undergroups and overgroups in which the original oformula was contained. Example:

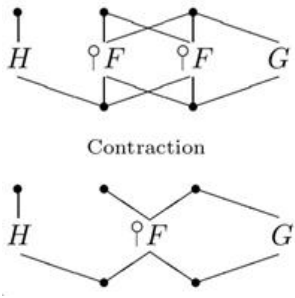

**4.** *Duplication* This rule comes in two versions: ***Undergroup Duplication*** and ***Overgroup Duplication***. The conclusion of Undergroup Duplication is the result of replacing, in the premise, some undergroup *U* with two adjacent undergroups whose contents are identical to that of *U*. Similarly for Overgroup Duplication. Examples:

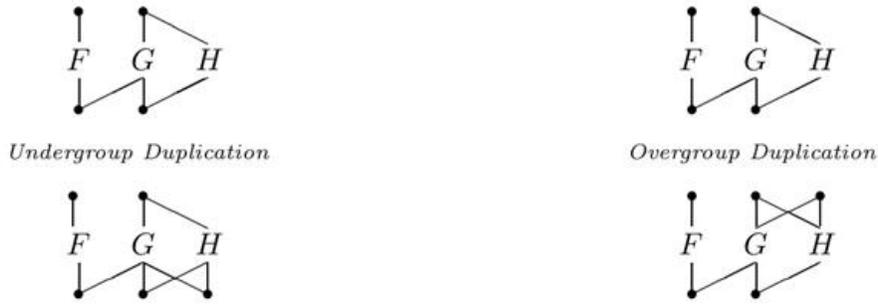

**5.** *Merging* In the top-down view, this rule merges any two adjacent overgroups, as illustrated below.

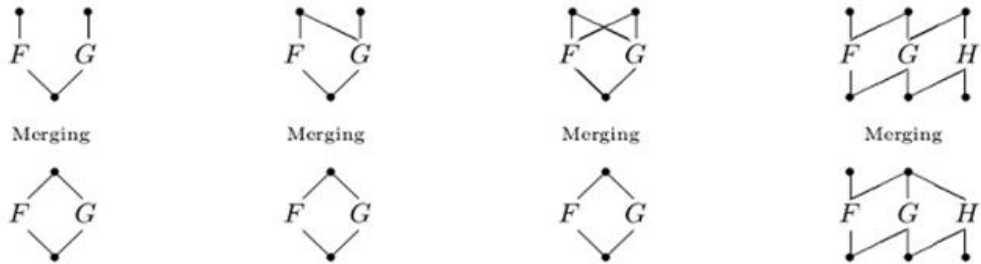

**6.** *Disjunction Introduction* The premise of this rule is obtained from the conclusion by replacing an oformula *F*∨*G* by two adjacent oformulas *F*,*G*, and including both of them in exactly the same undergroups and overgroups in which the original oformula was contained, as illustrated below:

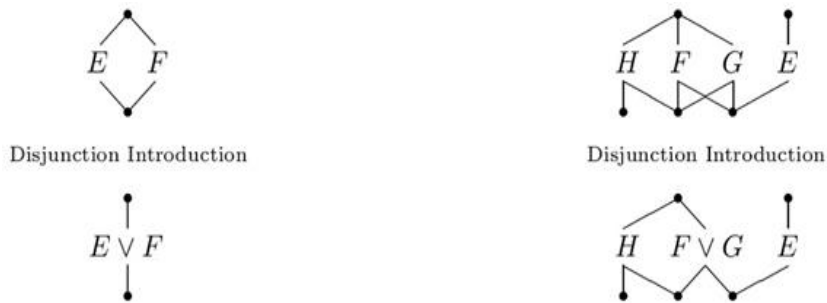

**7.** *Conjunction Introduction* The premise of this rule is obtained from the conclusion by applying the following two steps:

- Replace an oformula *F*∧*G* by two adjacent oformulas *F*,*G*, and include both of them in exactly the same undergroups and overgroups in which the original oformula was contained.

- Replace each undergroup *U* originally containing the oformula *F*∧*G* (and now containing *F*,*G* instead) by the two adjacent undergroups *U*-{*G*} and *U*-{*F*}.

Below we see three examples.

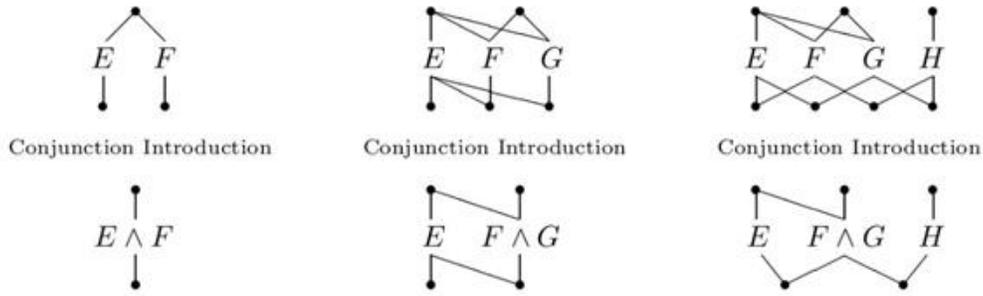

**8. *Recurrence Introduction*** The premise of this rule is obtained from the conclusion through replacing an oformula ↓$F$ by $F$ (while preserving all arcs), and inserting, anywhere in the cirquent, a new overgroup that contains $F$ as its only oformula. Examples:

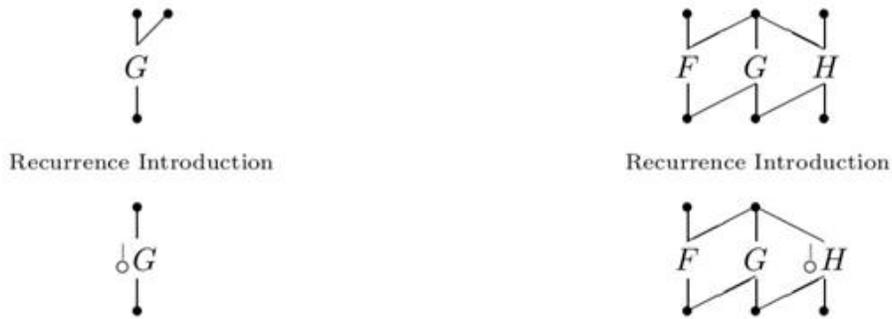

**9. *Corecurrence Introduction*** The premise of this rule is obtained from the conclusion through replacing an oformula ↑$F$ by $F$, and including $F$ in any (possibly zero) number of the already existing overgroups in addition to those in which the original oformula ↑$F$ was already present. Examples:

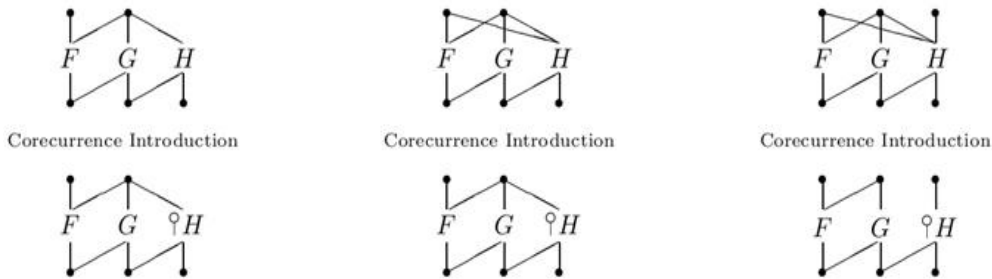

A **CL15-*proof*** (or simply a proof in this subsection) of a cirquent $C$ is a sequence of cirquents ending in $C$ such that the first cirquent is an axiom, and every subsequent cirquent follows from the immediately preceding cirquent by one of the rules of **CL15**. A **CL15-*proof*** of a formula $F$ is understood as a **CL15**-proof of the cirquent (⟨F⟩,⟨{F}⟩,⟨{F}⟩), i.e. the cirquent

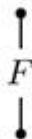

Let us look at some examples. The following is a proof of ↓$F \to F$, i.e. ↑¬$F \lor F$:

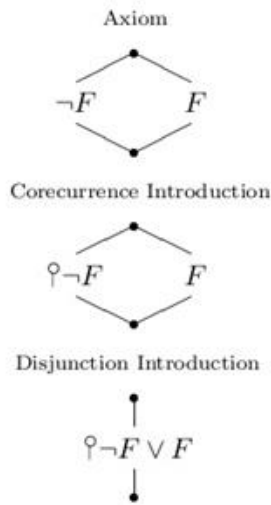

The following example shows a proof of $F \wedge F \rightarrow F$:

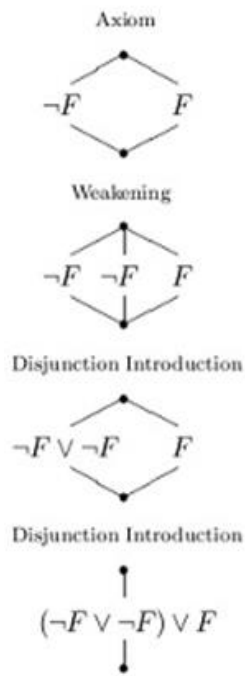

At the same time, the converse $F \rightarrow F \wedge F$ of the above formula has no proof. However, the latter becomes provable after prefixing each occurrence of $F$ with with ♭. Below is a proof of ♭$F \rightarrow$ ♭$F \wedge$ ♭$F$. The meaning of the principle expressed by this formula can be characterized by saying that solving two copies of a problem of the form ♭$F$ does not take any more resources (is not any harder) than solving just a single copy.

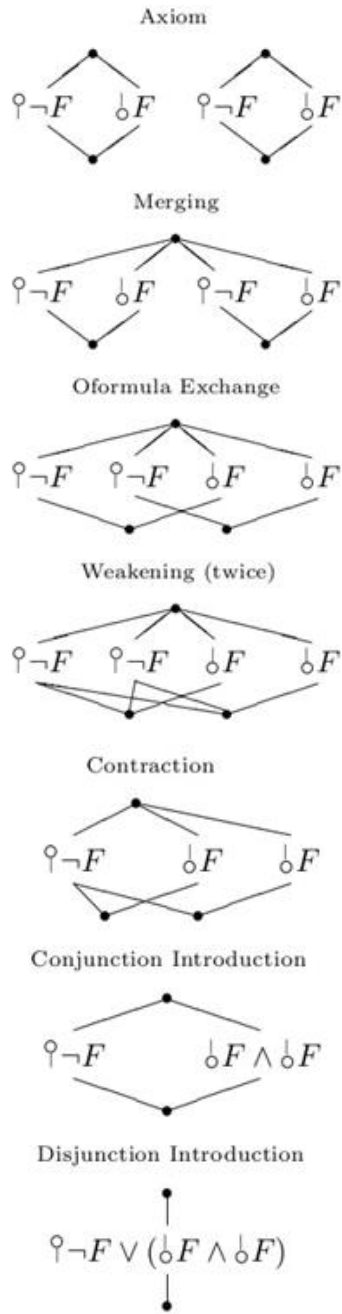

Below is a proof of $\flat F \to \flat\flat F$. Unlike the previously seen examples, proving this formula requires using Duplication.

Axiom

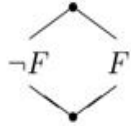

Overgroup Duplication

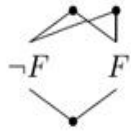

Corecurrence Introduction

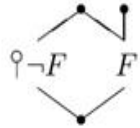

Overgroup Duplication

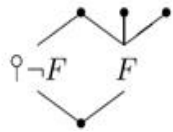

Recurrence Introduction

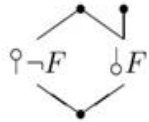

Recurrence Introduction

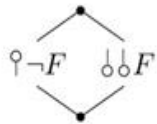

Disjunction Introduction

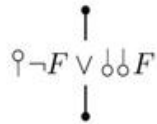

Now we prove ↓$E$ ∨ ↓$F$ → ↓($E$ ∨ $F$). The converse ↓($E$ ∨ $F$) → ↓$E$ ∨ ↓$F$, on the other hand, can be shown to be unprovable.

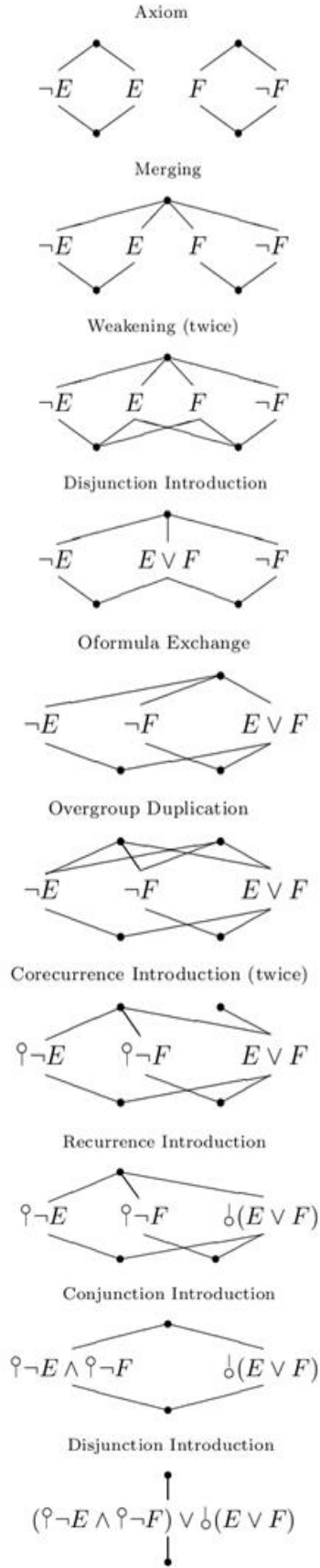

The formulas proven so far are also provable in affine logic (with ∧,∨ understood as multiplicatives, ↓,↑ as exponentials, and ¬F as $F^\perp$). In general, **CL15** is a proper extension of the corresponding (additive-free) fragment of affine logic. The following example shows the **CL15**-provability of the formula $↑↓F→↓↑F$, which is not provable in affine logic. The converse $↓↑F→↑↓F$, on the other hand, is unprovable in either system.

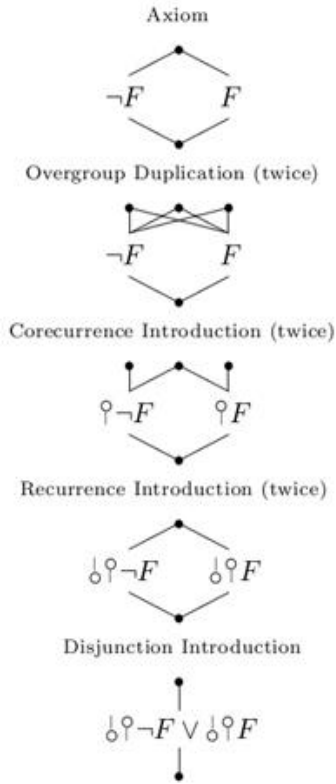

Another --- longer but recurrence-free --- example separating **CL15** from affine logic is $(E\wedge F) \vee (G\wedge H) \rightarrow (E \vee G) \wedge (F \vee H)$ (the already mentioned Blass principle), or even the special case $(E\wedge E) \vee (E\wedge E) \rightarrow (E\vee E) \wedge (E\vee E)$ of it; constructing a **CL15**-proof of this formula is left as an exercise.

**CL15** was introduced and proven uniform-constructively adequate in [Jap13a, Jap13b]. W. Xu in [Xu13] proved that **CL15** remains sound with ⋏,⋎ instead of ⩔,⩓. Completeness, however, is lost in this case because, as already mentioned, the formula $F\wedge \lambda(F\rightarrow F\wedge F) \rightarrow \lambda F$ is valid while $A\wedge \circ(A\rightarrow A\wedge A) \rightarrow \circ A$ is not.

**Open problems 6.4.2**

1. Is **CL15** decidable?
2. Extend the language of **CL15** so as to include the choice connectives ⊓ and ⊔, and axiomatize (if possible) the set of valid formulas in this extended language.
3. Replace ⩔, ⩓ with ⋏, ⋎ in the language of **CL15**. Is the set of valid formulas in this new language axiomatizable (and how)?
4. Does **CL15** remain complete with respect to multiform (rather than uniform) validity?

## 6.5 The brute force system **CL13**

**CL13-*formulas*** are formulas of the language of CoL that do not contain any function letters, do not contain any non-nullary game letters, and do not contain any operators other than ¬, ∧, ∨, ⊓, ⊔, △, ▽, ⋀, ⋁. As in the case of **CL15**, officially ¬ is only allowed to be applied to atoms, otherwise understood as the corresponding DeMorgan abbreviation. $E \rightarrow F$ is also understood as an abbreviation of its standard meaning. To define the system axiomatically, we need certain terminological conventions.

- A *literal* means an atom with or without negation ¬.

- As in the previous case, we often need to differentiate between *subformulas* as such, and particular *occurrences* of subformulas. We will be using the term *osubformula* to mean a subformula together with a particular occurrence. The prefix "o" will be used with a similar meaning in terms such as *oatom*, *oliteral*, etc.
- An osubformula is *positive* iff it is not in the scope of ¬. Otherwise it is *negative*. According to our conventions regarding the usage of ¬, only oatoms may be negative.
- A *politeral* is a positive oliteral.
- A ∧-*(sub)formula* is a (sub)formula of the form $E \wedge F$. Similarly for the other connectives of the language.
- A *sequential (sub)formula* is one of the form $E △ F$ or $E ▽ F$. We say that $E$ is the *head* of such a (sub)formula, and $F$ is its *tail*.
- Similarly, a *parallel (sub)formula* is one of the form $E \wedge F$ or $E \vee F$, a *toggling (sub)formula* is one of the form $E ⩕ F$ or $E ⩖ F$, and a *choice (sub)formula* is one of the form $E \sqcap F$ or $E \sqcup F$.
- A formula is said to be *quasielementary* iff it contains no general atoms and no operators other than ¬, ∧, ∨, ⩕, ⩖.
- A formula is said to be *elementary* iff it is a formula of classical propositional logic, i.e., contains no general atoms and no operators other than ¬, ∧, ∨.
- A *semisurface osubformula* (or *occurrence*) is an osubformula (or occurrence) which is not in the scope of a choice connective.
- A *surface osubformula* (or *occurrence*) is an osubformula (or occurrence) which is not in the scope of any connectives other than ¬, ∧, ∨.
- The *quasielementarization* of a formula $F$, denoted by $|F|$, is the result of replacing in $F$ every sequential osubformula by its head, every ⊓-osubformula by ⊤, every ⊔-osubformula by ⊥, and every general politeral by ⊥ (the order of these replacements does not matter). For instance, $\big|\big((P ⩖ q) \vee ((p \wedge \neg P) △ (Q \wedge R))\big) \wedge \big(q \sqcap (r \sqcup s)\big)\big| = \big((\bot ⩖ q) \vee (p \wedge \bot)\big) \wedge \top$.
- The *elementarization* of a quasielementary formula $F$, denoted by $\|F\|$, is the result of replacing in $F$ every ⩕-osubformula by ⊤ and every ⩖-osubformula by ⊥ (again, the order of these replacements does not matter). For instance, $\|(s \wedge (p ⩕ (q ⩖ r))) \vee (\neg s \vee (p ⩖ r))\| = (s \wedge \top) \vee (\neg s \vee \bot)$.
- A quasielementary formula $F$ is said to be *stable* iff its elementarization $\|F\|$ is a tautology of classical logic. Otherwise $F$ is *unstable*.

We now define **CL13** by the following six rules of inference, where $\mathcal{P} \succ F$ means "from premise(s) $\mathcal{P}$ conclude $F$". Axioms are not explicitly stated, but the set of premises of the (⩕) rule can be empty, in which case (the conclusion of) this rule acts like an axiom.

**Rule** (⩕): $H \succ F$, where $F$ is a stable quasielementary formula, and $H$ is the smallest set of formulas satisfying the following condition:

- Whenever $F$ has a surface osubformula $E_0 ⩕ E_1$, for both $i \in \{0,1\}$, $H$ contains the result of replacing in $F$ that osubformula by $E_i$.

**Rule** (⩖): $H \succ F$, where $F$ is a quasielementary formula, and $H$ is the result of replacing in $F$ a surface osubformula $E ⩖ G$ by $E$ or $G$.

**Rule** (△⊓): $|F|, H \succ F$, where $F$ is a non-quasielementary formula (note that otherwise $F = |F|$), and $H$ is the smallest set of formulas satisfying the following two conditions:

- Whenever $F$ has a semisurface osubformula $G_0 \sqcap G_1$, for both $i \in \{0,1\}$, $H$ contains the result of replacing in $F$ that osubformula by $G_i$.
- Whenever $F$ has a semisurface osubformula $E △ G$, $H$ contains the result of replacing in $F$ that osubformula by $G$.

**Rule** (⊔): $H \succ F$, where $H$ is the result of replacing in $F$ a semisurface osubformula $E \sqcup G$ by $E$ or $G$.

**Rule** (▽): $H \succ F$, where $H$ is the result of replacing in $F$ a semisurface osubformula $E \triangledown G$ by $G$.

**Rule** (M): $H \succ F$, where $H$ is the result of replacing in $F$ two --- one positive and one negative --- semisurface occurrences of some general atom $P$ by a nonlogical elementary atom $p$ which does not occur in $F$.

A **CL13-*proof*** of a formula $F$ can be defined as a sequence of formulas ending with $F$ such that every formula follows from some (possibly empty) set of earlier formulas by one of the rules of the system.

**Example 6.5.1** In view of the discussions of Section 3.10, the formula $(\neg p \lor\!\!\!\lor p) \land (\neg q \lor\!\!\!\lor q) \to \neg(p \land q) \lor\!\!\!\lor (p \land q)$ expresses --- or rather implies --- the known fact that, if two predicates $p$ and $q$ are recursively approximable, then so is their intersection $p \land q$. The following sequence is a **CL13**-proof of this formula, re-written into its official form $((p \land \neg p) \lor (q \land \neg q)) \lor\!\!\!\lor ((\neg p \lor \neg q) \lor\!\!\!\lor (p \land q))$:

1. $(p \lor q) \lor (\neg p \lor \neg q)$                        (∧): (no premises)
2. $(p \lor q) \lor ((\neg p \lor \neg q) \lor\!\!\!\lor (p \land q))$         (⩔): 1
3. $(p \lor \neg q) \lor (\neg p \lor \neg q)$                        (∧): (no premises)
4. $(p \lor \neg q) \lor ((\neg p \lor \neg q) \lor\!\!\!\lor (p \land q))$        (⩔): 3
5. $(p \lor (q \land \neg q)) \lor ((\neg p \lor \neg q) \lor\!\!\!\lor (p \land q))$      (∧): 2,4
6. $(\neg p \lor q) \lor (\neg p \lor \neg q)$                      (∧): (no premises)
7. $(\neg p \lor q) \lor ((\neg p \lor \neg q) \lor\!\!\!\lor (p \land q))$       (⩔): 6
8. $(\neg p \lor \neg q) \lor (p \land q)$                          (∧): (no premises)
9. $(\neg p \lor \neg q) \lor ((\neg p \lor \neg q) \lor\!\!\!\lor (p \land q))$      (⩔): 8
10. $(\neg p \lor (q \land \neg q)) \lor ((\neg p \lor \neg q) \lor\!\!\!\lor (p \land q))$    (∧): 7,9
11. $((p \land \neg p) \lor q) \lor ((\neg p \lor \neg q) \lor\!\!\!\lor (p \land q))$      (∧): 2,7
12. $((p \land \neg p) \lor \neg q) \lor ((\neg p \lor \neg q) \lor\!\!\!\lor (p \land q))$    (∧): 4,9
13. $((p \land \neg p) \lor (q \land \neg q)) \lor ((\neg p \lor \neg q) \lor\!\!\!\lor (p \land q))$   (∧): 5,10,11,12

**Example 6.5.2** Pick any two distinct connectives $\&_1$ and $\&_2$ from the list ∧,⩓,△,⊓. Then **CL13** proves the formula $P \&_1 Q \to P \&_2 Q$ if and only if $\&_1$ is to the left of $\&_2$ in the list. Similarly for the list ⊔,▽,⩔,∨. Below we verify this fact only for the case $\{\&_1, \&_2\} = \{⩓, △\}$. The reader may want to try some other combinations as exercises.

Here is a **CL13**-proof of $P ⩓ Q \to P △ Q$:

1. $\neg p \lor p$              (∧): (no premises)
2. $(\neg p \lor\!\!\!\lor \bot) \lor p$     (⩔): 1
3. $\neg q \lor q$             (∧): (no premises)
4. $(\neg p \lor\!\!\!\lor \neg q) \lor q$    (⩔): 3
5. $(\neg p \lor\!\!\!\lor \neg Q) \lor Q$   (M): 4
6. $(\neg p \lor\!\!\!\lor \neg Q) \lor (p △ Q)$   (△⊓): 2,5
7. $(\neg P \lor\!\!\!\lor \neg Q) \lor (P △ Q)$   (M): 6

On the other hand, the formula $P △ Q \to P ⩓ Q$, i.e. $(\neg P ▽ \neg Q) \lor (P ⩓ Q)$, has no proof. This can be shown through attempting and failing to construct, bottom-up, a purported **CL15**-proof of the formula. Here we explore one of the branches of a proof-search tree. $(\neg P ▽ \neg Q) \lor (P ⩓ Q)$ is not quasielementary, so it could not be derived by (be the conclusion of) the (⩔) or (∧) rule. The (⊔) rule does not apply either, as there is no ⊔ in the formula. This leaves us with one of the rules (▽), (△⊓) and (M). Let us see what happens if our target formula is derived by (▽). In this case the premise should be ¬Q ∨ (P⋏Q). The latter can be derived only by (△⊓) or (M). Again, let us try (M). The premise in this subcase should be ¬q ∨ (P⋏q) for some elementary atom $q$. But the only way ¬q ∨ (P⋏q) can be derived is by (△⊓) from the premise ¬q ∨ (⊥⋏q). This formula, in turn, could only be derived by (⋏), in which case ¬q∨⊥ is one of the premises. Now we are obviously stuck, as ¬q∨⊥ is not the conclusion of any of the rules of the system. We thus hit a dead end. All remaining possibilities can be checked in a similar routine/analytic way, and the outcome in each case will be a dead end.

**Exercise 6.5.3** Construct a **CL15**-proof of (¬P▽P) ⋏ (P▽¬P) → (¬P⊔P).

**Exercise 6.5.4** For which of the four disjunctions ∪∈{∨,⊔,▽,∀} are the following formulas **CL13**-provable and for which are not? (1) ¬P∪P; (2) P∪Q → Q∪P; (3) P∪P → P; (4) p∪p → p.

System **CL13** was introduced and proven uniform-constructively adequate in [Jap11a].

**Open problems 6.5.4**

1. Consider the first-order version of the language of **CL15** with choice (to start with) quantifiers. Adequately axiomatize the corresponding logic.
2. What is the computational complexity of (the property of provability in) **CL15**? (Expectation: PSPACE-complete).
3. Consider the set of the theorems of **CL15** that do not contain nonlogical elementary letters. Does this set remain complete with respect to multiform (rather than uniform) validity?

## 6.6 The brute force system **CL4**

**CL4**-*formulas* are formulas of the language of CoL that do not contain any operators other than ¬, ⋏, ∨, ⊓, ⊔, ⊓, ⊔, ∀, ∃. For simplicity and safety, we additionally require that, in any formula, no quantifier binds a variable which also has a free occurrence in that formula. As in the previous cases, negation applied to non-atoms is understood as the corresponding DeMorgan abbreviation, and $E→F$ is also the standard abbreviation. An *elementary* **CL4**-*formula* is a **CL4**-formula which contains no choice operators and no general game letters. Our axiomatization of **CL4** relies on the following terminology and notation.

- A *literal* is $L(t_1,…,t_n)$ or $¬L(t_1,…,t_n)$, where $L(t_1,…,t_n)$ is an atom. Such a literal is said to be *elementary* or *general* depending on whether the letter $L$ is elementary or general.
- A *surface occurrence* of a subformula is an occurrence which is not in the scope of ¬, ⊓, ⊔, ⊓, and/or ⊔.
- The *elementarization* $||F||$ of a formula $F$ is the result of replacing in $F$ all surface occurrences of subformulas of the form $G⊔H$ or ⊔$xG$ by ⊥, all surface occurrences of subformulas of the form $G⊓H$ or ⊓$xG$ by ⊤, and all surface occurrences of general literals by ⊥.
- A formula is said to be *stable* iff its elementarization is classically valid; otherwise it is *unstable*. By "classical validity", in view of Gödel's completeness theorem, we mean provability in some fixed standard version of classical first-order calculus with constants, function letters and =, where = is the identity predicate (so that, say, $x=x$, $x=y → (E(x)↔E(y))$, etc. are provable).
- We will be using the notation $F[E_1,…,E_n]$ to mean a formula $F$ together with some $n$ fixed surface occurrences of subformulas $E_1,…,E_n$. Using this notation sets a context, in which $F[G_1,…,G_n]$ will mean the result of replacing in $F[E_1,…,E_n]$ those occurrences of $E_1,…,E_n$ by $G_1,…,G_n$, respectively. Note that here we are talking about *occurrences* of $E_1,…,E_n$. Only those occurrences get replaced when moving from $F[E_1,…,E_n]$ to $F[G_1,…,G_n]$, even if the formula also had some other occurrences of $E_1,…,E_n$.

We now define **CL4** by the following four *rules* of inference, where $\mathcal{P} \succ F$ means "from premise(s) $\mathcal{P}$ conclude $F$". Axioms are not explicitly stated, but the set of premises of the Wait rule can be empty, in which case (the conclusion of) this rule acts like an axiom.

⊔-**Choose:** $F[H_i] \succ F[H_0 \sqcup H_1]$, where $i \in \{0,1\}$.

⊔-**Choose:** $F[H(t)] \succ F[\sqcup x H(x)]$, where $t$ is either a constant or a variable with no bound occurrences in the premise.

**Wait:** $\boldsymbol{E} \succ F$ ($n \geq 0$), where $F$ is stable, and $\boldsymbol{E}$ is a set of formulas satisfying the following two conditions:
     1. Whenever $F$ has the form $F[G \sqcap H]$, $\boldsymbol{E}$ contains both $F[G]$ and $F[H]$.
     2. Whenever $F$ has the form $F[\sqcap x G(x)]$, $\boldsymbol{E}$ contains $F[G(y)]$ for some variable $y$ which does not occur in the conclusion.

**Match:** $F[p(t), \neg p(s)] \succ F[P(t), \neg P(s)]$, where $P$ is an $n$-ary ($n \geq 0$) general game letter, $t$ and $s$ are $n$-tuples of terms, and $p$ is an $n$-ary elementary game letter which does not occur in the conclusion.

The definitions of proof and provability in **CL4** are standard. The following is a **CL4**-proof of $\sqcap x \sqcup y (\neg P(x) \vee P(y))$:

1. $\neg p(x) \vee p(y)$         Wait: (no premises)
2. $\neg P(x) \vee P(y)$        Match: 1
3. $\sqcup y (\neg P(z) \vee P(y))$     ⊔-Choose: 2
4. $\sqcap x \sqcup y (\neg P(x) \vee P(y))$   Wait: 3

On the other hand, **CL4** does not prove $\sqcup y \sqcap x (\neg P(x) \vee P(y))$. Indeed, obviously this unstable formula cannot be the conclusion of any rule but ⊔-Choose. If it is derived by this rule, the premise should be $\sqcap x (\neg P(x) \vee P(t))$ for some constant or variable $t$ different from $x$. $\sqcap x (\neg P(x) \vee P(t))$, in turn, could only be derived by Wait where, for some variable $z$ different from $t$, $\neg P(z) \vee P(t)$ is a (the) premise. The latter is an unstable formula and does not contain choice operators, so the only rule by which it can be derived is Match, where the premise is $\neg p(z) \vee p(t)$ for some elementary game letter $p$. Now we deal with a classically non-valid and hence unstable elementary formula, and it cannot be derived by any of the four rules of **CL4**.

Note that, in contrast, the "blind version" $\exists y \forall x (\neg P(x) \vee P(y))$ of $\sqcup y \sqcap x (\neg P(x) \vee P(y))$ *is* provable:

1. $\exists y \forall x (\neg p(x) \vee p(y))$    Wait: (no premises)
2. $\exists y \forall x (\neg P(x) \vee P(y))$   Match: 1

'There is $y$ such that, for all $x$, $\neg P(x) \vee P(y)$' is true yet not in a constructive sense, thus belonging to the kind of principles that have been fueling controversies between the classically- and constructivistically-minded. As noted [earlier](#), CoL is offering a peaceful settlement, telling the arguing parties: "There is no need to fight at all. It appears that you simply have two different concepts of '*there is*'/'*for all*'. So, why not also use two different names: . $\exists / \forall$ and $\sqcup / \sqcap$. Yes, $\exists y \forall x (\neg P(x) \vee P(y))$ is indeed right; and yes, $\sqcup y \sqcap x (\neg P(x) \vee P(y))$ is indeed wrong." Clauses 1 and 2 of [Exercise 6.6.1](#) illustrate a similar solution for the law of the excluded middle, the most controversial principle of classical logic.

The above-said remains true with the elementary letter $p$ instead of the general letter $P$, for what is relevant there is the difference between the constructive and non-constructive versions of logical operators rather than how atoms are understood. Then how about the difference between the elementary and non-elementary versions of atoms? This distinction allows CoL to again act in its noble role of a reconciliator/integrator, but this time between classical and linear logics, telling them: "It appears that you have two different concepts of the objects that logic is meant to study. So, why not also use two different sorts of atoms to represent such objects: elementary atoms *p*,*q*,… and general atoms *P*,*Q*,.... Yes, *p*→*p*∧*p* is indeed right; and yes, *P*→*P*∧*P* (Exercise 6.6.1, clause 4) is indeed wrong". However, the term "linear logic" in this context should be understood in a very generous sense, referring not to the particular deductive system proposed by Girard in [Gir87] but rather to the general philosophy and intuitions traditionally associated with it. The formula of clause 3 of the following exercise separates **CL4** from linear logic. That formula is provable in affine logic though. Switching to affine logic, i.e. restoring the deleted (from classical logic) rule of weakening, does not however save the case: the **CL4**-provable formulas of clauses 10, 11 and 18 of the following exercise are provable in neither linear nor affine logics.

**Exercise 6.6.1** Verify the following provabilities (Yes:) or unprovabilities (No:) in **CL4**. In clauses 14 and 15 below, provability of *E*⇔*F* means provability of both *E*→*F* and *F*→*E*.

1. Yes:  *P*∨¬*P*.
2. No:  *P*⊔¬*P*. Compare with 1.
3. Yes:  *P*∧*P* → *P*.
4. No:  *P* → *P*∧*P*. Compare with 3,5.
5. Yes:  *P* → *P*⊓*P*.
6. Yes:  (*P*⊔*Q*) ∧ (*P*⊔*R*) → *P* ⊔ (*Q*∧*R*).
7. No:   *P* ⊔ (*Q*∧*R*) → (*P*⊔*Q*) ∧ (*P*⊔*R*). Compare with 6,8.
8. Yes:  *p* ⊔ (*Q*∧*R*) → (*p*⊔*Q*) ∧ (*p*⊔*R*).
9. No:   *p*⊓ (*Q*∧*R*) → (*p*⊓*Q*) ∧ (*p*⊓*R*). Compare with 8.
10. Yes: (*P*∧*P*) ∨ (*P*∧*P*) → (*P*∨*P*) ∧ (*P*∨*P*).
11. Yes: (*P* ∧ (*R*⊓*S*)) ⊓ (*Q* ∧ (*R*⊓*S*)) ⊓ ((*P*⊓*Q*) ∧ *R*) ⊓ ((*P*⊓*Q*) ∧ *S*) → (*P*⊓*Q*) ∧ (*R*⊓*S*).
12. Yes: ∀*xP*(*x*) → ⊓*xP*(*x*).
13. No:  ⊓*xP*(*x*) → ∀*xP*(*x*). Compare with 12.
14. Yes: ∃*xP*(*x*)⊓∃*xQ*(*x*) ⇔ ∃*x*(*P*(*x*)⊓*Q*(*x*)). Similarly for ⊔ instead of ⊓, and/or ∀ instead of ∃.
15. Yes: ⊓*x*∃*yP*(*x*,*y*) ⇔ ∃*y*⊓*xP*(*x*,*y*). Similarly for ⊔ instead of ⊓, and/or ∀ instead of ∃.
16. Yes: ∀*x*(*P*(*x*) ∧ *Q*(*x*)) → ∀*xP*(*x*) ∧ ∀*xQ*(*x*).
17. No:  ⊓*x*(*P*(*x*) ∧ *Q*(*x*)) → ⊓*xP*(*x*) ∧ ⊓*xQ*(*x*). Compare with 16.
18. Yes: ⊓*x*((*P*(*x*) ∧ ⊓*xQ*(*x*)) ⊓ (⊓*xP*(*x*) ∧ *Q*(*x*))) → ⊓*xP*(*x*) ∧ ⊓*xQ*(*x*).
19. Yes: ∀*x*∃*y*(*y*=*f*(*x*)).
20. No:  ⊓*x*⊔*y*(*y*=*f*(*x*)). Compare with 21.

Taking into account that classical validity and hence stability is recursively enumerable, from the way **CL4** is axiomatized it is obvious that the set of theorems of **CL4** is recursively enumerable. Furthermore, it is known from [Jap07a] that the ∀,∃-free fragment (of the set of theorems) of **CL4** is decidable in polynomial space. Later M. Bauer [Bau14] proved that this fragment is in fact PSPACE-complete. Next, it is a straightforward observation that elementary formulas are derivable in **CL4** (in particular, from the empty set of premises by Wait) exactly when they are classically valid. Hence **CL4** is a conservative extension of classical predicate logic: the latter is nothing but the elementary fragment (i.e. the set of all elementary theorems) of the former.

System **CL4** (albeit without = and function letters) was introduced and proven uniform-constructively adequate in [Jap07a].

**Open problems 6.6.2**

1. Is the propositional fragment of **CL4** PSPACE-complete?
2. Consider the set of the theorems of **CL4** that do not contain nonlogical elementary letters. Does this set remain complete with respect to multiform (rather than uniform) validity?

3. Try to axiomatize the set of multiformly valid **CL4**-formulas.

## 6.7 The brute force system **CL12**

**CL12**-*formulas* are formulas of the language of CoL that do not contain any general game letters, and do not contain any operators other than ¬, ∧, ∨, ⊓, ⊔, ⊓, ⊔, ∀, ∃. As usual, ¬ applied to non-atoms, as well as →, are understood as abbreviations of their official counterparts.

**CL12**-*sequents* are expressions of the form $E_1,\ldots,E_n \circ\!\!-\, F$ where $E_1,\ldots,E_n$ ($n{\geq}0$) and $F$ are **CL12**-formulas. Semantically, such a sequent is understood the same way as the (non-**CL12**) formula $E_1 \wedge \ldots \wedge E_n \circ\!\!-\, F$, so for instance, when we say that the former is *logically valid*, we mean that so is the latter. And a *logical solution* of the former means a logical solution of the latter.

The language of **CL12** is thus incomparable with that of **CL4**. Namely, **CL4** allows general game letters while **CL12** does not; on the other hand, **CL12** allows (uniterated) ∘− while **CL4** does not. Due to the presence of ∘−, **CL12** is a very powerful tool for constructing CoL-based applied theories. It is the logical basis for all systems of clarithmetic (CoL-based arithmetic) constructed so far. For this reason, **CL12** is also the best-studied fragment of CoL, especially from the complexity-theoretic point of view.[Jap15] Below is some terminology used in our axiomatization of **CL12**. Some terms have the same meaning as in the case of **CL4**, and the meanings of some other terms are "slightly" readjusted.

   A *surface occurrence* of a subformula is an occurrence that is not in the scope of any choice operators.
   A formula not containing choice operators --- i.e., a formula of the language of classical first order logic --- is said to be *elementary*. A sequent is *elementary* iff all of its formulas are so. The *elementarization* $\|F\|$ of a formula $F$ is the result of replacing in $F$ all surface occurrences of ⊔- and ⊔-subformulas by ⊥, and all ⊓- and ⊓-subformulas by ⊤. Note that $\|F\|$ is (indeed) an elementary formula. The *elementarization* $\|G_1,\ldots,G_n \circ\!\!-\, F\|$ of a sequent $G_1,\ldots,G_n \circ\!\!-\, F$ is the elementary formula $\|G_1\| \wedge \ldots \wedge \|G_n\| \to \|F\|$.
   A sequent is said to be *stable* iff its elementarization is classically valid (i.e. provable in some standard version of first-order calculus with constants, function letters and =); otherwise it is *unstable*.
   We will be using the notation $F[E]$ to mean a formula $F$ together with some (single) fixed surface occurrence of a subformula $E$. Using this notation sets a context, in which $F[H]$ will mean the result of replacing in $F[E]$ that occurrence of $E$ by $H$.
   Bold-face letters $G,K,\ldots$ will usually stand for finite sequences of formulas. The standard meaning of an expression such as $G,F,K$ should also be clear: this is the sequence consisting of the formulas (order respected) of $G$, followed by the formula $F$, and followed by the formulas of $K$.

We now define **CL12** by the following six *rules* of inference, where $\mathcal{P} \succ S$ means "from premise(s) $\mathcal{P}$ conclude $S$". Axioms are not explicitly stated, but the set of premises of the Wait rule can be empty, in which case (the conclusion of) this rule acts like an axiom.

   1. ⊔-*Choose:* $G \circ\!\!-\, F[H_i] \;\succ\; G \circ\!\!-\, F[H_0 \sqcup H_1]$, where $i \in \{0,1\}$.
   2. ⊓-*Choose:* $G, E[H_i], K \circ\!\!-\, F \;\succ\; G, E[H_0 \sqcap H_1], K \circ\!\!-\, F$, where $i \in \{0,1\}$.
   3. ⊔-*Choose:* $G \circ\!\!-\, F[H(t)] \;\succ\; G \circ\!\!-\, F[\sqcup x H(x)]$, where $t$ is either a constant or a variable with no bound occurrences in the premise.
   4. ⊓-*Choose:* $G, E[H(t)], K \circ\!\!-\, F \;\succ\; G, E[\sqcap x H(x)], K \circ\!\!-\, F$, where $t$ is either a constant or a variable with no bound occurrences in the premise.
   5. *Replicate:* $G, E, K, E \circ\!\!-\, F \;\succ\; G, E, K \circ\!\!-\, F$.

6. **Wait:** $G_1 \circ\!\!-\! F_1, \ldots, G_n \circ\!\!-\! F_n \succ K \circ\!\!-\! E$ ($n \geq 0$), where $K \circ\!\!-\! E$ is stable and the following four conditions are satisfied:
- Whenever the conclusion has the form $K \circ\!\!-\! E[H_0 \sqcap H_1]$, both $K \circ\!\!-\! E[H_0]$ and $K \circ\!\!-\! E[H_1]$ are among the premises.
- Whenever the conclusion has the form $L, J[H_0 \sqcup H_1], M \circ\!\!-\! E$, both $L, J[H_0], M \circ\!\!-\! E$ and $L, J[H_1], M \circ\!\!-\! E$ are among the premises.
- Whenever the conclusion has the form $K \circ\!\!-\! E[\sqcap x H(x)]$, for some variable $y$ not occurring in the conclusion, $K \circ\!\!-\! E[H(y)]$ is among the premises.
- Whenever the conclusion has the form $L, J[\sqcup x H(x)], M \circ\!\!-\! E$, for some variable $y$ not occurring in the conclusion, $L, J[H(y)], M \circ\!\!-\! E$ is among the premises.

Each rule --- seen bottom-up --- encodes an action that a winning strategy should take in a corresponding situation, and the name of each rule is suggestive of that action. For instance, Wait (indeed) prescribes the strategy to wait till the adversary moves. This explains why we have called "Replicate" the rule which otherwise is nothing but what is commonly known as Contraction.

A **CL12-proof** of a sequent $S$ is a sequence $S_1, \ldots, S_n$ of sequents, with $S_n = S$, such that each $S_i$ follows by one of the rules of **CL12** from some (possibly empty in the case of Wait, and certainly empty in the case of $i=1$) set $\mathcal{P}$ of premises such that $\mathcal{P} \subseteq \{S_1, \ldots, S_{i-1}\}$. A **CL12-proof** of a formula $F$ is understood as a **CL12**-proof of the empty-antecedent sequent $\circ\!\!-\! F$. Whether it be a sequent or a formula, "***provable*** in **CL12**", of course, means "has a **CL12**-proof".

The uniform-constructive adequacy theorem stated earlier for all systems of this section extends without changes from **CL12**-formulas to **CL12**-sequents, understanding each such sequent $E_1, \ldots, E_n \circ\!\!-\! F$ as the (non-**CL12**) formula $E_1 \wedge \ldots \wedge E_n \circ\!\!-\! F$.

Just like all other logics of this section, **CL12** is a conservative extension of classical logic. That is, an elementary sequent $E_1, \ldots, E_n \circ\!\!-\! F$ is provable in **CL12** iff the formula $E_1 \wedge \ldots \wedge E_n \to F$ is valid in the classical sense. This is an immediate corollary of the adequacy theorem for **CL12**. But this fact can also be easily verified directly. Indeed, assume $E_1, \ldots, E_n, F$ are elementary formulas. Note that then the elementarization of $E_1, \ldots, E_n \circ\!\!-\! F$ is $E_1 \wedge \ldots \wedge E_n \to F$. If the latter is classically valid, then $E_1, \ldots, E_n \circ\!\!-\! F$ follows from the empty set of premises by Wait. And if $E_1 \wedge \ldots \wedge E_n \to F$ is not classically valid, then $E_1, \ldots, E_n \circ\!\!-\! F$ cannot be the conclusion of any of the rules of **CL12** except Replicate. However, applying (bottom-up) Replicate does not take us any closer to finding a proof of the sequent, as the premise still remains an unstable elementary sequent.

**Example 6.7.1** In this example, $\times$ is a binary function letter and $^3$ is a unary function letter. We write $x \times y$ and $x^3$ instead of $\times(x,y)$ and $^3(x)$, respectively. The following sequence of sequents is a CL12-proof of its last sequent.

1. $\forall x\, (x^3 = (x \times x) \times x),\ t = s \times s,\ r = t \times s \circ\!\!-\! r = s^3$      Wait: (no premises)
2. $\forall x\, (x^3 = (x \times x) \times x),\ t = s \times s,\ r = t \times s \circ\!\!-\! \sqcup y(y = s^3)$      $\sqcup$-Choose: 1
3. $\forall x\, (x^3 = (x \times x) \times x),\ t = s \times s,\ \sqcup z(z = t \times s) \circ\!\!-\! \sqcup y(y = s^3)$      Wait: 2
4. $\forall x\, (x^3 = (x \times x) \times x),\ t = s \times s,\ \sqcap y \sqcup z(z = t \times y) \circ\!\!-\! \sqcup y(y = s^3)$      $\sqcap$-Choose: 3
5. $\forall x\, (x^3 = (x \times x) \times x),\ t = s \times s,\ \sqcap x \sqcap y \sqcup z(z = x \times y) \circ\!\!-\! \sqcup y(y = s^3)$      $\sqcap$-Choose: 4
6. $\forall x\, (x^3 = (x \times x) \times x),\ \sqcup z(z = s \times s),\ \sqcap x \sqcap y \sqcup z(z = x \times y) \circ\!\!-\! \sqcup y(y = s^3)$      Wait: 5
7. $\forall x\, (x^3 = (x \times x) \times x),\ \sqcap y \sqcup z(z = s \times y),\ \sqcap x \sqcap y \sqcup z(z = x \times y) \circ\!\!-\! \sqcup y(y = s^3)$      $\sqcap$-Choose: 6
8. $\forall x\, (x^3 = (x \times x) \times x),\ \sqcap x \sqcap y \sqcup z(z = x \times y),\ \sqcap x \sqcap y \sqcup z(z = x \times y) \circ\!\!-\! \sqcup y(y = s^3)$      $\sqcap$-Choose: 7
9. $\forall x\, (x^3 = (x \times x) \times x),\ \sqcap x \sqcap y \sqcup z(z = x \times y) \circ\!\!-\! \sqcup y(y = s^3)$      Replicate: 8
10. $\forall x\, (x^3 = (x \times x) \times x),\ \sqcap x \sqcap y \sqcup z(z = x \times y) \circ\!\!-\! \sqcap x \sqcup y(y = s^3)$      Wait: 9

**Exercise 6.7.2** To see the resource-consciousness of **CL12**, show that it does not prove $p⊓q → (p⊓q)∧(p⊓q)$, even though this formula has the form $F→F∧F$ of a classical tautology. Then show that, in contrast, **CL12** proves the *sequent* $p⊓q ∘\!\!- (p⊓q)∧(p⊓q)$ because, unlike the antecedent of a →-combination, the antecedent of a ∘\!\!- -combination is reusable (trough Replicate).

**Exercise 6.7.3** Show that **CL12** proves $⊔x⊓y\, p(x,y) ∘\!\!- ⊔x(⊓y\, p(x,y) ∧ ⊓y\, p(x,y))$. Then observe that, on the other hand, **CL12** does not prove any of the formulas

$$⊔x⊓y\, p(x,y) → ⊔x(⊓y\, p(x,y) ∧ ⊓y\, p(x,y));$$
$$⊔x⊓y\, p(x,y) ∧ ⊔x⊓y\, p(x,y) → ⊔x(⊓y\, p(x,y) ∧ ⊓y\, p(x,y));$$
$$⊔x⊓y\, p(x,y) ∧ ⊔x⊓y\, p(x,y) ∧ ⊔x⊓y\, p(x,y) → ⊔x(⊓y\, p(x,y) ∧ ⊓y\, p(x,y));$$
$$…$$

Intuitively, this contrast is due to the fact that, even though both ↓$A$ and ⋏$A = A∧A∧A$... are resources allowing to reuse $A$ any number of times, the "branching" form of reusage offered by ↓$A$ is substantially stronger than the "parallel" form of reusage offered by ⋏$A$. ↓$⊔x⊓y\, p(x,y) → ⊔x(⊓y\, p(x,y) ∧ ⊓y\, p(x,y))$ is a valid principle of CoL while ⋏$⊔x⊓y\, p(x,y) → ⊔x(⊓y\, p(x,y) ∧ ⊓y\, p(x,y))$ is not.

The uniform-constructive adequacy theorem for **CL12**, generalized from **CL12**-formulas to **CL12**-sequents, holds in the following strong form:

**Theorem 6.7.4**
    1. A **CL12**-sequent is provable in **CL12** iff it is logically valid.
    2. There is an effective procedure which takes an arbitrary **CL12**-proof of an arbitrary **CL12**-sequent $G ∘\!\!- F$ and constructs a linear amplitude, constant space and linear time logical solution of $G ∘\!\!- F$.

Thus, every cirquent either has a very efficient logical solution (as efficient as it gets), or no logical solution solution at all.

We say that a **CL12**-formula $F$ is a *logical consequence* of **CL12**-formulas $E_1,…,E_n$ ($n≥0$) iff **CL12** proves the sequent $E_1,…,E_n ∘\!\!- F$.

The following rule, which we (also) call *Logical Consequence* (*LC*), will be the only logical rule of inference in **CL12**-based applied systems:

$$E_1,…,E_n\; ➢\; F,\;\; \text{where } F \text{ is a logical consequence of } E_1,…,E_n.$$

Remember that, philosophically speaking, computational *resources* are symmetric to computational problems: what is a problem for one player to solve is a resource that the other player can use. Namely, having a problem $A$ as a computational resource intuitively means having the (perhaps externally provided) ability to successfully solve/win $A$. For instance, as a resource, $⊓x⊔y(y=x^2)$ means the ability to tell the square of any number. According to the following thesis, logical consequence lives up to its name:

**Thesis 6.7.5** Assume $E_1,…,E_n$ and $F$ are **CL12**-formulas such that there is a *-independent (whatever interpretation *) intuitive description and justification of a winning strategy for $F^*$, which relies on the availability and "recyclability" --- in the strongest sense possible --- of $E_1^*,…,E_n^*$ as computational resources. Then $F$ is a logical consequence of $E_1,…,E_n$.

**Example 6.7.6** Imagine a **CL12**-based applied formal theory, in which we have already proven two facts: ∀$x(x^3=(x×x)×x)$ (the meaning of "cube" in terms of multiplication) and $⊓x⊓y⊔z(z=x×y)$ (the computability of multiplication), and now we want to derive ⊓x⊔y(y=x³) (the computability of "cube"). This is how we can reason to justify ⊓x⊔y(y=x³):

*Consider any s (selected by Environment for x in ⊓x⊔y(y=x³)). We need to find s³. Using the resource ⊓x⊓y⊔z(z=x×y), we first find the value t of s×s, and then the value r of t×s. According to ∀x(x³=(x×x)×x), such an r is the sought s³.*

Thesis 6.7.5 promises that the above intuitive argument will be translatable into a **CL12**-proof of

$$\forall x(x^3=(x\times x)\times x), \sqcap x\sqcap y\sqcup z(z=x\times y) \circ\!\!-\, \sqcap x\sqcup y(y=x^3)$$

(and hence the succedent ⊓x⊔y(y=x³) will be derivable in the theory by LC as the formulas of the antecedent are already proven). Such a proof indeed exists --- see Example 6.7.1.

While Thesis 6.7.5 is about the *completeness* of LC, the following Theorems 6.7.7 and 6.7.8 are about *soundness*, establishing that LC preserves computability, and does so in a certain very strong sense. It is these two theorems to which **CL12** (LC, that is) owes its appeal as a logical basis for complexity-oriented applied theories. Theorem 6.7.7 carries good news for the cases where time efficiency is of main concern, and otherwise we are willing to settle for at least linear space. Theorem 6.7.8 does the same but for the cases where the primary concern is space efficiency --- namely, when we want to keep the latter sublinear. These theorems use the term "**native magnitude**", by which, for a sequent S, we mean the greatest of all constants (here seen as the corresponding natural numbers) that appear in S, or 0 is there are no constants in S.

**Theorem 6.7.7** There is an effective procedure that takes an arbitrary **CL12**-proof $\mathcal{P}$ of an arbitrary **CL12**-sequent $E_1,\ldots,E_n \circ\!\!-\, F$, arbitrary HPMs $N_1,\ldots,N_n$ and constructs an HPM $M$ such that the following holds:

   Assume * is an interpretation and $a,s,t$ are unary arithmetical functions such that:

- For each $i \in \{1,\ldots,n\}$, $N_i$ is an $a$ amplitude, $s$ space and $t$ time solution of $E_i$*.
- For any $x$, $a(x) \geq \max(x,c)$, where $c$ is the native magnitude of $E_1,\ldots,E_n \circ\!\!-\, F$.
- For any $x$, $s(x) \geq x$.
- For any $x$, $t(x) \geq x$.

Then thre is a number $b$ which only depends on $\mathcal{P}$ such that, with $R$ abbreviating $a^b(y)$, $M$ is a solution of $F$* which runs in amplitude $R$, space $O(s(R))$ and time $O(t(R))$.

**Theorem 6.7.8** There is an effective procedure that takes an arbitrary **CL12**-proof $\mathcal{P}$ of an arbitrary **CL12**-sequent $E_1,\ldots,E_n \circ\!\!-\, F$, arbitrary HPMs $N_1,\ldots,N_n$ and constructs an HPM $M$ such that the following holds:
   Assume * is an interpretation and $a,s,t$ are unary arithmetical functions such that:

- For each $i \in \{1,\ldots,n\}$, $N_i$ is an $a$ amplitude, $s$ space and $t$ time solution of $E_i$*.
- For any $x$, $a(x) \geq \max(x,c)$, where $c$ is the native magnitude of $E_1,\ldots,E_n \circ\!\!-\, F$.
- For any $x$, $s(x) \geq \log(x)$.
- For any $x$, $t(x) \geq x$ and $t(x) \geq s(x)$.

Then there is are numbers $b,d$ which only depend on $\mathcal{P}$ such that, with $R$ abbreviating $a^b(y)$, $M$ is a solution of $F$* which runs in amplitude $R$, space $O(s(R))$ and time $O\big((t(R))^d\big)$.

**CL12** was introduced in [Jap10], and proven to be uniform-constructively adequate in [Jap12c]. Theorems 6.7.7 and 6.7.8 were proven in [Jap15].

**Open problems 6.7.9**

1. Add the sequential connectives △, ▽ to the language of **CL12** and adequately axiomatize the corresponding logic.
2. Try to axiomatize the set of  multiformly valid **CL12**-sequents.
3. Along with elementary game letters, allow also general game letter in the language of **CL12**, and try to axiomatize the set of valid sequents of this extended language.

# 7 Clarithmetic (CoL-based arithmetic)

## 7.1 Introduction

As we agreed earlier, the ultimate purpose of logic is providing a tool for navigating the real life. As such, first or foremost it should be able to serve as a basis for *applied* ("*substantial*") *theories*, one of the best known examples of which is Peano arithmetic **PA**. Pure logics, such as classical logic or **CL12**, are "about everything", with no specific interpretation applied to their otherwise meaningless nonlogical atoms. For this reason, a formula of a pure logic can be valid or invalid, but not true or false. In contrast, an applied theory always comes with a particular interpretation, which makes all its atoms meaningful, and all sentences "true" or "false" (in our case meaning computable or incomputable). To make this point more clear to a computer scientist, pure logics can be compared with programming languages, and applied theories based on them with application programs written in those languages. A programming language created for its own sake, mathematically or esthetically appealing but otherwise unusable as a general-purpose, comprehensive and convenient tool for writing application programs, will hardly be of much interest.

In this respect, CoL is a reasonable alternative to classical logic. One can base applied theories on CoL just as meaningfully as on classical logic (and certainly more meaningfully than on intuitionistic or linear logics which offer no concept of truth), but with significant advantages, especially if we care about constructive, computational and complexity-theoretic aspects of the theory. Number theories based on CoL are called "*Clarithmetics*".

The nonlogical axioms of a clarithmetic generally will be a collection of (formulas expressing) problems whose algorithmic solutions are known. Often, together with nonlogical axioms, we also have nonlogical rules of inference, preserving the property of computability. Then, the soundness of the corresponding underlying axiomatization of CoL --- which usually comes in the strong, constructive sense --- guarantees that every theorem *T* of the theory also has an algorithmic solution and that, furthermore, such a solution can be mechanically extracted from a proof of *T*. Does not this look like exactly what the constructivists have been calling for?

Unlike the mathematical or philosophical constructivism, however, and even unlike the early-day theory of computation, modern computer science has long understood that what really matters is not just computability but rather efficient computability. And here comes the good news: CoL, without any modifications whatsoever, can be used for studying efficient or complexity-conscious computability just as successfully as for studying computability-in-principle. This is no surprise. Let us take the fragment **CL12** of CoL for specificity. As we saw from Theorem 6.7.4, **CL12** is a sound and complete logic of not only (1) computability-in-principle, but also of (2) linear amplitude, constant space and linear time computability. This means it is also a logic of any sort of computability between these two extremes. If different sorts of computability required different versions of CoL, then CoL would not be universal enough, and thus would not quite be a Logic with a capital "L". It would be no fun to develop a separate "logic" for each of the so many interesting classes from the complexity zoo.

But if CoL does not differentiate between various sorts of computational complexity, then how could it be useful for studying them? The whole point is that the differences between clarithmetical theories tailored to various complexity classes should be and is achieved due to differences in the non-logical postulates of those theories rather than the underlying logic. Roughly speaking, if the nonlogical axioms of a clarithmetic represent problems computable within a given complexity constraints and if the nonlogical inference rules preserve this property, then CoL as a pure logic is gentle enough to guarantee that all theorems of the theory also enjoy the same property. Of course, any very weak logic --- take the empty logic as an extreme --- would also have this "virtue". But CoL is as strong as a logic could be and, as we are going to see later, it allows us to achieve not only soundness but also completeness of clarithmetics based on it. Here we should differentiate between two --- *extensional* and *intensional* --- sorts of completeness for clarithmetical theories. Extensional completeness with respect to a given complexity class $C$ means that every *problem* with a $C$-complexity solution is represented (expressed) by some theorem of the theory, while intensional completeness means that every *sentence* expressing a problem with a $C$-complexity solution is a theorem of the theory. Note that intensional completeness implies extensional completeness but not vice versa, because a problem may be expressed via different sentences, some of which might be provable and some not. Gödel's celebrated theorem is about intensional rather than extensional incompleteness. In fact, in the context of **PA** or other classical-logic-based theories, extensional completeness is not interesting at all. It is trivially achieved there because the provable sentence 0=0 represents all true sentences. In clarithmetics usually both sorts of completeness can be achieved, even though, due to the Gödel incompleteness phenomenon, intensional completeness can only be achieved at the expense of sacrificing recursive enumerability (but not simplicity or elegance).

By now eleven clarithmetical theories, named **CLA1** through **CLA11**, have been introduced and studied [Jap10, Jap11c, Jap14, Jap16a, Jap16b, Jap16c]. These theories are notably simple: most of them happen to be conservative extensions of **PA** whose only non-classical axiom is the sentence ⊓x⊔y(y = x')  asserting computability of the successor function ', and whose only non-logical rule of inference is "constructive induction", the particular form of which varies from system to system. The diversity of such theories is typically related to different complexity conditions imposed on the underlying concept of interactive computability. For instance, **CLA4** soundly and completely captures the set of polynomial time solvable interactive number-theoretic problems,[Jap11c] **CLA5** does the same for polynomial space,[Jap16a] **CLA6** for elementary recursive time (=space),[Jap16a] and **CLA7** for primitive recursive time (=space).[Jap16a] Most interesting is the system **CLA11**, which, due to having four parameters, is in fact a scheme of clarithmetical theories rather than a particular theory. Tuning its parameters in an essentially mechanical way, **CLA11** yields sound and complete systems for an infinite and diverse class of computational complexities, including those to which the earlier disparate systems of clarithmetic were tailored. Parameters #1, #2 and #3 are sets of terms or pseudoterms used as bounds in certain postulates, and parameter #4 is a (typically empty yet "expandable") set of formulas used as supplementary axioms. The latter determines the intensional strength of the theory, while parameters #1, #2 and #3 respectively govern the amplitude, space and time complexities of the problems represented by the theorems of the theory.

The logical basis of all clarithmetical theories studied so far is **CL12**.  More expressive fragments of CoL can be considered in the future for this purpose, if the syntax of such fragments is sufficiently developed.

## 7.2 Clarithmetic vs. bounded arithmetic

It has been long noticed that many complexity classes can be characterized by certain versions of arithmetic. Of those, systems of **bounded arithmetic**[Haj93, Kra95] should be named as the closest predecessors of our systems of clarithmetic. Just like our clarithmetical systems, they control computational complexity by explicit resource bounds attached to quantifiers, usually in induction or similar postulates. The best known alternative line of research, primarily developed by recursion theorists, controls computational complexity via type information instead. Here we will not attempt any comparison with these alternative approaches because of big differences in the defining frameworks.

The story of bounded arithmetic starts with Parikh's 1971 work [Par71], where the first system $\mathbf{I\Delta_0}$ of bounded arithmetic was introduced. Paris and Wilkie, in [Par85] and a series of other papers, advanced the study of $\mathbf{I\Delta_0}$ and of how it relates to complexity theory. Interest towards the area dramatically intensified after the appearance of Buss' 1986 influential work [Bus86], where systems of bounded arithmetic for polynomial hierarchy, polynomial space and exponential time were introduced. Clote and Takeuti [Clo92], Cook and Nguyen [Coo10] and others introduced a host of theories related to other complexity classes.

All theories of bounded arithmetic are weak subtheories of **PA**, typically obtained by imposing certain syntactic restrictions on the induction axiom or its equivalent, and then adding some old theorems of **PA** as new axioms to partially bring back the baby thrown out with the bath water. Since the weakening of the deductive strength of **PA** makes certain important functions or predicates no longer definable, the non-logical vocabularies of these theories typically have to go beyond the original vocabulary {0, ',+,×} of **PA**. These theories achieve soundness and extensional completeness with respect to the corresponding complexity classes in the sense that a function $f(x)$ belongs to the target class if and only if it is provably total in the system --- that is, if there is a $\Sigma_1$-formula $F(x,y)$ that represents the graph of $f(x)$, such that the system proves $\forall x\exists!yF(x,y)$.

Here we want to point out several differences between clarithmetic and bounded arithmetic.

**7.2.1. Generality** While bounded arithmetic and the other approaches are about functions, clarithmetics are about interactive problems, with functions being nothing but special cases of the latter. This way, clarithmetics allow us to systematically study not only computability in its narrow sense, but also many other meaningful properties and relations, such as, for instance, various sorts of enumerabilities, approximabilities or reducibilities. As we remember, just like function effectiveness, such relations happen to be special cases of our broad concept of computability. Having said that, the differences discussed in the subsequent paragraphs of this subsection hold regardless of whether one keeps in mind the full generality of clarithmetics or restricts attention back to functions only, the "common denominators" of the two approaches.

**7.2.2. Intensional strength** Our systems extend rather than restrict **PA**. Furthermore, instead of **PA**, as a classical basis one can take anything from a very wide range of sound theories, beginning from certain weak fragments of **PA** and ending with the absolute-strength theory **Th(N)** of the standard model of arithmetic (the "truth arithmetic"). It is exactly due to this flexibility that we can achieve not only extensional but also intensional completeness --- something inherently unachievable within the traditional framework of bounded arithmetic, where computational soundness by its very definition entails deductive weakness.

**7.2.3. Language** Due to the fact that our theories are no longer weak, there is no need to have any new *non-logical* primitives in the language and the associated new axioms in the theory: all recursive or arithmetical relations and functions can be expressed through 0, ',+,× in the standard way. Instead, the languages of the theories **CLA1-CLA11** of clarithmetic only have the two additional *logical* connectives ⊓,⊔ and two additional quantifiers ⊓,⊔. It is CoL's constructive semantics for these operators that allows us to express nontrivial computational problems. Otherwise, formulas not containing these operators --- formulas of the traditional language of **PA**, that is --- only express elementary problems (moveless games). This explains how our approach makes it possible to reconcile unlimited deductive strength with computational soundness. For instance, the formula $\forall x\exists yF(x,y)$ may be provable even if $F(x,y)$ is the graph of a function which is "too hard" to compute. This does not have any relevance to the complexity class characterized by the theory because the formula $\forall x\exists yF(x,y)$, unlike its "constructive counterpart" ⊓x⊔yF(x,y), carries no nontrivial computational meaning.

### 7.2.4. Quantifier alternation
Our approach admits arbitrarily many alternations of bounded quantifiers in induction or whatever similar postulates, whereas the traditional bounded arithmetics are typically very sensitive in this respect, with different quantifier complexities yielding different computational complexity classes.

### 7.2.5. Uniformity
As already pointed out, our system **CLA11** offers uniform treatments of otherwise disparate systems for various complexity classes. It should be noted that the same holds for the bounded-arithmetic approach of [Coo10], albeit for a rather different spectrum of complexity classes. The ways uniformity is achieved, however, are drastically different. In the case of [Coo10], the way to build your own system is to add, to the base theory, an axiom expressing a complete problem of the target complexity class. Doing so thus requires quite some nontrivial complexity-theoretic knowledge. In our case, on the other hand, adequacy is achieved by straightforward, brute force tuning of the corresponding parameters of **CLA11**. E.g., for linear space, we simply need to take parameter #2 to be the set of $(0, ',+)$-combinations of variables, i.e., the set of terms that "canonically" express the linear functions. If we (simultaneously) want to achieve adequacy with respect to polynomial time, we shall (simultaneously) take parameter #3 to be the set of $(0, ',+,\times)$-combinations of variables, i.e., the set of terms that express the polynomial functions. And so on.

## 7.3 Motivations

The main motivations for studying clarithmetics are as follows, somewhat arbitrarily divided into the categories "general", "theoretical" and "practical".

### 7.3.1. General
Increasingly loud voices are being heard[Gol06] that, since the real computers are interactive, it might be time in theoretical computer science to seriously consider switching from Church's narrow understanding of computational problems as functions to more general, interactive understandings. Clarithmetics serve the worthy job of lifting "efficient arithmetics" to the interactive level. Of course, the already existing results are only CoL's first modest steps in this direction, and there is still a long way to go. In any case, our generalization from functions to interaction appears to be beneficial even if, eventually, one is only interested in functions, because it allows a smoother treatment and makes our systems easy-to-understand in their own rights. Imagine how awkward it would be if one had tried to restrict the language of classical logic only to formulas with at most one alternation of quantifiers because more complex formulas seldom express things that we comprehend or care about, and, besides, things can always be Skolemized anyway. Or, if mankind had let the Roman-European tradition prevail in its reluctance to go beyond positive integers and accept 0 as a legitimate quantity, to say nothing about the negative, fractional, or irrational numbers.

The "smoothness" of our approach is related to the fact that, in it, all formulas --- rather than only those of the form $\forall x \exists ! y F(x,y)$ with $F \in \Sigma_1$ --- have clearly defined meanings as computational problems. This allows us to apply certain systematic and scalable methods of analysis that otherwise would be inadequate. For instance, the soundness proofs for various clarithmetical theories go semantically by induction on the lengths of proofs, by showing that all axioms have given complexity solutions, and that all rules of inference preserve the property of having such solutions. Doing the same is impossible in the traditional approaches to bounded arithmetic (at least those based on classical logic), because not all intermediate steps in proofs will have the form $\forall x \exists ! y F(x,y)$ with $F \in \Sigma_1$. It is no accident that, to prove computational soundness, such approaches usually have to appeal to syntactic arguments that are around "by good luck", such as cut elimination.

As mentioned, our approach extends rather than restricts **PA**. This allows us to safely continue relying on our standard arithmetical intuitions when reasoning within clarithmetic, without our hands being tied by various constraints, without the caution necessary when reasoning within weak theories. Generally, a feel for a formal theory and a "sixth sense" that it takes for someone to comfortably reason within the theory require time and efforts to develop. Many of us have such a "sixth sense" for **PA** but not so many have it for weaker theories. This is so because weak theories, being artificially restricted and thus forcing us to pretend that we do not know certain things that we actually do know, are farther from a mathematician's normal intuitions than **PA** is. Even if this was not the case, mastering the one and universal theory **PA** is still easier and promises a greater payoff than trying to master tens of disparate yet equally important weak theories that are out there.

**7.3.2. Theoretical** Among the main motivations for studying bounded arithmetics has been a hope that they can take us closer to solving some of the great open problems in complexity theory, for, quoting Cook and Nguen [Coo10], "it ought to be easier to separate the theories corresponding to the complexity classes than to separate the classes themselves". The same applies to our systems of clarithmetic and **CLA11** in particular which allows us to capture, in a uniform way, a very wide and diverse range of complexity classes.

While the bounded arithmetic approach has been around and extensively studied since long ago, the progress towards realizing the above hope has been very slow. This fact alone justifies all reasonable attempts to try something substantially new and so far not well explored. The clarithmetics line of research qualifies as such. Specifically, studying "nonstandard models" of clarithmetics, whatever they may mean, could be worth the effort.

Among the factors which might be making clarithmetics more promising in this respect than their traditional alternatives is that the former achieves intensional completeness while the latter inherently have to settle for merely extensional completeness. Separating theories intensionally is generally easier than separating them extensionally, yet intensional completeness implies that the two sorts of separation mean the same.

Another factor relates to the ways in which theories are axiomatized in uniform treatments, namely, the approach of **CLA11** versus that of [Coo10]. As noted earlier, the uniform method of [Coo10] achieves extensional completeness with respect to a given complexity class by adding to the theory an axiom expressing a complete problem of that class. The same applies to the method used in [Clo92]. Such axioms are typically long formulas as they carry nontrivial complexity-theoretic information. They talk --- through encoding and arithmetization --- about graphs, computations, etc. rather
than about numbers. This makes such axioms hard to comprehend directly as number-theoretic statements, and makes the corresponding theories hard to analyze. This approach essentially means translating our complexity-theoretic knowledge into arithmetic. For this reason, it is likely to encounter the same kinds of challenges as the ordinary, informal theory of computation does when it comes to separating complexity classes. Also, oftentimes we may simply fail to know a complete problem of a given, not very well studied, complexity class.

The uniform way in which **CLA11** axiomatizes its instances is very different from the above. Here all axioms and rules are "purely arithmetical", carrying no direct complexity-theoretic information. This means that the number-theoretic contents of such theories are easy to comprehend, which, in turn, carries a promise that their model theories might be easier to successfully study, develop and use in proving independence/separation results.

**7.3.3. Practical** More often than not, the developers of complexity-bound arithmetics have also been motivated by the potential of practical applications in computer science. Here we quote Schwichtenberg's [Sch06] words: 'It is well known that it is undecidable in general whether a given program meets its specification. In contrast, it can be checked easily by a machine whether a formal proof is correct, and from a constructive proof one can automatically extract a corresponding program, which by its very construction is correct as well. This at least in principle opens a way to produce correct software, e.g. for safety-critical applications. Moreover, programs obtained from proofs are "commented" in a rather extreme sense. Therefore it is easy to apply and maintain them, and also to adapt them to particular situations.' In a more ambitious and, at this point, somewhat fantastic perspective, after developing reasonable theorem-provers, CoL-based efficiency-oriented systems can be seen as declarative programming languages in an extreme sense, where human "programming" just means writing a formula expressing the problem whose efficient solution is sought for systematic usage in the future. That is, a program simply coincides with its specification. The compiler's job would be finding a proof (the hard part) and translating it into a machine-language code (the easy part). The process of compiling could thus take long but, once compiled, the program would run fast ever after.

What matters in applications like the above, of course, is the intensional rather than extensional strength of a theory. The greater that strength, the better the chances that a proof/program will be found for a declarative, ad hoc or brute force specification of the goal. Attempts to put an intensionally weak theory (regardless of its extensional strength) to practical use would usually necessitate some pre-processing of the goal, such as expressing it through a certain standard-form $\Sigma_1$-formula. But this sort of pre-processing often essentially means already finding --- outside the formal system --- a solution of the target problem or, at least, already finding certain insights into such a solution.

In this respect, **CLA11** is exactly what we need. Firstly, because it is easily, "mechanically" adjustable to a potentially infinite variety of target complexities that one may come across in real life. It allows us to adequately capture a complexity class from that variety without any preliminary complexity-theoretic knowledge about the class, such as knowledge of some complete problem of the class (yet another sort of "pre-processing") as required by the approaches in the style of [Clo92] or [Coo10]. All relevant knowledge about the class is automatically extracted by the system from the definition (ad hoc description) of the class, without any need to look for help outside the formal theory itself. Secondly, and more importantly, **CLA11** fits the bill because of its intensional strength, which includes the full deductive power of **PA** and which is only limited by the Gödel incompleteness phenomenon. Even when the theory possesses no arithmetical knowledge on top of what is implied by the Peano axioms, is still provides "practically full" information about computability within the corresponding complexity constraints. This is in the same sense as **PA**, despite Gödel's incompleteness, provides "practically full" information about arithmetical truth. Namely, if a formula $F$ is not provable in the theory, it is unlikely that anyone would find an algorithm solving the problem expressed by $F$ within the required complexity restrictions: either such an algorithm does not exist, or showing its correctness requires going beyond ordinary combinatorial reasoning formalizable in **PA**.

## 7.4 Common preliminaries for all our theories of clarithmetic

All systems of clarithmetic presented on this website have the same ***language***, obtained from the language of CL12 by removing all nonlogical predicate letters, removing all constants but 0, and removing all but the following three function letters:

- *successor*, unary. We write $t'$ for *successor*$(t)$.
- *sum*, binary. We write $t_1+t_2$ for *sum*$(t_1,t_2)$.
- *product*, binary. We write $t_1 \times t_2$ for *product*$(t_1,t_2)$.

Let us call this language **L**. Throughout the rest of Section 7, unless otherwise specified or implied by the context, when we say "***formula***", it is to be understood as formula of **L**. As always, ***sentences*** are formulas with no free occurrences of variables. An **L**-***sequent*** is a sequent all of whose formulas are sentences of **L**. For a formula $F$, $\forall F$ means the $\forall$-closure of $F$, i.e., $\forall x_1 \ldots \forall x_n F$, where $x_1, \ldots, x_n$ are the free variables of $F$ listed in their lexicographic order. Similarly for $\exists F, \sqcap F, \sqcup F$.

A formula is said to be ***elementary*** iff it contains no choice operators. We will be using the lowercase $p, q, \ldots$ as metavariables for elementary formulas. This is as opposed to the uppercase letters $E, F, G, \ldots$, which will be used as metavariables for any (elementary or nonelementary) formulas.

As one can see, **L** is an extension of the language of **PA** --- namely, the extension obtained by adding the choice operators $\sqcap, \sqcup, \sqcap, \sqcup$. The language of **PA** is the *elementary fragment* of **L**, in the sense that formulas of the former are nothing but elementary formulas of the latter. We remind the reader that, deductively, **PA** is the theory based on classical first-order logic with the following nonlogical axioms, that we shall refer to as the ***Peano axioms***:

1. $\forall x\, (0 \neq x')$
2. $\forall x \forall y\, (x' = y' \rightarrow x = y)$
3. $\forall x\, (x + 0 = x)$
4. $\forall x \forall y\, (x + y' = (x + y)')$
5. $\forall x\, (x \times 0 = 0)$
6. $\forall x \forall y\, (x \times y' = (x \times y) + x)$
7. $\forall \big(p(0) \wedge \forall x\, (p(x) \rightarrow p(x')) \rightarrow \forall x\, p(x)\big)$ for each elementary formula $p(x)$

The concept of an interpretation explained in Section 5.2 can now be restricted to interpretations that are only defined (other than the word "*Universe*") on $'$, $+$ and $\times$, as the present language **L** has no other nonlogical function or predicate letters. Of such interpretations, the ***standard interpretation*** † is the one whose domain **Dm**† is the set of natural numbers, whose denotation function **Dn**† is the function which sends each constant to the number represented by it in the decimal notation, and which interprets the letter $'$ as the standard successor function $var_1 + 1$, interprets $+$ as the sum function $var_1 + var_2$, and interprets $\times$ as the product function $var_1 \times var_2$.

We often terminologically identify a formula $F$ with the game $F^\dagger$, and typically write $F$ instead of $F^\dagger$ unless doing so may cause ambiguity. Correspondingly, whenever we say that an elementary sentence is ***true***, it is to be understood as that the sentence is true under the standard interpretation, i.e., is true in what is more commonly called the ***standard model of arithmetic***.

Terminologically we will further identify natural numbers with the corresponding decimal numerals (constants). Usually it will be clear from the context whether we are talking about a number or a binary numeral. For instance, if we say that $x$ is greater than $y$, then we obviously mean $x$ and $y$ as numbers; on the other hand, if we say that $x$ is longer than $y$, then $x$ and $y$ are seen as numerals. Thus, 999 is greater but not longer than 100.

All theories of clarithmetic presented in the following subsections share not only the language **L** but also logical postulates. Namely, the only **logical rule of inference** of each theory is our old friend LC (Logical Consequence). There are no logical axioms, but the instances of LC that take no premises can be considered as such. All theories also share the *Peano axioms*. Having said this, in our following descriptions of those theories we will only present their (nonlogical) ***extra-Peano axioms*** and ***nonlogical rules***.

A sentence $F$ is considered provable in such a theory iff there is a sequence of sentences, called a ***proof*** of $F$, where each sentence is either a (Peano or extra-Peano) axiom, or follows from some previous sentences by one of the (logical or nonlogical) rules of inference, and where the last sentence is $F$. An ***extended proof*** is defined in the same way, only, with the additional requirement that each application of LC should come together with an attached **CL12**-proof of the corresponding sequent. With some fixed, effective, sound and complete axiomatization **A** of classical first order logic in mind, a ***superextended proof*** is an extended proof with the additional requirement that every application of Wait in the justification of a **CL12**-derivation in it comes with an **A**-proof of the elementarization of the conclusion. The property of being a superextended proof is (efficiently) decidable, while the properties of being an extended proof or just a proof are only recursively enumerable.

A weaker property than provability is what we call "prepresentability" ("pr" for "provably"). We say that a sentence $F$ is ***prepresentable*** in a given theory **T** of clarithmetic iff there is a **T**-provable sentence $X$ with $X^\dagger = F^\dagger$, where, as we remember, † is the standard interpretation. The earlier discussed intensional completeness theorems establish the provability of all "good" sentences, while the extensional completeness theorems merely establish the prepresentability of such sentences.

Generally, as in the above definition of provability and proofs, in all clarithmetical theories we will only be interested in proving *sentences*. So, for technical convenience, we agree that, if a formula $F$ is not a sentence but we say that it is provable in such a theory, it simply means that ⊓F is provable. Correspondingly, for readability, when formulating an inference rule

$$\frac{E_1 \quad \cdots \quad E_n}{F}$$

of the theory where $E_1,\ldots,E_n,F$ are not required to be sentences, it is always to be understood as a lazy way to write

$$\frac{\sqcap E_1 \quad \cdots \quad \sqcap E_n}{\sqcap F}$$

Keep this convention in mind when reading the subsequent subsectios of this section.

An *n*-ary ($n \geq 0$) pseudoterm, or **pterm** for short, is an elementary formula $p(y,x_1,\ldots,x_n)$ with all free variables as shown and one of such variables --- $y$ in the present case --- designated as what we call the *value variable* of the pterm, such that **PA** proves $\forall x_1\ldots\forall x_n \exists! y\, p(y,x_1,\ldots,x_n)$. Here, as always, $\exists! y$ means "there is a unique $y$ such that". If $p(y, x_1,\ldots,x_n)$ is a pterm, we shall usually refer to it as **p**$(x_1,\ldots,x_n)$ or just **p**, making $p$ boldface and dropping the value variable $y$ or dropping all variables. Correspondingly, where $F(y)$ is a formula, we write $F(\mathbf{p}(x_1,\ldots,x_n))$ or just $F(\mathbf{p})$ to denote the formula $\exists y(p(y, x_1,\ldots,x_n) \wedge F(y))$, which, in turn, is equivalent to $\forall y(p(y, x_1,\ldots,x_n) \rightarrow F(y))$. These sort of expressions, allowing us to syntactically treat pretms as if they were genuine terms of the language, are unambiguous in that all "disabbreviations" of them are provably equivalent in the system. Terminologically, genuine terms of **L**, such as $(x+y)\times z$, will also count as pterms. Every *n*-ary pterm $\mathbf{p}(x_1,\ldots,x_n)$ represents --- in the obvious sense --- some **PA**-provably total *n*-ary function $f(x_1,\ldots,x_n)$. For further notational and terminological convenience, in many contexts we shall identify pterms with the functions that they represent.

In our metalanguage, $|x|$ will refer to the length of the binary representation of $x$. As in the case of other standard functions, the expression $|x|$ will be simultaneously understood as a pterm naturally representing the function $|x|$. The delimiters "$|\ldots|$" will automatically also be treated as parentheses, so, for instance, when $f$ is a unary function or pterm, we usually write "$f|x|$" to mean the same as the more awkward expression "$f(|x|)$". Further generalizing this notational convention, if **x** stands for an *n*-tuple $(x_1,\ldots,x_n)$ ($n \geq 0$) and we write **p**|**x**|, it is to be understood as $\mathbf{p}(|x_1|,\ldots,|x_n|)$.

# 7.5 Clarithmetics for polynomial time, polynomial space, elementary and primitive recursive computability

For a variable $x$, by a **polynomial sizebound** for $x$ we shall mean a standard formula of the language of **PA** saying that $|x| \leq t(|y_1|,\ldots,|y_n|)$, where $y_1,\ldots,y_n$ are any variables different from $x$, and $t(|y_1|,\ldots,|y_n|)$ is any $(0,',+,\times)$-combination of the pterms $|y_1|,\ldots,|y_n|$. An **exponential sizebound** is defined the same way, only with "$|x| \leq t(y_1,\ldots,y_n)$" instead of "$|x| \leq t(|y_1|,\ldots,|y_n|)$". So, for instance, $|x| \leq |y|+|z|$ is a polynomial sizebound for $x$ while $|x| \leq y+z$ is an exponential sizebound.

We say that a formula $F$ is **polynomially bounded** iff the following condition is satisfied: Whenever $\sqcap xG(x)$ [resp. $\sqcup xG(x)$] is a subformula of $F$, $G(x)$ has the form $S(x) \rightarrow H(x)$ [resp. $S(x) \wedge H(x)$], where $S(x)$ is a polynomial sizebound for $x$ none of whose free variables is bound by $\forall$ or $\exists$ within $F$. "**Exponentially bounded**" is defined the same way, with the only difference that $S(x)$ is required to be an exponential rather than a polynomial sizebound.

Where *t* is a term, we will be using *t*0 and *t*1 as abbreviations for the terms 0″×*t* and (0″×*t*)′, respectively.

**Theory CLA4** only has two extra-Peano axioms: ⊓x⊔y(y=x′), ⊓x⊔y(y=x0) and a single nonlogical rule (of *Induction*), where *F(x)* is any polynomially bounded formula:

$$\frac{F(0) \qquad F(x) \to F(x0) \qquad F(x) \to F(x1)}{F(x)}$$

**Theory CLA5** has a single extra-Peano axiom: ⊓x⊔y(y=x′) and a single nonlogical rule (of Induction), where *F(x)* is any polynomially bounded formula:

$$\frac{F(0) \qquad F(x) \to F(x')}{F(x)}$$

**Theory CLA6** only differs from **CLA5** in that *F(x)* in Induction is required to be an exponentially bounded formula.

**Theory CLA7** only differs from **CLA5** and **CLA6** in that there are no restrictions at all on *F(x)* in Induction.

**Fact 7.5.1** Every (elementary **L** -) sentence provable in **PA** is also provable in any of our clarithemtical theories, including the ones defined in subsequent subsections.

This fact allows us to construct "lazy" proofs where some steps can be justified by simply indicating their provability in **PA**. That is, we can treat theorems of **PA** as if they were axioms of our clarithmetical theory.

**Example 7.5.2** The following sequence is a lazy proof of ⊓x(x=0 ⊔ x≠0) in **CLA4**. This sentence says that the "zeroness" predicate is decidable:

I. 0=0 ⊔ 0≠0                  LC: (no premises)
II. ∀x (x=0 → x0=0)            **PA**
III. ∀x (x≠0 → x0≠0)            **PA**
IV. ⊓x (x=0 ⊔ x≠0 → x0=0 ⊔ x0≠0)    LC: II,III
V. ∀x (x1≠0)                     **PA**
VI. ⊓x (x=0 ⊔ x≠0 → x0=0 ⊔ x0≠0)    LC: V
VII. ⊓x (x=0 ⊔ x≠0)              Induction: I,IV,VI

An extended version of the above proof will include the following three additional justifications (CL12-proofs):

      A justification for Step I:
1. ∘− 0=0            Wait: (no premises)
2. ∘− 0=0 ⊔ 0≠0     ⊔-Choose: 1

      A justification for Step IV:
1. ∀x (x=0 → x0=0), ∀x(x≠0 → x0≠0) ∘− s=0 → s0=0           Wait: (no premises)
2. ∀x (x=0 → x0=0), ∀x(x≠0 → x0≠0) ∘− s=0 → s0=0 ⊔ s0≠0    ⊔-Choose: 1
3. ∀x (x=0 → x0=0), ∀x(x≠0 → x0≠0) ∘− s≠0 → s0≠0           Wait: (no premises)
4. ∀x (x=0 → x0=0), ∀x(x≠0 → x0≠0) ∘− s≠0 → s0=0 ⊔ s0≠0    ⊔-Choose: 3
5. ∀x (x=0→ x0=0), ∀x(x≠0 → x0≠0) ∘− s=0 ⊔ s≠0 → s0= 0 ⊔ s0≠0    Wait: 2,4
6. ∀x (x=0 → x0= 0), ∀x(x≠0 → x0≠0) ∘− ⊓x(x=0 ⊔ s≠0 → x0=0 ⊔ x0≠0)    Wait: 5

      A justification for Step VI:
1. ∀x (x1≠0) ∘− s=0 → s1≠0              Wait: (no premises)

2. $\forall x\,(x\mathbf{1}{\neq}0)$  ∘–  $s{\neq}0 \to s\mathbf{1}{\neq}0$     Wait: (no premises)
3. $\forall x\,(x\mathbf{1}{\neq}0)$  ∘–  $s{=}0 \sqcup s{\neq}0 \to s\mathbf{1}{\neq}0$     Wait: 1,2
4. $\forall x\,(x\mathbf{1}{\neq}0)$  ∘–  $s{=}0 \sqcup s{\neq}0 \to s\mathbf{1}{=}0 \sqcup s\mathbf{1}{\neq}0$     ⊔-Choose: 3
5. $\forall x\,(x\mathbf{1}{\neq}0)$  ∘–  $\sqcap x(x{=}0 \sqcup x{\neq}0 \to x\mathbf{1}{=}0 \sqcup x\mathbf{1}{\neq}0)$     Wait: 4

As we just saw, (additionally) justifying an application of LC takes more space than the (non-extended) proof itself. And this is a typical case for clarithmetical proofs. Luckily, however, there is no real need to formally justify LC. Firstly, this is so because **CL12** is an analytic system, and proof-search in it is a routine (even if sometimes long) syntactic exercise. Secondly, thanks to Thesis 6.7.5, there is no need to generate formal **CL12**-proofs anyway: instead, we can use intuition on games and strategies.

Below we write **CLA4!** for the extension of **CLA4** obtained from the latter by taking all true (in the standard model) sentences of the language of **PA** as additional axioms. Similarly for **CLA5!**, **CLA6!**, **CLA7!**.

**Theorem 7.5.3** (adequacy of **CLA4**) Consider an arbitrary sentence $S$ of the language **L**.

 **Constructive soundness**: If $S$ is provable in **CLA4** or **CLA4!**, then $S^\dagger$ has a polynomial time solution. Such a solution can be automatically extracted from an extended **CLA4**- or **CLA4!**-proof of $S$.

 **Extensional completeness**: If $S^\dagger$ has a polynomial time solution, then $S$ is prepresentable in **CLA4**.

 **Intensional completeness**: If $S^\dagger$ has a polynomial time solution, then $S$ is provable in **CLA4!**.

**Theorem 7.5.4** The same adequacy theorem holds for **CLA5** (resp. **CLA6**, resp. **CLA7**), only with "polynomial space" (resp. "elementary recursive time (=space)", resp. "primitive recursive time (=space)") instead of "polynomial time".

 **CLA4** was introduced and proven adequate in [Jap11c]. The same for **CLA5**, **CLA6** and **CLA7** was done in [Jap16a].

## 7.6 Clarithmetics for provable computability

 **Theory CLA8** is obtained from the earlier seen CLA7 by taking the following rule of *Finite Search* as an additional rule of inference, where $p(x)$ is any elementary formula:

$$\frac{p(x) \sqcup \neg p(x) \qquad \exists x\, p(x)}{\sqcup x\, p(x)}$$

 A justification for the above rule is that, if we know how to decide the predicate $p(x)$ (left premise), and we also know that the predicate is true of at least one number (right premise), then we can apply the decision procedure to $p(0), p(1), p(2),\ldots$ until a number $n$ is hit such that the procedure finds $p(n)$ true, after which the conclusion $\sqcup x\, p(x)$ can be solved by choosing $n$ for $x$ in it.

**Theory CLA9** is obtained from CLA7 by taking the following rule of *Infinite Search* as an additional rule of inference, where $p(x)$ is any elementary formula:

$$\frac{p(x) \sqcup \neg p(x)}{\exists x\, p(x) \to \sqcup x\, p(x)}$$

 Note that this rule merely "modifies" Finite Search by changing the status of $\exists x\, p(x)$ from being a premise of the rule to being an antecedent of the conclusion. A justification for Infinite Search is that, if we know how to decide the predicate *p(x)*, then we can apply the decision procedure to *p*(0), *p*(1), *p*(2), ... until (if and when) a number *n* is hit such that the procedure finds *p(n)* true, after which the conclusion can be solved by choosing *n* for *x* in its consequent. Note that, unlike the earlier-outlined strategy for Finite Search, the present strategy may look for *n* forever, and thus never make a move. This, however, only happens when ∃*x p(x)* is false, in which case the conclusion is automatically won.

**Theory CLA10** is obtained from [CLA7](#) by adding to it the above rule of [Infinite Search](#) and the following rule of *Constructivization*, where *q(x)* is any [elementary formula](#) containing no free variables other than *x*:

$$\frac{\exists x \, q(x)}{\sqcup x \, q(x)}$$

The admissibility of this rule, simply allowing us to change ∃*x* to ⊔*x*, is obvious in view of the restriction that ∃*x p(x)* is a sentence (as *p(x)* contains no free variables other than *x*). Indeed, if an *x* satisfying *p(x)* exists, then it can as well be "computed" (generated), even if we do not know what particular machine "computes" it. Note the non-constructive character of this justification.

The three theories **CLA8-CLA10** are adequate with respect to the three natural concepts of computability explained below: **PA**-provably recursive time (=space) computability, constructively **PA**-provable computability, and (simply) **PA**-provable computability.

A natural extreme beyond primitive recursive time is **PA**-*provably recursive time*. That means considering **PA**-provably recursive functions instead of primitive recursive functions as time complexity bounds for computational problems. It can be easily seen to be equivalent to **PA**-provably recursive space, just like there is no difference between time and space at the level of elementary recursive or primitive recursive functions. Theory [CLA8](#) turns out to be adequate with respect to **PA**-provably recursive time computability in the [same sense](#) as **CLA7** is (constructively) sound and complete with respect to primitive recursive time computability.

Not all computable problems have recursive (let alone provably so) time complexity bounds though. In other words, unlike the situation in traditional theory of computation, in our case not all computable problems are *recursive time computable*. An example is ⊓*x*(∃*y p(x,y)* → ⊔*y p(x,y)*), where *p(x,y)* is a decidable binary predicate such that the unary predicate ∃*y p(x,y)* is undecidable (for instance, *p(x,y)* means "Turing machine *x* halts within *y* steps"). The problem ⊓*x*(∃*y p(x,y)* → ⊔*y p(x,y)*) is solved by the following effective strategy: Wait till Environment chooses a value *m* for *x*. After that, for *n*=0,1,2,..., figure out whether *p(m,n)* is true. If and when you find an *n* such that *p(m,n)* is true, choose *n* for *y* in the consequent and [retire](#). On the other hand, if there was a recursive bound *t* for the time complexity of a solution *M* of ⊓*x*(∃*y p(x,y)* → ⊔*y p(x,y)*), then the following would be a decision procedure for (the undecidable) ∃*y p(x,y)*: Given an input *m* (in the role of *x*), run *M* for *t*(|*m*|+1) steps in the scenario where Environment chooses *m* for *x* at the very beginning of the play of ⊓*x*( ∃*y p(x,y)* → ⊔*y p(x,y)*), and does not make any further moves. If, during this time, *M* chooses a number *n* for *y* in the consequent such that *p(m,n)* is true, accept; otherwise reject. So, a next natural step on the road of constructing incrementally strong clarithmetical theories for incrementally weak concepts of computability is to go beyond **PA**-provably recursive time computability and consider the weaker concept of *constructively* **PA**-*provable computability* of (the problem represented by) a sentence *X*. The latter means existence of a machine *M* such that **PA** proves that *M* computes *X*, even if the running time of *M* is not bounded by any recursive function. Theory [CLA9](#) turns out to be sound and complete with respect to this sort of computability, in the sense that a sentence *X* is provable in **CLA9** if and only if *X* is constructively **PA**-provably computable.

An even weaker concept of (simply) **PA**-*provable computability* is obtained from that of constructively **PA**-provable computability by dropping the "constructiveness" condition. Namely, **PA**-provable computability of a sentence *X* means that **PA** proves that a machine *M* solving *X* exists, yet without necessarily being able to prove "*M* solves *X*" for any particular machine *M*. An example of a sentence that is **PA**-provably computable yet not constructively so is *S* ⊔ ¬*S*, where *S* is an [elementary sentence](#) such that **PA** proves neither *S* nor ¬*S*, such as, for instance, Gödel's sentence. Let $L$ be a machine that chooses the left disjunct of $S\sqcup\neg S$ and [retires](). Similarly, let $R$ be a machine that chooses the right disjunct and retires. One of these two machines is a solution of $S\sqcup\neg S$, and, of course, **PA** "knows" this. Yet, **PA** does not "know" which one of them is a solution (otherwise either $S$ or $\neg S$ would be provable); nor does it have a similar sort of "knowledge" for any other particular machine. System **CLA10** turns out to be sound and complete with respect to the present sort of computability, in the sense that a sentence $X$ is provable in **CLA10** if and only if $X$ is **PA**-provably computable.

**CLA8, CLA9** and **CLA10** were introduced and proven adequate in [[Jap14]]().

## 7.7 Tunable clarithmetic

Among the pterms/functions that we use in this section is $(x)_y$, standing for the $y$th least significant bit of the binary representation of $x$, i.e. the $y$th bit from the right, where the bit count starts from 0 rather than 1. When $y \geq |x|$, "the $y$th least significant bit of the binary representation of $x$", by convention, is 0. Another abbreviation is *Bit*, defined by $Bit(y,x) =_{\text{def}} (x)_y = 1$.

By a ***bound*** we shall mean a [pterm]() $p(x_1,\ldots,x_n)$ with all free variables as shown --- which may as well be written simply as $p(\mathbf{x})$ or $p$ --- satisfying the monotonicity condition $\forall(x_1 \leq y_1 \wedge \ldots \wedge x_n \leq y_n \rightarrow p(x_1,\ldots,x_n) \leq p(y_1,\ldots,y_n))$. A ***boundclass*** means a set of bounds closed under variable renaming. A ***boundclass triple*** is a triple $R = (R_{amplitude}, R_{space}, R_{time})$ of boundclasses.

Where $p$ is a [pterm]() and $F$ is a formula, we use the abbreviation $\sqcap x \leq p F$ for $\sqcap x(x \leq p \rightarrow F)$, $\sqcup x \leq p F$ for $\sqcup x(x \leq p \wedge F)$, $\sqcap |x| \leq p F$ for $\sqcap x(|x| \leq p \rightarrow F)$, and $\sqcup |x| \leq p F$ for $\sqcup x(|x| \leq p \wedge F)$. Similarly for the blind quantifiers, and similarly for $<$ instead of $\leq$.

Let $F$ be a formula and $B$ a boundclass. We say that $F$ is $B$-***bounded*** iff every $\sqcap$-subformula (resp. $\sqcup$-subformula) of $F$ has the form $\sqcap z \leq b|\mathbf{s}| H$ (resp. $\sqcup z \leq b|\mathbf{s}| H$), where $z$ and all items of the tuple $\mathbf{s}$ are pairwise distinct variables not bound by $\forall$ or $\exists$ in $F$, and $b|\mathbf{s}|$ is a bound from $B$.

Every [boundclass triple]() $R$ and set $A$ of sentences induces the ***theory*** $\mathbf{CLA11}^R_A$ that we deductively define as follows.

The extra-Peano ***axioms*** of $\mathbf{CLA11}^R_A$, with $x$ and $y$ below being arbitrary two distinct variables, are:

- $\sqcap x \sqcup y(y=x')$
- $\sqcap x \sqcup y(y=|x|)$
- $\sqcap x \sqcap y(Bit(y,x) \sqcup \neg Bit(y,x))$
- All sentences of $A$.

In the naturally occurring cases, the above set $A$ will typically be either empty or consisting of all true sentences of **PA**.

$\mathbf{CLA11}^R_A$ has two nonlogical ***rules*** of inference: $R$-Induction and $R$-Comprehension. These rules are meant to deal exclusively with sentences, and correspondingly, in our schematic representations of them below, as in the earlier cases, each premise or conclusion $H$ should be understood as its $\sqcap$-closure $\sqcap H$, with the prefix $\sqcap$ dropped merely for readability.

The rule of *R*-***Induction*** is

$$\frac{F(0) \qquad F(x) \to F(x')}{x \leq \boldsymbol{b}|\boldsymbol{s}| \to F(x)}$$

where $x$ and all items of the tuple $\boldsymbol{s}$ are pairwise distinct variables, $F(x)$ is an $R_{space}$-bounded formula, and $\boldsymbol{b}(\boldsymbol{s})$ is a bound from $R_{time}$.

The rule of *R*-***Comprehension*** is

$$\frac{p(y) \sqcup \neg p(y)}{\sqcup |x| \leq \boldsymbol{b}|\boldsymbol{s}| \forall y < \boldsymbol{b}|\boldsymbol{s}| \big(Bit(y,x) \leftrightarrow p(y)\big)}$$

where $x,y$ and all items of the tuple $\boldsymbol{s}$ are pairwise distinct variables, $p(y)$ is an elementary formula not containing $x$, and $\boldsymbol{b}(\boldsymbol{s})$ is a bound from $R_{amplitude}$.

Let $B$ be a set of bounds. We define the **linear closure** of $B$ as the smallest boundclass $C$ such that $B \subseteq C$, $0 \in C$ and, whenever bounds $\boldsymbol{b}$ and $\boldsymbol{c}$ are in $C$, so are the bounds $\boldsymbol{b}'$ and $\boldsymbol{b}+\boldsymbol{c}$. The **polynomial closure** of $B$ is defined the same way but with the additional condition that, whenever $\boldsymbol{b}$ and $\boldsymbol{c}$ are in $C$, so is $\boldsymbol{b} \times \boldsymbol{c}$. Correspondingly, we say that $B$ is **linearly closed** (resp. **polynomially closed**) iff $B$ is the same as its linear (resp. polynomial) closure.

Let $\boldsymbol{b}=\boldsymbol{b}(x_1,\ldots,x_m)$ and $\boldsymbol{c}=\boldsymbol{c}(y_1,\ldots,y_n)$ be two functions which depend exactly on the corresponding tuples of displayed variables, or pterms understood as such functions. We write $\boldsymbol{b} \leq \boldsymbol{c}$ iff $m=n$ and $\boldsymbol{b}(a_1,\ldots,a_m) \leq \boldsymbol{c}(a_1,\ldots,a_m)$ is true for all natural numbers $a_1,\ldots,a_m$. Next, where $B$ and $C$ are boundclasses, we write $\boldsymbol{b} \leq C$ to mean that $\boldsymbol{b} \leq \boldsymbol{c}$ for some $\boldsymbol{c} \in C$, and write $B \leq C$ to mean that $\boldsymbol{b} \leq C$ for all $\boldsymbol{b} \in B$. Finally, where $a_1,s_1,t_1,a_2,s_2,t_2$ are bounds, we write $(a_1,s_1,t_1) \leq (a_2,s_2,t_2)$ to mean that $a_1 \leq a_2$, $s_1 \leq s_2$ and $t_1 \leq t_2$.

**Definition 7.7.1** We say that a boundclass triple $R$ is ***regular*** iff the following conditions are satisfied:

1. For every bound $\boldsymbol{b}(\boldsymbol{s}) \in R_{amplitude} \cup R_{space} \cup R_{time}$ and any (=some) variable $z$ not occurring in $\boldsymbol{b}(\boldsymbol{s})$, the game $\sqcap\sqcup(z=\boldsymbol{b}|\boldsymbol{s}|)$ has an $R$ tricomplexity solution, and such a solution can be effectively constructed from $\boldsymbol{b}(\boldsymbol{s})$.
2. $R_{amplitude}$ is *at least linear*, $R_{space}$ is *at least logarithmic*, and $R_{time}$ is *at least polynomial*. This is in the sense that, for any variable $x$, we have $x \leq R_{amplitude}$, $|x| \leq R_{space}$ and $x, x^2, x^3, \ldots \leq R_{time}$.
3. All three components of $R$ are linearly closed and, in addition, $R_{time}$ is also polynomially closed.
4. For each component $B \in \{R_{amplitude}, R_{space}, R_{amplitude}\}$ of $R$, whenever $\boldsymbol{b}(x_1,\ldots,x_n)$ is a bound in $B$ and $c_1,\ldots,c_n \in R_{amplitude} \cup R_{space}$, we have $\boldsymbol{b}(c_1,\ldots,c_n) \leq B$.
5. For every triple $(a_1(\mathbf{x}), s_1(\mathbf{x}), t_1(\mathbf{x}))$ of bounds in $R_{amplitude} \times R_{space} \times R_{time}$ there is a triple $(a_2(\mathbf{x}), s_2(\mathbf{x}), t_2(\mathbf{x}))$ in $R_{amplitude} \times R_{space} \times R_{time}$ such that $(a_1(\mathbf{x}), s_1(\mathbf{x}), t_1(\mathbf{x})) \leq (a_2(\mathbf{x}), s_2(\mathbf{x}), t_2(\mathbf{x}))$ and $|t_2(\mathbf{x})| \leq s_2(\mathbf{x}) \leq a_2(\mathbf{x}) \leq t_2(\mathbf{x})$.

**Definition 7.7.2** We say that a theory $\mathbf{CLA11}^R_A$ is ***regular*** iff the boundclass triple $R$ is regular and, in addition, the following conditions are satisfied:

1. Every sentence of $A$ has an $R$ tricomplexity solution. Here, if $A$ is infinite, we additionally require that there is an effective procedure that returns an $R$ tricomplexity solution for each sentence of $A$.
2. For every bound $\boldsymbol{b}(\mathbf{x})$ from $R_{amplitude} \cup R_{space} \cup R_{time}$ and every (=some) variable $z$ not occurring in $\boldsymbol{b}(\mathbf{x})$, $\mathbf{CLA11}^R_A$ proves $\sqcap\sqcup(z=\boldsymbol{b}|\mathbf{x}|)$.

We agree that, whenever $A$ is a set of sentences, $A!$ is an abbreviation of $A\cup\mathbf{Th(N)}$, where $\mathbf{Th(N)}$ is the set of all true (in the standard model) sentences of **PA**.

**Theorem 7.7.3** (**Adequacy** of **CLA11**) Assume a theory $\mathbf{CLA11}^R_A$ is regular, and consider an arbitrary sentence $S$ of its language. Recall that $^\dagger$ is the standard interpretation.

1. **Constructive soundness:** If $S$ is provable in $\mathbf{CLA11}^R_A$ or $\mathbf{CLA11}^R_{A!}$, then $S^\dagger$ has an $R$ tricomplexity solution. Such a solution can be automatically extracted from an extended $\mathbf{CLA11}^R_A$- or $\mathbf{CLA11}^R_{A!}$-proof of $S$.
2. **Extensional completeness:** If $S^\dagger$ has an $R$ tricomplexity solution, then $S$ is prepresentable in $\mathbf{CLA11}^R_A$.
3. **Intensional completeness:** If $S^\dagger$ has an $R$ tricomplexity solution, then $S$ is provable in $\mathbf{CLA11}^R_{A!}$.

Below we are going to see an infinite yet incomplete series of natural theories that are regular and thus adequate in the sense of the above theorem. All these theories look like $\mathbf{CLA11}^R_\emptyset$, with the subscript $\emptyset$ indicating that the $A$ (supplemental axioms) parameter is empty.

Given a set $S$ of bounds, by $S^\heartsuit$ (resp. $S^\spadesuit$) we shall denote the linear closure (resp. polynomial closure) of $S$. We define the series $B_1^1, B_1^2, B_1^3, \ldots, B_2, B_3, B_4, B_5, B_6, B_7, B_8$ of boundclasses as follows:

1. $B_1^1 = \{|x|\}^\heartsuit$ (*logarithmic* boundclass); $B_1^2 = \{|x|^2\}^\heartsuit$; $B_1^3 = \{|x|^3\}^\heartsuit$; ... ;
2. $B_2 = \{|x|\}^\spadesuit$ (*polylogarithmic* boundclass);
3. $B_3 = \{x\}^\heartsuit$ (*linear* boundclass);
4. $B_4 = \{x\times|x|, x\times|x|^2, x\times|x|^3,\ldots\}^\heartsuit$ (*quasilinear* boundclass);
5. $B_5 = \{x\}^\spadesuit$ (*polynomial* boundclass);
6. $B_6 = \{2^{|x|}, 2^{|x|^2}, 2^{|x|^3},\ldots\}^\spadesuit$ (*quasipolynomial* boundclass);
7. $B_7 = \{2^x\}^\spadesuit$ (*exponential-with-linear-exponent* boundclass);
8. $B_8 = \{2^x, 2^{x^2}, 2^{x^3},\ldots\}^\spadesuit$ (*exponential-with-polynomial-exponent* boundclass).

**Fact 7.7.4** For any boundclass triple $R$ listed below, the theory $\mathbf{CLA11}^R$ is regular:

$(B_3,B_1^1,B_5)$;  $(B_3,B_1^2,B_5)$;  $(B_3,B_1^3,B_5)$;  ...;  $(B_3,B_2,B_5)$;  $(B_3,B_2,B_6)$;  $(B_3,B_2,B_7)$;  $(B_3,B_3,B_5)$;  $(B_3,B_3,B_6)$;  $(B_3,B_3,B_7)$;  $(B_4,B_1^1,B_5)$;  $(B_4,B_1^2,B_5)$;  $(B_4,B_1^3,B_5)$;  ...;  $(B_4,B_2,B_5)$;  $(B_4,B_2,B_6)$;  $(B_4,B_4,B_5)$;  $(B_4,B_4,B_6)$;  $(B_4,B_4,B_7)$;  $(B_5,B_1^1,B_5)$;  $(B_5,B_1^2,B_5)$;  $(B_5,B_1^3,B_5)$;  ...;  $(B_5,B_2,B_5)$;  $(B_5,B_2,B_6)$;  $(B_5,B_5,B_5)$;  $(B_5,B_5,B_6)$;  $(B_5,B_5,B_7)$;  $(B_5,B_5,B_8)$.

**CLA11** was introduced and the above results were proven in [Jap16b, Jap16c].

# 8 CoL-based knowledgebase and resourcebase systems

The reason for the failure of ⊓$x$($p(x)$ ⊔ ¬$p(x)$) as a computability-theoretic principle is that the problem represented by this formula may have no effective solution --- that is, the predicate $p(x)$ may be undecidable. The reason why this principle would fail in the context of knowledgebase systems, however, is much simpler. A knowledgebase system may fail to solve the problem ⊓$x$(*Married*($x$) ⊔ ¬*Married*($x$)) not because the latter has no effective solution (of course it has one), but because the system simply lacks sufficient knowledge to tell any given person's marital status. On the other hand, any system would be able to "solve" the problem ∀$x$(*Married*($x$) ∨ ¬*Married*($x$)) as this is an automatically won elementary game so that there is nothing to solve at all. Similarly, while ∀$y$∃$x$*Father*($x,y$) is an automatically solved elementary problem expressing the trivial knowledge that every person has a father, ability to solve the problem ⊓$y$⊔$x$*Father*($x,y$) implies the nontrivial knowledge of everyone's actual father. Obviously the knowledge expressed by $A$⊔$B$ or ⊔$xA(x)$ is generally stronger than the knowledge expressed by $A$∨$B$ or ∃$xA(x)$, yet the formalism of classical logic fails to capture this difference --- the difference whose relevance hardly requires any explanation. The traditional approaches to knowledgebase systems try to mend this gap by augmenting the language of classical logic with special epistemic constructs, such as the modal "know that" operator □, after which probably □$A$∨□$B$ would be suggested as a translation for $A$⊔$B$ and ∀$y$∃$x$□$A(x,y)$ for ⊓$y$⊔$xA(x,y)$. Leaving it for the philosophers to argue whether, say, ∀$y$∃$x$□$A(x,y)$ really expresses the constructive meaning of ⊓$y$⊔$xA(x,y)$, and forgetting that epistemic constructs often yield unnecessary and unpleasant complications such as messiness and non-semidecidability of the resulting logics, some of the major issues still do not seem to be taken care of. Most of the actual knowledgebase and information systems are interactive, and what we really need is a logic of *interaction* rather than just a logic of knowledge. Furthermore, a knowledgebase logic needs to be *resource-conscious*. The informational resource expressed by ⊓$x$(*Pregnant*($x$) ⊔ ¬*Pregnant*($x$)) is not as strong as the one expressed by ⊓$x$(*Pregnant*($x$) ⊔ ¬*Pregnant*($x$)) ∧ ⊓$x$(*Pregnant*($x$) ⊔ ¬*Pregnant*($x$)): the former implies the resource provider's commitment to tell only one (even though an arbitrary one) woman's pregnancy status, while the latter is about doing the same for any two Women. The former is the informational resource provided by a single disposable pregnancy test device, while the latter is the resource provided by two such devices. Neither classical logic nor its standard epistemic extensions have the ability to account for such differences. But CoL promises to be adequate. It *is* a logic of interaction, it *is* resource-conscious, and it *does* capture the relevant differences between truth and actual ability to find/compute/know truth.

When CoL is used as a logic of knowledgebases, its formulas represent interactive queries. A formula whose main operator is ⊔ or ⊔ can be understood as a question asked by the user, and a formula whose main operator is ⊓ or ⊓ as a question asked by the system. Consider the problem ⊓$x$⊔$y$*Has*($x,y$), where *Has*($x,y$) means "patient $x$ has disease $y$" (with *Healthy* counting as one of the possible "diseases"). This formula is the following question asked by the system: "Who do you want me to diagnose?" The user's response can be "Laura". This move brings the game down to ⊔$y$*Has*(*Laura,y*). This is now a question asked by the user: "What does Laura have?". The system's response can be "flu", taking us to the terminal position *Has*(*Laura,Flu*). The system has been successful iff Laura really has the flu.

Successfully solving the above problem ⊓$x$⊔$y$*Has*($x,y$) requires having all relevant medical information for each possible patient, which in a real diagnostic system would hardly be the case. Most likely, such a system, after receiving a request to diagnose $x$, would make counterqueries regarding $x$'s symptoms, blood pressure, test results, age, gender, etc., so that the query that the system will be solving would have a higher degree of interactivity than the two-step query ⊓$x$⊔$y$*Has*($x,y$) does, with questions and counterquestions interspersed in some complex fashion. Here is when other CoL operations come into play. Negation turns queries into counterqueries; parallel operations generate combined queries; recurrence operations allow repeated queries; implications and rimplications act as various sorts of query reduction operations; etc. Here we are expanding our example. Let *Sympt*($x,s$) mean "patient $x$ has (set of) symptoms $s$", and *Pos*($x,t$) mean "patient $x$ tests positive for test $t$". Imagine a diagnostic system which can diagnose any particular patient $x$, but needs some additional information. Specifically, it needs to know $x$'s symptoms; plus, the system may require to have $x$ taken a test $t$ which it selects dynamically in the course of a dialogue with the user depending on what responses it received.

The interactive task/query that such a system is performing/solving can then be expressed by the following formula, which we shall call *Diagn* for further references:

$$Diagn \equiv \sqcap x\bigl(\sqcup s Sympt(x,s) \wedge \sqcap t(Pos(x,t) \sqcup \neg Pos(x,t)) \to \sqcup y Has(x,y)\bigr).$$

A possible scenario of playing the above game is the following. At the beginning, the system waits until the user specifies a patient *x* to be diagnosed. We can think of this stage as systems's requesting the user to select a particular (value of) *x*, remembering that the presence of $\sqcap x$ automatically implies such a request. After a patient *x* --- say *x=X* --- is selected, the system requests to specify *X*'s symptoms. Notice that our game rules make the system successful if the user fails to provide this information, i.e. specify a (the true) value for *s* in $\sqcup s Sympt(X,s)$. Once a response --- say, *s=S* --- is received, the system selects a test *t=T* and asks the user to perform it on *X*, i.e. to choose the true disjunct of $Pos(X,T) \sqcup \neg Pos(X,T)$. Finally, provided that the user gave correct answers to all counterqueries (and if not, the user has lost), the system makes a diagnostic decision, i.e. specifies a value *Y* for *y* in $\sqcup y Has(X,y)$ for which *Has(X,Y)* is true.

The presence of a single "copy" of $\sqcap t(Pos(x,t) \sqcup \neg Pos(x,t))$ in the antecedent of *Diagn* means that the system may request testing a given patient only once. If up to *n* tests were potentially needed instead, this would be expressed by taking the ∧-conjunction of *n* identical conjuncts $\sqcap t(Pos(x,t) \sqcup \neg Pos(x,t))$. And if the system potentially needed an unbounded number of tests, then we would write $\wedge \sqcap t(Pos(x,t) \sqcup \neg Pos(x,t))$, thus further weakening *Diagnoser*: a system that performs this weakened task is not as good as the one performing the original *Diagn*, as it requires stronger external (user-provided) informational resources. Replacing the main quantifier $\sqcap x$ by $\forall x$, on the other hand, would strengthen *Diagn*, signifying the system's ability to diagnose a patent purely on the basis of his/her symptoms and test result without knowing who the patient really is. However, if in its diagnostic decisions the system uses some additional information on patients such their medical histories stored in its knowledgebase and hence needs to know the patient's identity, $\sqcap x$ cannot be upgraded to $\forall x$. Replacing $\sqcap x$ by $\wedge x$ would be a yet another way to strengthen *Diagn*, signifying the system's ability to diagnose all patients rather than any particular one; obviously effects of at least the same strength would be achieved by just prefixing *Diagn* with ⋏ or ⋎.

As we've just mentioned system's **knowledgebase**, let us make clear what it means. Formally, this is a finite ∧-conjunction *KB* of formulas, which can also be thought of as the (multi)set of its conjuncts. We call the elements of this set the **internal informational resources** of the system. Intuitively, *KB* represents all of the nonlogical knowledge available to the system, so that (with a fixed built-in logic in mind) the strength of the former determines the query-solving power of the latter. Conceptually, however, we do not think of *KB* as a part of the system properly. The latter is just "pure", logic-based problem-solving software of universal utility that initially comes to the user without any nonlogical knowledge whatsoever. Indeed, built-in nonlogical knowledge would make it no longer universally applicable: Laura can have the flu in the world of one potential user while be healthy in the world of another user, and $\forall x \forall y(x \times y = y \times x)$ can be false to a user who understands × as Cartesian rather than number-theoretic product. It is the user who selects and maintains *KB* for the system, putting into it all informational resources that (s)he believes are relevant, correct and maintainable. Think of the formalism of CoL as a highly declarative programming language, and the process of creating *KB* as programming in it.

The knowledgebase *KB* of the system may include atomic elementary formulas expressing factual knowledge, such as *Married(Laura)*, or non-atomic elementary formulas expressing general knowledge, such as $\forall x(\exists y Father(x,y) \to Male(x))$ and $\forall x \forall y(x \times (y+1) = (x \times y)+x)$; it can also include nonclassical formulas such as $\text{⋎} \sqcap x(Female(x) \sqcup Male(x))$, expressing potential knowledge of everyone's gender, or $\text{⋎} \sqcap x \sqcup y(x^2=y)$, expressing ability to repeatedly compute the square function, or something more complex and more interactive, such as the earlier seen formula *Diagn*. With each resource $R \in KB$ is associated (if not physically, at least conceptually) its **provider** --- an agent that solves the query *R* for the system, i.e. plays the game *R* against the system. Physically the provider could be a computer program allocated to the system, or a network server having the system as a client, or another knowledgebase system to which the system has querying access, or even human personnel servicing the system. E.g., the provider for as $\text{⋎} \sqcap x \sqcup y Bloodpressure(x,y)$ would probably be a team of nurses repeatedly performing the task of measuring the blood pressure of a patient specified by the system and reporting the outcome back to the system. Again, we do not think of providers as a part of the system itself. The latter only sees *what* resources are available to it, without knowing or caring about *how* the corresponding providers do their job; furthermore, the system does not even care *whether* the providers really do their job right. The system's responsibility is only to correctly solve queries for the user *as long as* none of the providers fails to do their job. Indeed, if the system misdiagnoses a patient because a nurse-provider gave it wrong information about that patient's blood pressure, the hospital (ultimate user) is unlikely to fire the system and demand refund from its vendor; more likely, it would fire the nurse. Of course, when $R$ is elementary, the provider has nothing to do, and its successfully playing $R$ against the system simply means that $R$ is true. Note that in the picture that we have just presented, the system plays each game $R \in KB$ in the role of ⊥, so that, from the system's perspective, the game that it plays against the provider of $R$ is ¬$R$ rather than $R$.

The most typical internal informational resources, such as factual knowledge or queries solved by computer programs, can be reused an arbitrary number of times and with unlimited branching capabilities, i.e. in the strong sense captured by ♭, and thus they would be prefixed with a ♭ as we did with ⊓$x$(*Female*($x$) ⊔ *Male*($x$)). There was no point in ♭-prefixing *Married*(*Laura*), ∀$x$(∃$y$*Father*($x$,$y$) → *Male*($x$)) or ∀$x$∀$y$($x$×($y$+1) = ($x$×$y$)+$x$) because every elementary game $A$ is equivalent to ♭$A$ and hence remains "recyclable" even without recurrence operators. As noted in [Section 3.8](#) for the corresponding rimplications, there is no difference between ♭ and ⩑ as long as "simple" resources are concerned such as such ♭⊓$x$⊔$y$($x^2$=$y$). However, in some cases --- say, when a resource with a high degree of interactivity is supported by an unlimited number of independent providers each of which however allows to run only one single "session" --- the weaker operator ⩑ will have to be used instead of ♭. Yet, some of the internal informational resources could be essentially non-reusable. A provider possessing a single piece of litmus paper would be able to support the resource ⊓$x$(*Acid*($x$) ⊔ ¬*Acid*($x$)) but not ♭⊓$x$(*Acid*($x$) ⊔ ¬*Acid*($x$)) and not even ⊓$x$(*Acid*($x$) ⊔ ¬*Acid*($x$)) ∧ ⊓$x$(*Acid*($x$) ⊔ ¬*Acid*($x$)). Most users, however, would try to refrain from including this sort of a resource into *KB*, and rather make it a part (antecedent) of possible queries. Indeed, knowledgebases with non-recyclable resources would tend to weaken from query to query and require more careful maintainance and updates. Whether recyclable or not, all of the resources of *KB* can be used independently and in parallel. This is exactly what allows us to identify *KB* with the ∧-conjunction of its elements.

Assume $KB = R_1 \wedge ... \wedge R_n$, and let us now try to visualize a system solving a query $F$ for the user. The designer would probably select an interface where the user only sees the moves made by the system in $F$, and hence gets the illusion that the system is just playing $F$. But in fact the game that the system is really playing is $KB \rightarrow F$, i.e. ¬$R_1$∨...∨¬$R_n$∨$F$. Indeed, the system is not only interacting with the user in $F$, but --- in parallel --- also with its resource providers against whom, as we already know, it plays ¬$R_1$,...,¬$R_n$. As long as those providers do not fail to do their job, the system loses each of the games ¬$R_1$,...,¬$R_n$. Then our semantics for ∨ implies that the system wins its play over the "big game" ¬$R_1$∨...∨¬$R_n$∨$F$ if and only if it wins it in the $F$ component, i.e. successfully solves the query $F$.

Thus, the system's ability to solve a query $F$ reduces to its ability to generate a solution for $KB \rightarrow F$, i.e. generate a reduction of $F$ to $KB$. What would give the system such an ability is built-in knowledge of CoL --- in particular, a [***uniform-constructively sound*** **axiomatization**](#) of it. According to the uniform-constructive soundness property, it would be sufficient for the system to find a proof of $KB \rightarrow F$, which would allow it to automatically construct a machine $M$ and then run it on $KB \rightarrow F$ with a guaranteed success.

Notice that it is uniform-constructive soundness rather than simple soundness of the built-in (axiomatization of the) logic that allows the knowledgebase system to function. By *simple soundness* here we mean that every provable formula is [nonlogically valid](#). This is not sufficient for two reasons.

One reason is that nonlogical validity of a formula $E$ only implies that, for every interpretation *, a solution for the problem $E$* exists. It may be the case, however, that different interpretations require different solutions, so that choosing the right solution requires knowledge of the actual interpretation, i.e. the *meaning*, of the atoms of $E$. Our assumption is that the system has no nonlogical knowledge, which, in more precise terms, means nothing but that it has no knowledge of the interpretation *. Thus, a solution that the system generates for $E^*$ should be successful for any possible interpretation *. In other words, it should be a uniform solution for $E$. This is where uniform-constructive soundness of the underlying logic becomes critical, by which every provable formula is not only nonlogically, but also logically valid.

The other reason why simple soundness of the built-in logic would not be sufficient for a knowledgebase system to function --- even if every provable formula was known to be logically valid --- is the following. With simple soundness, after finding a proof of $E$, even though the system would know that a solution for $E^*$ exists, it might have no way to actually find such a solution. On the other hand, uniform-constructive soundness guarantees that a (uniform) solution for every provable formula not only exists, but can be effectively extracted from a proof.

As for completeness of the built-in logic, unlike uniform-constructive soundness, it is a desirable but not necessary condition. So far complete axiomatizations have been found only for various limited fragments of the language of CoL. We hope that the future will bring completeness results for more and more expressive fragments as well. But even if not, we can still certainly succeed in finding ever stronger axiomatizations that are uniform-constructively sound even if not necessarily complete. It should be remembered that, when it comes to practical applications in the proper sense, the logic that will be used is likely to be far from complete anyway. For example, the popular classical-logic-based systems and programming languages are incomplete, and the reason is not that a complete axiomatization for classical logic is not known, but rather the unfortunate fact of life that often efficiency only comes at the expense of completeness.

But even CL4, despite the absence of recurrence operators in it, is already very powerful. Why don't we see a simple example to get the taste of it as a query-solving logic. Let, as before, $Acid(x)$ mean "solution $x$ contains acid", and $Red(x)$ mean "litmus paper turns red in solution $x$". Assume that the knowledgebase $KB$ of a CL4-based system contains $\forall x(Red(x) \leftrightarrow Acid(x))$ and $\sqcap x(Red(x) \sqcup \neg Red(x))$, accounting for knowledge of the fact that a solution contains acid iff the litmus paper turns red in it, and for availability of a provider who possesses a piece of litmus paper that it can dip into any solution and report the paper's color to the system. Then the system can solve the acidity query $\sqcap x(Acid(x) \sqcup \neg Acid(x))$. This follows from the fact, left as an exercise for the reader to verify, that CL4 proves $KB \rightarrow \sqcap x(Acid(x) \sqcup \neg Acid(x))$.

The context of knowledgebase systems can be further extended to systems for planning and action. Roughly, the formal semantics in such applications remains the same, and what changes is only the underlying philosophical assumption that the truth values of predicates and propositions are fixed or predetermined. Rather, those values in CoL-based planning systems will be viewed as something that interacting agents may be able to manage. That is, predicates or propositions there stand for *tasks* rather than *facts*. E.g., $Pregnant(Laura)$ --- or, perhaps, $Impregnate(Laura)$ instead --- can be seen as having no predetermined truth value, with Laura or her mate being in control of whether to make it true or not. And the nonelementary formula $\sqcap xHit(x)$ describes the task of hitting any one target $x$ selected by the environment/commander/user. Note how naturally resource-consciousness arises here: while $\sqcap xHit(x)$ is a task accomplishable with one ballistic missile, the stronger task $\sqcap xHit(x) \wedge \sqcap xHit(x)$ would require two missiles instead. All of the other operators of CoL, too, have natural interpretations as operations on physical (as opposed to just informational) tasks, with $\rightarrow$ acting as a task reduction operation. To get a feel of this, let us look at the task

*Give me a wooden stake $\sqcap$ Give me a silver bullet $\rightarrow$ Destroy the vampire $\sqcap$ Kill the werewolf.*

This is a task accomplishable by an agent who has a mallet and a gun as well as sufficient time, energy and bravery to chase and eliminate any one (but not both) of the two monsters, and only needs a wooden stake and/or a silver bullet to complete his noble mission. Then the story told by the legal run ⟨1.1, 0.1⟩ of the above game is that the environment asked the agent to kill the werewolf, to which the agent replied by the counterrequest to give him a silver bullet. The task will be considered eventually accomplished by the agent iff he indeed killed the werewolf as long as a silver bullet was indeed given to him.

# 9 Literature

## 9.1 Selected papers on CoL by Japaridze

- [Jap11a] G.Japaridze. *Toggling operators in computability logic*. **Theoretical Computer Science** 412 (2011), pp. 971-1004. Preprint

- [Jap11b] G.Japaridze. *From formulas to cirquents in computability logic*. **Logical Methods in Computer Science** 7 (2011), Issue 1, Paper 1, pp. 1-55.

- [Jap11c] G.Japaridze. *Introduction to clarithmetic I*. **Information and Computation** 209 (2011), pp. 1312-1354. Preprint

- [Jap12a] G.Japaridze. *Separating the basic logics of the basic recurrences*. **Annals of Pure and Applied Logic** 163 (2012), pp. 377-389. Preprint

- [Jap12b] G.Japaridze. *A new face of the branching recurrence of computability logic*. **Applied Mathematics Letters** 25 (2012), pp. 1585-1589. Preprint

- [Jap12c] G.Japaridze. *A logical basis for constructive systems*. **Journal of Logic and Computation** 22 (2012), pp.605-642.

- [Jap13a] G.Japaridze. *The taming of recurrences in computability logic through cirquent calculus, Part I*. **Archive for Mathematical Logic** 52 (2013), pp. 173-212. Preprint

- [Jap13b] G.Japaridze. *The taming of recurrences in computability logic through cirquent calculus, Part II*. **Archive for Mathematical Logic** 52 (2013), pp. 213-259. Preprint

- [Jap14] G.Japaridze. *Introduction to clarithmetic III*. **Annals of Pure and Applied Logic** 165 (2014), pp. 241-252. Preprint

- [Jap15] G.Japaridze. *On the system CL12 of computability logic*. **Logical Methods in Computer Science** 11 (2015), Issue 3, paper 1, pp.1-71.

- [Jap16a] G.Japaridze. *Introduction to clarithmetic II*. **Information and Computation** 247 (2016), pp.290-312. Preprint

- [Jap16b] G.Japaridze. *Build your own clarithmetic I: Setup and completeness*. **Logical Methods in Computer Science** 12 (2016), Issue 3, pp. 1-59.

- [Jap16c] G.Japaridze. *Build your own clarithmetic II: Soundness*. **Logical Methods in Computer Science** 12 (2016), Issue 3, pp. 1-59.

## 9.2 Selected (SCI-indexed) papers on CoL by other authors

- [Bau14] M.Bauer. *A PSPACE-complete first order fragment of computability logic*. **ACM Transactions on Computational Logic** 15 (2014), No 1, Article 1, 12 pages.

- [Bau15] M.Bauer. *The computational complexity of propositional cirquent calculus*. **Logical Methods is Computer Science** 11 (2015), Issue 1, Paper 12, pp.1-16.

- [Kwo14] K.Kwon. *Expressing algorithms as concise as possible via computability logic*. **IEICE Transactions on Fundamentals of Electronics, Communications and Computer Sciences**, vol. E97-A (2014), pp.1385-1387.

- [Mez10] I.Mezhirov and N.Vereshchagin. *On abstract resource semantics and computability logic*. **Journal of Computer and Systems Sciences** 76 (2010), pp. 356-372.

- [Qu13] M.Qu, J.Luan, D.Zhu and M.Du. *On the toggling-branching recurrence of computability logic*. **Journal of Computer Science and Technology** 28 (2013), pp. 278-284.

- [Xu12a] W.Xu and S.Liu. *The countable versus uncountable branching recurrences in computability logic*. **Journal of Applied Logic** 10 (2012), pp. 431-446.

- [Xu12b] W.Xu and S.Liu. *Soundness and completeness of the cirquent calculus system CL6 for computability logic*. **Logic Journal of the IGPL** 20 (2012), pp. 317-330.

- [Xu13] W.Xu and S.Liu. *The parallel versus branching recurrences in computability logic*. **Notre Dame Journal of Formal Logic** 54 (2013), pp. 61-78.

- [Xu14] W.Xu. *A propositional system induced by Japaridze's approach to IF logic*. **Logic Journal of the IGPL** 22 (2014), pp. 982-991.

- [Xu16] W.Xu. *A cirquent calculus system with clustering and ranking*. **Journal of Applied Logic** 16 (2016), pp.37-49.

## 9.3  PhD theses, MS theses and externally funded projects on CoL

- M.Qu. **Research on the toggling-branching recurrence of Computability Logic**. PhD Thesis. Shandong University, 2014.
- Y.Zhang. **Time and Space Complexity Analysis for the System CL2 of Computability Logic**.MS Thesis. Shandong University, 2013.
- W.Xu. **On Some Operators and Systems of Computability Logic**. PhD Thesis. Xidian University, 2012.
- W.Xu. **A Study of Cirquent Calculus Systems for Computability Logic**. Research project funded by the National Science Foundation of China (61303030) and the Fundamental Research Funds for the Central Universities of China (K50513700). Xidian University, 2013-2016.
- G.Japaridze. **A Logical Study of Interactive Computational Problems Understood as Games**. Research project funded by the National Science Foundation of US (CCR-0208816). Villanova University, 2002-2006.

## 9.4 Lecture notes, presentations and other links

- A list of open problems in CoL
- LaTeX macros for the operators of CoL
- Course on CoL (taught in 2007 at Xiamen University. Includes lecture notes in PowerPoint)
- Three-hour tour of CoL (PowerPoint slides)
- Four-hour tour of CoL (PowerPoint slides)
- On abstract resource semantics and computabilty logic (video lecture by N.Vereshchagin)
- Proofs as games
- Computability logic-1
- Lambda the Ultimate: Introduction to computability logic
- Lambda the Ultimate: In the beginning was game semantics
- Giorgi Japaridze: Research and Publications

## 9.5 Additional references